\documentclass[12pt,reqno,a4paper]{amsart}
\usepackage{amsmath,amssymb,amsthm,amsfonts}

\numberwithin{equation}{section}
\newtheorem{thm}{Theorem}[section]
\newtheorem{lem}[thm]{Lemma}
\newtheorem{prop}[thm]{Proposition}
{\theoremstyle{definition}
}
\newtheorem{theorem}{Theorem}[section]
\newtheorem{proposition}[theorem]{Proposition}

\newtheorem{lemma}[theorem]{Lemma}

{\theoremstyle{definition}
{
\newtheorem{remark}[theorem]{Remark}

}}

\def\QQ{{\mathbb Q}}
\def\ZZ{{\mathbb Z}}

\newcommand{\cal}{\mathcal}

\newcommand{\A}{{\cal A}}
\newcommand{\BB}{{\cal B}}

\newcommand{\DD}{{\cal D}}

\newcommand{\FF}{{\cal F}}
\newcommand{\GG}{{\cal G}}
\newcommand{\II}{{\cal I}}
\newcommand{\JJ}{{\cal J}}

\newcommand{\LL}{{\cal L}}

\newcommand{\OO}{{\cal O}}
\newcommand{\PP}{{\cal P}}

\newcommand{\RR}{{\cal R}}

\newcommand{\TTT}{{\cal T}}
\newcommand{\UU}{{\cal U}}
\newcommand{\VV}{{\cal V}}

\newcommand{\fG}{{\mathfrak G}}

\newcommand{\fU}{{\mathfrak U}}

\newcommand{\Cc}{{\mathbb{C}}}

\newcommand{\Ee}{{\mathbb{E}}}

\newcommand{\Ii}{{\mathbb{I}}}
\newcommand{\Jj}{{\mathbb{J}}}

\newcommand{\Nn}{{\mathbb{N}}}

\newcommand{\Rr}{{\mathbb{R}}}

\newcommand{\Tt}{{\mathbb{T}}}

\newcommand{\Zz}{{\mathbb{Z}}}

\def\e{\mathrm{e}}
\def\i{\mathrm{i}}

\def\diag{\operatorname{diag}}

\def\Mat{\operatorname{Mat}}
\def\GL{\operatorname{GL}}

\def\SL{\operatorname{SL}}
\def\SO{\operatorname{SO}}

\def\GamG{\Gamma\backslash G}
\def\GLZ{\GL(d,\ZZ)}
\def\SLZ{\SL(d,\ZZ)}
\def\SLR{\SL(d,\Rr)}

\def\veca{{\text{\boldmath$a$}}}
\def\vecb{{\text{\boldmath$b$}}}

\def\veck{{\text{\boldmath$k$}}}

\def\vecm{{\text{\boldmath$m$}}}

\def\vecp{{\text{\boldmath$p$}}}

\def\vect{{\text{\boldmath$t$}}}
\def\vecu{{\text{\boldmath$u$}}}
\def\vecv{{\text{\boldmath$v$}}}

\def\vecx{{\text{\boldmath$x$}}}
\def\vecy{{\text{\boldmath$y$}}}
\def\vecz{{\text{\boldmath$z$}}}
\def\vecalf{{\text{\boldmath$\alpha$}}}
\def\vecbeta{{\text{\boldmath$\beta$}}}

\def\vecnu{{\text{\boldmath$\nu$}}}

\def\vecomega{{\text{\boldmath$\omega$}}}

\def\vecxi{{\text{\boldmath$\xi$}}}

\def\vecnull{{\text{\boldmath$0$}}}

\def\scrC{{\mathcal C}}
\def\scrF{{\mathcal F}}

\def\scrS{{\mathcal S}}

\newcommand{\norm}[1]{\left\| #1\right\|}
\newcommand{\trans} {\,^\top\!}

\newcommand{\id}  {\operatorname{Id}}
\newcommand{\im}  {\operatorname{Im}}

\newcommand{\Diff} {\operatorname{Diff}}  
\newcommand{\vf}   {\operatorname{Vect}} 
  
\begin{document}

\title[Multidimensional renormalization]{Multidimensional continued fractions, dynamical renormalization and KAM theory}

\author[K. Khanin]{Kostya Khanin}
\address{Department of Mathematics, University of Toronto,
100 St George Street, Toronto, Ontario M5S 3G3, Canada}
\email{khanin@math.toronto.edu} 

\author[J. Lopes Dias]{Jo\~ao Lopes Dias}
\address{Departamento de Matem\'atica, ISEG, 
Universidade T\'ecnica de Lisboa,
Rua do Quelhas 6, 1200-781 Lisboa, Portugal}
\email{jldias@iseg.utl.pt}

\author[J. Marklof]{Jens Marklof}
\address{School of Mathematics, University of Bristol, Bristol BS8 1TW, U.K.}
\email{j.marklof@bristol.ac.uk}

\thanks{}
\date{4/9/05}  

\begin{abstract}
The disadvantage of `traditional' multidimensional continued fraction algorithms is that it is not known whether they provide simultaneous rational approximations for generic vectors. Following ideas of Dani, Lagarias and Kleinbock-Margulis we describe a simple algorithm based on the dynamics of flows on the homogeneous space $\SLZ\backslash\SLR$ (the space of lattices of covolume one) that indeed yields best possible approximations to any irrational vector. The algorithm is ideally suited for a number of dynamical applications that involve small divisor problems. We explicitely construct renormalization schemes for (a) the linearization of vector fields on tori of arbitrary dimension and (b) the construction of invariant tori for Hamiltonian systems.
\end{abstract}
 
\maketitle
\tableofcontents


\section{Introduction}

This paper has two main goals. The first one is to introduce
a new multidimensional continued fraction algorithm that is ideally suited for
different dynamical applications. The algorithm can be used in order to effectively deal
with small divisors whenever quasi-periodicity with several frequencies is an essential
feature of a problem. Our second goal is to demonstrate the strength of the algorithm 
by applying it to KAM theory. We use a renormalization approach to prove two theorems of
KAM type. The method, being conceptually very simple, is also very general, and allows us to consider a wide class of frequency vectors. For reasons of clarity we restrict our attention to vectors satisfying an explicit Diophantine condition (valid for a set of vectors of full Lebesgue measure); generalizations to more general frequency vectors follow straightforwardly from the approach presented here, and will be detailed in a separate publication. 

The classical continued fraction algorithm produces, for every irrational 
$\alpha\in\Rr$, a sequence of rational numbers $p_n/q_n$ that approximate 
$\alpha$ up to an error of order $1/q_n^2$. The first objective of this 
paper is to develop a multidimensional analogue that allows us to 
approximate any irrational $\vecalf\in\Rr^{d-1}$ by rational vectors. 
The drawback of all `traditional' multidimensional continued fraction 
algorithms is that it is not even known whether they provide simultaneous rational approximations for almost Lebesgue almost all $\vecalf$. 
In the case $d \geq 4$ the only result in this direction is a recent computer-assisted proof 
of the almost everywhere strong convergence for ordered Jacobi-Perron 
algorithm \cite{Hardcastle1,Hardcastle2}. However, even in this case an explicit description of the set of bad vectors seems difficult. For example, the existence of `noble' vectors, that is vectors corresponding to a periodic continued-fraction expansion, for which approximations do not converge, is rather unsatisfactory.

The algorithm we employ here does not suffer from such pathologies. Following Lagarias' seminal ideas in \cite{Lagarias94}, our approach is based on the dynamics of the geodesic\footnote{The term `geodesic' is slightly inaccurate when $d>2$. The orbits of the flow on $\GamG$ we are discussing here correspond in fact to geodesics on the unit cotangent bundle of the space $\GamG/\SO(d)$ for a certain family of initial conditions. Only for $d=2$ the cotangent bundle can be identified with $\GamG$.} flow on the homogeneous space $\GamG$ with $G=\SLR$ and $\Gamma=\SLZ$. Notice that $\GamG$ may be identified with the space of lattices in $\Rr^d$ of covolume one or, equivalently, with the Teichm\"uller space of flat $d$-dimensional tori.

The problem of multidimensional continued fractions may be formulated in the following way. Given a vector $\vecalf\in\Rr^{d-1}$ find a sequence of matrices $T^{(n)} \in \GLZ$, $n\in \Nn$, such that the `cocycle' corresponding to the products $P^{(n)}=T^{(n)}T^{(n-1)}\dots T^{(1)}$ exponentially contracts in the direction of the vector $\vecomega= (\begin{smallmatrix} \vecalf \\ 1 \end{smallmatrix})\in\Rr^d$ and exponentially expands in all other directions. Thus, the cocycle should have one negative Lyapunov exponent and $d-1$ positive. In this spirit, our algorithm comprises the following steps:
\begin{enumerate}
\item With every $\vecalf\in\Rr^{d-1}$ associate the orbit 
$\{C(t):t\geq 0\}\subset\SLR$, where
\begin{equation}
C(t)= \begin{pmatrix} 1_{d-1} & \vecalf \\ 0 & 1 \end{pmatrix} 
\begin{pmatrix} 1_{d-1} \e^{-t} & 0 \\ 0 & \e^{(d-1) t} \end{pmatrix},
\end{equation}
$1_{d-1}$ denotes the $(d-1)\times(d-1)$ unit matrix.
\item Fix a fundamental domain $\FF$ of $\SLZ$ in $\SLR$. Given a sequence of times $t_1<t_2<\ldots\to\infty$, use classical reduction theory to find the matrices $P^{(n)}\in\SLZ$ that map the points $C(t_n)$ to $\FF$. 
\item 
Define the $n$th continued fraction map by
\begin{equation}
\vecalf^{(n-1)}\mapsto\vecalf^{(n)} = \frac{T^{(n)}_{11} \vecalf^{(n-1)} + \vect^{(n)}_{12}}
{\trans\vect^{(n)}_{21}\vecalf^{(n-1)} + t^{(n)}_{22}},
\end{equation}
where $\vecalf^{(0)}=\vecalf$ and
\begin{equation}
T^{(n)}=\begin{pmatrix} T^{(n)}_{11} & \vect^{(n)}_{12} \\ \trans\vect^{(n)}_{21} & t^{(n)}_{22} \end{pmatrix}  \in\SLZ
\end{equation}
is the $n$th transfer matrix defined by $P^{(n)}=T^{(n)} P^{(n-1)}$.
\end{enumerate}  
Since the action of $\SLR$ on $\Rr^{d-1}$ by fractional linear transformation defines a group action, we have
\begin{equation}\label{alfP}
\vecalf^{(n)} = \frac{P^{(n)}_{11} \vecalf + \vecp^{(n)}_{12}}
{\trans\vecp^{(n)}_{21}\vecalf + p^{(n)}_{22}},
\end{equation}
where
\begin{equation}
P^{(n)}=T^{(n)}T^{(n-1)}\cdots T^{(1)}=\begin{pmatrix} P^{(n)}_{11} & \vecp^{(n)}_{12} \\ \trans\vecp^{(n)}_{21} & p^{(n)}_{22} \end{pmatrix}  \in\SLZ.
\end{equation}

Dani \cite{Dani85} and Kleinbock-Margulis \cite{Kleinbock98} observed that Diophantine properties of $\vecalf$ translate to divergence properties of the corresponding orbit $\{ \Gamma C(t): t\geq 0\}\subset\GamG$ in the cusps of $\GamG$. We exploit these results to show that, under mild Diophantine conditions on $\vecalf$ (satisfied by a set of $\vecalf$ of full Lebesgue measure, cf. Sec. \ref{diocondi}), there is a sequence of times $t_n$ such that the transfer matrices $T^{(n)}$ are uniformly hyperbolic in a sense made precise in Sec. \ref{hyp}. This fact allows us to develop renormalization schemes for vector fields and Hamiltonian flows that had previously been constructed only in dimension one \cite{jld2} or for very special choices of $\vecalf$ \cite{Abad2,Koch,Koch2002,jld}. In particular, we obtain renormalization-based proofs of the following theorems.

We say $\vecomega\in\Rr^d$ is {\em Diophantine} if there are constants $\epsilon>0$, $C>0$ such that
\begin{equation} \label{dio33}
\|\veck\|^{(d-1)(1+\epsilon)} |\veck\cdot\vecomega | > C,
\end{equation}
for all $\veck\in\ZZ^{d}-\{\vecnull\}$. Note that we may assume without loss of generality that $\vecomega$ is of the form $\vecomega = (\begin{smallmatrix} \vecalf \\ 1 \end{smallmatrix})$ with $\vecalf\in\Rr^{d-1}$. Condition \eqref{dio33} then translates to a standard Diophantine condition on $\vecalf$, see Sec. \eqref{diocondi} for details.

\begin{theorem}\label{theorem: main. intro}
For any real analytic vector field $\vecv$ on $\Tt^d$, $d\geq2$, sufficiently close to a constant
vector field with Diophantine vector $\vecomega\in\Rr^d$, there is $\epsilon>0$, an analytic curve $(-\epsilon,\epsilon)\ni s\mapsto\vecp^s\in\Rr^d$ and an analytic conjugacy $h$ isotopic to the identity between
the flow generated by $\vecv+\vecp^s$ and the linear
flow $\phi_t(\vecx)=\vecx+t(1+s)\vecomega$ on $\Tt^d$, $t\geq0$, for each $|s|<\epsilon$.
Moreover, the maps $\vecv\mapsto h$ and $\vecv\mapsto \vecp$ are analytic.
\end{theorem}

Let us emphasize that the result holds for all analytic vector fields close to a constant one without any additional conditions, such as preservation of volume etc. 
The second theorem deals with the case of Hamiltonian vector fields.
Let $B\subset\Rr^d$, $d\geq2$, be an open set containing the origin, and let $H^0$ be a real-analytic Hamiltonian function
\begin{equation}\label{def H0 intro}
H^0(\vecx,\vecy)=\vecomega\cdot\vecy+\frac12 \trans\vecy Q\vecy,
\quad
(\vecx,\vecy)\in\Tt^d\times B,
\end{equation}
with $\vecomega\in\Rr^d$ and a real symmetric $d\times d$ matrix $Q$.
It is said to be non-degenerate if $\det Q\not=0$.

\begin{theorem}\label{theorem: main. intro for H}
Suppose $H^0$ is non-degenerate and $\vecomega$ is Diophantine.
If $H$ is a real analytic Hamiltonian on $\Tt^d\times B$ sufficiently close to $H^0$,
then the Hamiltonian flow of $H$ leaves invariant a Lagrangian $d$-dimensional torus where it is analytically conjugated to the linear flow $\phi_t(\vecx)=\vecx+t\vecomega$ on $\Tt^d$, $t\geq0$.
The conjugacy depends analytically on $H$.
\end{theorem}
 
\begin{proof}[Outline of the proof of Theorem \ref{theorem: main. intro}]
Consider a vector field $X(\vecx,\vecy)=\vecomega + f(\vecx) + \vecy$ where $\vecx\in \Tt^d$, and $\vecy\in \Rr^d$ is an auxiliary parameter. The vector field $f(\vecx)$ is a sufficiently small analytic perturbation
of a constant vector field. We may furthermore assume that $\vecomega = (\begin{smallmatrix} \vecalf \\ 1 \end{smallmatrix})$ for some Diophantine $\vecalf\in\Rr^{d-1}$; this achieved by a rescaling of time. The aim is to find a value of parameter $\vecy=\vecy_\vecomega$ such that the vector field $X(\vecx,\vecy_\vecomega)$ is linearizable to a constant vector field identically equal to $\vecomega$ by means of an
analytic transformation of the coordinates on $\Tt^d$. 

Renormalization is an iterative process, and we thus assume that after the $(n-1)$th renormalization step the vector field is of the slightly more general form
\begin{equation}
X_{n-1}(\vecx,\vecy)=\vecomega^{(n-1)} + f_{n-1}(\vecx,\vecy) 
\end{equation}
where 
$\vecomega^{(n-1)} = (\begin{smallmatrix} \vecalf^{(n-1)} \\ 1 \end{smallmatrix})$
and $\vecalf^{(n-1)}$ is given by the continued fraction algorithm, cf. \eqref{alfP}.
The Fourier modes of $f_{n-1}$ are smaller than in the previous step, and decay exponentially as $\left\|\veck\right\| \to +\infty$. We define a cone of resonant modes by a relation 
$I^+_{n-1}=\{\veck\in\Zz^d \colon |\veck\cdot\vecomega^{(n-1)}|\leq \sigma_{n-1}\|\veck \|\}$.
The $n$th step requires the following operations.
\begin{enumerate}
\item Eliminate all Fourier modes outside of the resonant cone $I^+_{n-1}$.
\item Apply a linear operator corresponding to a coordinate transformation given by the inverse transfer matrix ${T^{(n)}}^{-1}$.
\item Rescale time to ensure that the frequency vector is of the form $\vecomega^{(n)} = (\begin{smallmatrix} \vecalf^{(n)} \\ 1 \end{smallmatrix})$.
\end{enumerate} 
The conjugate action
on the Fourier modes is given by $\veck \mapsto \trans {T^{(n)}}^{-1}\veck$.
It follows from the hyperbolicity of $T^{(n)}$ that this transformation contracts for $\veck \in I^+_{n-1}$ if $\sigma_{n-1}$ is small enough. This gives a significant improvement of the analyticity domain which results in the decrease of the estimates for the corresponding Fourier modes.
As a result, all Fourier modes apart from the zero modes get smaller. To decrease the size of the latter, we choose a parameter $\vecy=\vecy_n$ in such a way that the corresponding zero modes vanish, and then consider a neighbourhood of $\vecy$-values centred at $\vecy_n$. That is, the auxiliary parameter $\vecy$ is used to eliminate an instability in the direction of constant vector fields. To get enough control on the parameter dependence we perform an affine rescaling of this parameter on every renormalization step. One can then show that the corresponding sequence of parameter domains is nested and converges to a single point $\vecy=\vecy_\vecomega$ for which the initial vector field is indeed linearizable.

In order for the scheme to be effective, the sequence of stopping times $t_n$ and the sizes of the resonant cones defined by the sequence of $\sigma_n$ must be chosen properly. Large intervals $\delta t_n = t_n - t_{n-1}$ improve hyperbolicity but, on the other hand, worsen estimates for the norms $\| T^{(n)} \|$, $\| {T^{(n)}}^{-1}\|$. Similarly, if  $\sigma_n$ is too small the elimination of non-resonant modes will give large contributions; on the other hand, for large values of $\sigma_n$ the multiplication by $\trans {T^{(n+1)}}^{-1}$ will not yield a contraction for $\veck \in I^+_{n}$. As we shall show below a right choice of sequences $\{(t_n, \sigma_n)\}$ can be made, depending on the Diophantine properties of the vector $\vecomega$.
\end{proof}

The idea of renormalization was introduced to the theory of dynamical
systems by Feigenbaum \cite{Feigenbaum} in the late 1970's. In the case of Hamiltonian systems with two degrees of freedom MacKay proposed in the early 1980's a renormalization scheme for the
construction of KAM invariant tori \cite{MacKayThesis}. The scheme was realized for the construction of invariant curves for two-dimensional conservative maps
of the cylinder. An important feature of MacKay's approach is the analysis of both smooth KAM invariant curves and so-called critical curves corresponding
to critical values of a parameter above which
invariant curves no longer exist. From the point of view of
renormalization theory the KAM curves correspond to a trivial
linear fixed point for the renormalization transformations, while critical curves give
rise to very complicated fixed points with nontrivial critical
behavior. MacKay's renormalization scheme was carried out only for a small class of Diophantine rotation numbers with periodic continued fraction expansion (such as the golden mean). Khanin and Sinai studied a different renormalization scheme for general Diophantine rotation numbers \cite{Khanin}. Both of the above early approaches were based on renormalization for maps or their generating functions. Essentially, the renormalization transformations are defined in the space of pairs of mappings which, being iterates of the same map, commute with each other. These commutativity conditions
cause difficult technical problems, and led MacKay \cite{MacKay} to propose the development of alternative renormalization schemes acting directly on vector fields. 
The same idea was realized by Koch \cite{Koch} who proves a KAM type result for analytic perturbations of linear Hamiltonians $H^0(\vecx,\vecy)=\vecomega\cdot\vecy$, for frequencies $\vecomega$ which are eigenvectors of hyperbolic matrices in $\SLZ$ with only one unstable direction. Notice that the set of such frequencies has zero Lebesgue measure and in the case $d=2$ corresponds to vectors with a quadratic irrational slope.
Further improvements and applications of Koch's techniques appeared in \cite{Abad2,Koch2002,jld,jld2,Gaidashev}, emphasizing the connection between KAM and renormalization theories.
The results of this paper illustrate that such a programme can indeed be carried out in considerable generality.
Another direction was followed in \cite{Koch2004}, presenting a computer-assisted proof of the existence of MacKay's golden mean critical renormalization fixed point in the context of Hamiltonian vector fields with two degrees of freedom.

Other renormalization ideas have appeared in the context of the stability of invariant tori for nearly integrable Hamiltonian systems (see e.g. \cite{Bricmont,Gallavotti,Gentile} and references therein). Inspired by quantum field theory and an analogy with KAM theory, this approach uses an iterative resummation of Poincar\'e's Lindstedt series to prove its convergence.

For the sake of transparency we have restricted our attention to Diophantine vectors $\vecomega$. A more detailed analysis under weaker Diophantine conditions is, in principle, possible within the present framework. It is however a fundamental open problem to state a sharp (i.e. the weakest possible) Diophantine condition under which the above conjugacy can be established. The answer to this question is known only in the classical case $d=2$ \cite{Yoccoz2} where the Diophantine condition is of Brjuno type. 

It would also be interesting to see whether the multidimensional continued fraction algorithm presented here will allow generalizations of other one-dimensional renormalization constructions. A concrete challenge is for instance the extension of the recent results on the reducibility of cocycles over irrational rotations by \'Avila and Krikorian \cite{Avila05}. 

In the next section we introduce the multidimensional continued fraction algorithm, and include a discussion of its hyperbolicity properties required in the renormalization schemes. The remainder of the paper provides a detailed account of two exemplary cases, the renormalization of vector fields (Section \ref{Section: Renormalization of vector fields}) and Hamiltonian flows (Section \ref{Section: Renormalization of Hamiltonian flows}).


\section{Multidimensional continued fractions and flows on homogeneous spaces}
\label{section:Multidimensional continued fractions}

\subsection{Flows on homogeneous spaces}
Let us set $G=\SLR$ and $\Gamma=\SLZ$, and define the
diagonal subgroup $\{E^t : t\in\Rr\}$ in $G$, where
\begin{equation}
E^t= \diag( \e^{r_1 t},\ldots,\e^{r_d t}) 
\end{equation}
with constants satisfying the conditions
\begin{equation}
r_1,\ldots, r_{d-1} < 0 < r_d, \qquad \sum_{j=1}^d r_j =0 .
\end{equation}
The right action of $E^t$ on the homogeneous space $\GamG\GamG$ generates
the flow
\begin{equation}
\Phi^t: \GamG \to \GamG, \qquad \Gamma M \mapsto \Gamma M E^t .
\end{equation}
Since $G$ is a simple Lie group with finite center, $\Phi^t$
is ergodic and mixing \cite{Moore66}.

Let $\scrF\subset G$ be a fundamental domain of the left action 
of $\Gamma$ on $G$.
Recall that, by definition of the fundamental domain of a free
group action,
\begin{equation}
\bigcup_{P\in\Gamma} P \scrF = G, \qquad
\scrF \cap P \scrF = \varnothing \text{ for all $P\in\Gamma-\{ 1 \}$,}
\end{equation}
and hence, for any given $M\in G$, 
there is a unique family of $P(t)\in\Gamma$ such that
\begin{equation} \label{Mt}
M(t) := P(t) M E^t \in \scrF
\end{equation} 
holds for all $t\in\Rr$.

\subsection{A convenient parametrization}
Let us consider those $M\in G$ which can be written as
\begin{equation}\label{Malf}
M= 
\begin{pmatrix}
1 & \vecalf \\ 0 & 1
\end{pmatrix} 
\begin{pmatrix}
A & \vecnull \\ \trans\vecbeta & \gamma
\end{pmatrix} 
\end{equation}
where $A\in \Mat_{d-1}(\Rr)$ 
(the space of real $(d-1)\times(d-1)$ matrices), 
$\vecalf,\vecbeta\in\Rr^{d-1}$ are column vectors, 
$\gamma\in\Rr$ with $\gamma>0$.
This yields a local parametrization of $G$ for the set
\begin{equation}
G_+ := \left\{
\begin{pmatrix}
1 & \vecalf \\ 0 & 1
\end{pmatrix} 
\begin{pmatrix}
A & \vecnull \\ \trans\vecbeta & \gamma
\end{pmatrix} \in G
: \;
A\in \Mat_{d-1}(\Rr), \,
\vecalf,\vecbeta\in\Rr^{d-1}, \, 
\gamma\in\Rr_{>0}
\right\} ,
\end{equation}
which is particularly convenient for our purposes.
All other matrices are either of the above form with $\gamma<0$ instead,
or may be written as
\begin{equation}
M= S
\begin{pmatrix}
1 & \vecalf \\ 0 & 1
\end{pmatrix} 
\begin{pmatrix}
A & \vecnull \\ \trans\vecbeta & \gamma
\end{pmatrix} 
\end{equation}
where $S\in\Gamma$ is a suitably chosen ``signed permutation matrix'',
i.e., every row and every column contains one and only one non-zero
coefficient, which is either $1$ or $-1$. 
In the following we will stay clear
of the parameter singularity at $\gamma=0$, and thus may assume 
without loss of generality $S=1$.

To work out the action of a general element $T\in G$ in the above
parametrization, consider
\begin{equation}
T: M \mapsto \tilde M:=TM 
\end{equation} 
where
\begin{equation}
T=\begin{pmatrix} T_{11} & \vect_{12} \\ 
\trans\vect_{21} & t_{22} \end{pmatrix},
\end{equation}
$M$ is as above and 
\begin{equation}
\tilde M= 
\begin{pmatrix}
1 & \tilde \vecalf \\ 0 & 1
\end{pmatrix} 
\begin{pmatrix}
\tilde A & \vecnull \\ \trans\tilde{\vecbeta} & \tilde \gamma
\end{pmatrix} .
\end{equation}
A short calculation yields the fractional linear action
\begin{equation}\label{fraclin1}
\vecalf\mapsto \tilde\vecalf = \frac{T_{11} \vecalf + \vect_{12}}
{\trans\vect_{21}\vecalf + t_{22}},
\end{equation}
and
\begin{equation}\label{fraclin2}
\gamma \mapsto \tilde\gamma = (\trans\vect_{21}\vecalf + t_{22})\gamma ,
\end{equation}
and more complicated expressions for $\tilde A,\tilde\beta$
which will not be needed in the following.

\subsection{Multidimensional continued fractions}

Let $t_0=0<t_1<t_2<\ldots\to\infty$ be sequence of times, with gaps
\begin{equation}
\delta t_n:=t_n-t_{n-1}
\end{equation}
chosen large enough so that $P(t_n)\neq P(t_{n-1})$,
where $P(t)$ is defined by (\ref{Mt}). The sequence $P^{(n)}:=P(t_n)$
of matrices in $\Gamma$ may be viewed as the continued fraction approximants
of the vector $\vecalf$, which are the best possible for
suitable choices of a fundamental domain $\scrF$ and times $t_n$,
see \cite{Lagarias94}.
Let us furthermore put $M^{(n)}:=M(t_n)$ with $M(t)$ as in (\ref{Mt}),
and define $\vecalf^{(n)}$, $\gamma^{(n)}$ by the decomposition (\ref{Malf}),
i.e., by
\begin{equation}\label{Malf2}
M^{(n)}= 
\begin{pmatrix}
1 & \vecalf^{(n)} \\ 0 & 1
\end{pmatrix} 
\begin{pmatrix}
A^{(n)} & \vecnull \\ \trans\vecbeta^{(n)} & \gamma^{(n)} 
\end{pmatrix} .
\end{equation}
 From $M^{(n)}=P^{(n)} M E^{t_n}$ and (\ref{fraclin1}), (\ref{fraclin2})
we deduce
\begin{equation}
\vecalf^{(n)} = \frac{P^{(n)}_{11} \vecalf + \vecp^{(n)}_{12}}
{\trans\vecp^{(n)}_{21}\vecalf + p^{(n)}_{22}}, 
\end{equation}
and
\begin{equation}\label{gamman}
\gamma^{(n)} = 
(\trans\vecp^{(n)}_{21}\vecalf + p^{(n)}_{22})\, \e^{r_d t_n}\, \gamma 
\end{equation}
where
\begin{equation}
P^{(n)}=\begin{pmatrix} P^{(n)}_{11} & \vecp^{(n)}_{12} \\ 
\trans\vecp^{(n)}_{21} & p^{(n)}_{22} \end{pmatrix}.
\end{equation}

It is evident that if the components of $(\trans\vecalf,1)$ are linearly
independent over $\QQ$, then $\gamma\neq 0$ implies $\gamma^{(n)}\neq 0$
for all $n\geq 0$.
 
We shall later employ the transfer matrices $T^{(n)}$ defined
by $P^{(n)}=T^{(n)}P^{(n-1)}$. Here, 
$M^{(n)}=T^{(n)} M^{(n-1)} E^{\delta t_n}$ implies
\begin{equation}
\vecalf^{(n)} = \frac{T^{(n)}_{11} \vecalf^{(n-1)} + \vect^{(n)}_{12}}
{\trans\vect^{(n)}_{21}\vecalf^{(n-1)} + t^{(n)}_{22}},
\end{equation}
and
\begin{equation}\label{gamman2}
\gamma^{(n)} = 
(\trans\vect^{(n)}_{21}\vecalf^{(n-1)} + t^{(n)}_{22})\, 
\e^{r_d \delta t_n}\, \gamma^{(n-1)} ,
\end{equation}
where
\begin{equation}
T^{(n)}=\begin{pmatrix} T^{(n)}_{11} & \vect^{(n)}_{12} \\ 
\trans\vect^{(n)}_{21} & t^{(n)}_{22} \end{pmatrix}.
\end{equation}

\subsection{Siegel sets}

In dimensions $d>2$ it is difficult to describe the geometry 
of a fundamental domain $\scrF$. To overcome this problem, 
C.~Siegel introduced simply connected sets $\scrS_d\subset G$
which have the property that they contain $\scrF$ and are contained
in a finite number of translates $P\scrF$, $P\in\Gamma$.
Consider the Iwasawa decomposition
\begin{equation}\label{Iwasawa}
M=n a k
\end{equation}
where
\begin{equation}
n=
\begin{pmatrix} 
 1   & u_{12} & \ldots & u_{1d}  \\
  &  \ddots  &   \ddots         & \vdots       \\
  &      &    \ddots   & u_{d-1,d} \\
 & & & 1             
\end{pmatrix} , \qquad 
a=
\begin{pmatrix} 
 a_1   &  &  & \\
  &  \ddots  & &       \\
  &  & \ddots   &       \\
  &  &  &  a_d            
\end{pmatrix} 
\end{equation}
and $k\in\SO(d)$, with $u_{ij},a_j\in\Rr$, $a_j>0$, $a_1 \cdots a_d=1$.
Then
\begin{equation}\label{Siegel}
\scrS_d = \{ nak :\; n\in{\mathcal F}_N , \;
a_j \geq \frac{\sqrt 3}{2} a_{j+1}>0
\; (j=1,\ldots,d-1), \;
k\in\SO(d) \}
\end{equation}
is an example of a {\em Siegel set} \cite{Raghunathan72};
here ${\mathcal F}_N$ denotes a compact fundamental region of 
$(\Gamma\cap N)\backslash N$, 
where $N$ is the upper triangular group of elements of the form
$n$ as above.

\subsection{Dani's correspondence}

We assume from now on that $r_1,\ldots,r_{d-1}=-1$, $r_d=d-1$, i.e.,
\begin{equation}
E^t=\diag( \e^{-t},\ldots,\e^{-t},\e^{(d-1)t}).
\end{equation}
Let us denote by $|\,\cdot\,|$ the maximum norm in $\Rr^{d-1}$.
A vector $\vecalf\in\Rr^{d-1}$ is called {\em badly approximable} or of {\em bounded type},
if one of the following equivalent conditions is satisfied.
\begin{itemize}
\item[(i)]
{\em There exists a 
constant $C>0$ such that 
\begin{equation}
|k \vecalf + \vecm|^{d-1} |k| > C
\end{equation}
for all $\vecm\in\ZZ^{d-1}$, $k\in\ZZ-\{0\}$.}
\item[(ii)]
{\em There exists a 
constant $C>0$ such that 
\begin{equation}
|\vecm|^{d-1} |\vecm\cdot \vecalf +k | > C
\end{equation}
for all $\vecm\in\ZZ^{d-1}-\{\vecnull\}$, $k\in\ZZ$.}
\end{itemize}

The statements (i) and (ii) 
are equivalent in view of Khintchine's transference 
principle (\cite{Cassels57} Chapter V).

We recall {\em Dani's correspondence} in the following proposition
(cf.~\cite{Dani85}, Theorem 2.20).

\begin{prop}
The orbit $\{ \Gamma M E^t: t\geq 0 \}$,
with $M$ as in {\rm (\ref{Malf})},
is bounded in $\GamG$ if and only if
the vector $\vecalf$ is of bounded type. 
\end{prop}

The reason why the parameters
$A,\vecbeta,\gamma$ are irrelevant in the statement is that the 
family of matrices
\begin{equation}\label{Wt}
W(t)=
E^{-t}
\begin{pmatrix}
A & \vecnull \\ \trans\vecbeta & \gamma
\end{pmatrix} 
E^t
\end{equation}
is bounded in $G$ for all $t\geq 0$.

The boundedness of the orbit $\{ \Gamma M E^t: t\geq 0 \}$ implies
of course that there is a compact set $\scrC\in G$ such that
$M(t)\in\scrC$ for all $t\geq 0$, with $M(t)$ as in (\ref{Mt}).

\subsection{Diophantine conditions} \label{diocondi}

A vector $\vecalf\in\Rr^{d-1}$ is called {\em Diophantine},
if there exist constants $\epsilon>0$, $C>0$ such that 
\begin{equation} \label{dio2}
|\vecm|^{(d-1)(1+\epsilon)} |\vecm\cdot \vecalf +k | > C
\end{equation}
for all $\vecm\in\ZZ^{d-1}-\{\vecnull\}$, $k\in\ZZ$.
It is well known that Diophantine vectors form a set of full Lebesgue measure
\cite{Cassels57}.

Let us show that (\ref{dio2}) implies the inequality
\begin{equation} \label{dio3}
\|\veck\|^{(d-1)(1+\epsilon)} |\veck\cdot\vecomega | > C,
\end{equation}
for all $\veck\in\ZZ^{d}-\{\vecnull\}$, where
$\vecomega = (\begin{smallmatrix} \vecalf \\ 1 \end{smallmatrix})$, cf. \eqref{dio33}.
With $\veck=(\begin{smallmatrix} \vecm \\ k \end{smallmatrix})$, 
(\ref{dio2}) yields
\begin{equation} 
|\veck\cdot\vecomega | 
> C |\vecm|^{-(d-1)(1+\epsilon)} 
\geq C |\veck|^{-(d-1)(1+\epsilon)}
\geq C \|\veck\|^{-(d-1)(1+\epsilon)}
\end{equation}
for all $\vecm\in\ZZ^{d-1}-\{\vecnull\}$, $k\in\ZZ$. In the case
when $\vecm=\vecnull$, we have $k\neq 0$ 
(since $\veck\neq\vecnull$) and thus (\ref{dio3}) holds 
trivially.\footnote{Note that every admissible constant in (\ref{dio2}) 
needs to satisfy $C<1/2$; to see this, choose $\vecm=(1,0,\ldots,0)$,
and $k\in\ZZ$ such that $|\alpha_1+k|\leq 1/2$.} Note also that (\ref{dio3}) evidently implies (\ref{dio2}), however with different choices for $C$ in both inequalities.

Following \cite{Kleinbock98} we define the following function on $G$, 
\begin{equation}\label{delta}
\delta(M) = \inf_{\veck\in\ZZ^d-\{\vecnull\}} | \trans\veck M | .
\end{equation}
It is easily checked that $\delta(M)$ is invariant under left
action of $\Gamma$, and may thus be viewed as a function on $\GamG$.
In terms of the Iwasawa parametrization (\ref{Iwasawa}) and the Siegel
set $\scrS_d$ defined in (\ref{Siegel}) we have the following estimate.

\begin{lem} \label{onetwo}
For $M=nak\in\scrS_d$ as in {\rm (\ref{Iwasawa})}, {\rm (\ref{Siegel})},
there are constants $0<C_1\leq C_2$ such that for all $0<a_d \leq 1$
\begin{equation}
C_1 a_d \leq \delta(M) \leq C_2 a_d .
\end{equation}
\end{lem}

\begin{proof}
Since $\|\vecx\| \ll |\vecx| \ll \|\vecx\|$ 
for all $\vecx\in\Rr^d$, 
we may prove the statement of the lemma for the function
\begin{equation}\label{delta2}
\tilde\delta_d(M) = \inf_{\veck\in\ZZ^d-\{\vecnull\}} \| \trans\veck M \| 
\end{equation}
instead.\footnote{In the following, $A \ll B$ means `there is a constant $C>0$ such that $A \leq CB$'. If $A\ll B\ll A$ we will also use the notation $A\asymp B$.}
Due to the rotational invariance of the Euclidean distance we may
assume that $k\in\SO(d)$ is the identity.

{\em Proof by induction.} The statement trivially holds for 
$d=1$. Therefore let us assume the assertion is true 
for dimension $d-1$.
The $j$th coefficient of the vector
$\trans\veck M$ is 
\begin{equation}
(\trans\veck M)_j = \bigg( k_j + \sum_{i=1}^{j-1} k_i \, u_{ij}\bigg)  a_j .
\end{equation} 
Since $a_1\to\infty$ when $a_d\to 0$, this implies that when taking the 
infimum in (\ref{delta}) we must take $k_1=0$ for all sufficiently small
$a_d$. Thus we now need to estimate
\begin{equation}\label{inf}
\inf_{\tilde\veck\in\ZZ^{d-1}-\{\vecnull\}} \max_{2\leq j \leq d} 
\bigg|\bigg( k_j + \sum_{i=1}^{j-1} k_i \, u_{ij}\bigg)  a_j \bigg|
=
a_1^{-1/(d-1)} 
\inf_{\tilde\veck\in\ZZ^{d-1}-\{\vecnull\}} \max_{2\leq j \leq d} 
\bigg|\bigg( k_j + \sum_{i=1}^{j-1} k_i \, u_{ij}\bigg)  \tilde a_j \bigg|
\end{equation}
where $\trans{\tilde\veck}=(k_2,\ldots,k_d)$,
$\tilde a_j = a_1^{1/(d-1)} a_j$ so that $\tilde a_2\cdots\tilde a_d=1$.
Now
\begin{equation}
\inf_{\tilde\veck\in\ZZ^{d-1}-\{\vecnull\}} \max_{2\leq j \leq d} 
\bigg|\bigg( k_j + \sum_{i=1}^{j-1} k_i \, u_{ij}\bigg)  \tilde a_j \bigg|
=:\delta_{d-1}(\tilde M) 
\asymp \tilde\delta_{d-1}(\tilde M) 
\end{equation}
where $\tilde M=\tilde n \tilde a$ with
\begin{equation}
\tilde n=
\begin{pmatrix} 
 1   & u_{23} & \ldots & u_{2d}  \\
  &  \ddots  &   \ddots         & \vdots       \\
  &      &    \ddots   & u_{d-1,d} \\
 & & & 1             
\end{pmatrix} , \qquad 
\tilde a=
\begin{pmatrix} 
 \tilde a_2   &  &  & \\
  &  \ddots  & &       \\
  &  & \ddots   &       \\
  &  &  &  \tilde a_d           
\end{pmatrix} .
\end{equation}
It is easily checked that $\tilde M\in\scrS_{d-1}$,
so by the induction hypothesis, for suitable constants
$0<C_{1,d-1}\leq C_{2,d-1}$, we have
\begin{equation}
C_{1,d-1}\; \tilde a_d \leq \tilde\delta_{d-1}(\tilde M) 
\leq C_{2,d-1}\; \tilde a_d ,
\end{equation}
provided $\tilde a_d=a_1^{1/(d-1)} a_d\leq 1$. So for $a_d$ sufficiently small
and $a_1^{1/(d-1)} a_d\leq 1$, we have 
\begin{equation}
C_{1,d-1}\; a_d \leq \tilde\delta_d(M) 
\leq C_{2,d-1}\; a_d .
\end{equation}

In the remaining case $\tilde a_d> 1$,
all $\tilde a_j$ are bounded from above and below
by positive constants, and hence $\tilde\delta_{d-1}(\tilde M)$ is 
bounded from above and below by positive constants.
Furthermore $\tilde a_d>1$ implies $a_1^{-1/(d-1)} < a_d$,
and, in view of our choice of the Siegel set,
$a_1^{-1}=a_2\cdots a_d \gg a_d^{d-1}$. So 
\begin{equation}
a_d \ll a_1^{-1/(d-1)} < a_d
\end{equation} 
and the required bound follows from (\ref{inf}) also for the case 
$\tilde a_d> 1$.
\end{proof}

\begin{lem} \label{prime}
Choose $M$ as in {\rm (\ref{Malf})}
and suppose $\vecalf$ satisfies condition {\rm (\ref{dio2})}.
Then there exists a constant $C'>0$ such that for all $t\geq 0$
\begin{equation}
\delta( M E^t ) > C' \e^{-\theta t}
\end{equation}
where
\begin{equation}\label{relation theta epsilon}
\theta= \frac{(d-1)\epsilon}{d+(d-1)\epsilon} .
\end{equation}
\end{lem}

\begin{proof}
Let us put $\trans\veck=(\trans\vecm, k)$ with $\vecm\in\ZZ^{d-1}$ 
and $k\in\ZZ$. Then 
\begin{multline}
\delta( M E^t ) = \inf_{(\vecm,k)\in\ZZ^d-\{\vecnull\}} 
\big|\big(\trans\vecm \e^{-t}, (\trans\vecm\vecalf+k) 
\e^{(d-1)t}\big) \;W(t) \big| \\
\gg
\inf_{(\vecm,k)\in\ZZ^d-\{\vecnull\}} 
|(\trans\vecm \e^{-t}, (\trans\vecm\vecalf+k) \e^{(d-1)t}) |
\end{multline}
since $W(t)$, as defined in (\ref{Wt}), is bounded in $G$ for all $t\geq 0$.
Furthermore for $t$ sufficiently large
\begin{equation}
\inf_{(\vecm,k)\in\ZZ^d-\{\vecnull\}} 
|(\trans\vecm \e^{-t}, (\trans\vecm\vecalf+k) \e^{(d-1)t}) |
=
\inf_{\vecm\in\ZZ^{d-1}-\{\vecnull\},\; k\in\ZZ} 
|(\trans\vecm \e^{-t}, (\trans\vecm\vecalf+k) \e^{(d-1)t}) |
\end{equation}
which, in view of the Diophantine condition (\ref{dio2}), is bounded
from below by
\begin{equation}
\geq \inf_{\vecm\in\ZZ^{d-1}-\{\vecnull\}} 
|(\trans\vecm \e^{-t}, C |\vecm|^{-(d-1)(1+\epsilon)} \e^{(d-1)t}) |
= \e^{-\theta t} 
\inf_{\vecm\in\ZZ^{d-1}-\{\vecnull\}} 
|(\trans\vecx, C |\vecx|^{-(d-1)(1+\epsilon)})|
\end{equation}
where $\vecx=\e^{(\theta-1)t}\vecm$. We conclude the proof by noting that
\begin{equation}
\inf_{\vecm\in\ZZ^{d-1}-\{\vecnull\}} 
|(\trans\vecx, C |\vecx|^{-(d-1)(1+\epsilon)})|
\geq
\inf_{\vecy\in\Rr^{d-1}-\{\vecnull\}} 
|(\trans\vecy, C |\vecy|^{-(d-1)(1+\epsilon)})| >0 .
\end{equation}
\end{proof}

The fact that $\epsilon=0$ implies $\theta=0$ is consistent with
Dani's correspondence. On the other hand, $\theta<1$ for any
$\epsilon<\infty$.

\subsection{Norm estimates}

Let $\|\,\cdot\,\|$ denote the usual matrix norm
\begin{equation}
\| M \| := \sup_{\vecx\neq\vecnull} \frac{\| M \vecx \|}{\|\vecx\|}.
\end{equation}

\begin{prop}\label{proposition norm estimates}
Choose $M=M^{(0)}$ as in {\rm (\ref{Malf})},
and suppose $\vecalf$ satisfies condition {\rm (\ref{dio2})}.
Then there are constants $c_1,c_2,c_3,c_4,c_5,c_6>0$ 
such that for all $n\in\Nn\cup\{0\}$
\begin{eqnarray}
\label{M1}
\| M^{(n)} \| &\leq& c_1\exp[(d-1)\theta t_{n}] , \\
\label{M2}
\| {M^{(n)}}^{-1} \|  &\leq& c_2 \exp(\theta t_n), \\
\label{P1}
\| P^{(n)} \|  &\leq& c_3\exp[(d\,\theta+1-\theta) t_{n}], \\
\label{P2}
\| {P^{(n)}}^{-1} \|  &\leq& c_4 \exp[(d-1+\theta) t_n], \\
\label{T1}
\| T^{(n)} \|&\leq& c_5\exp[(1-\theta)\delta t_n+d\,\theta\, t_n], \\
\label{T2}
\| {T^{(n)}}^{-1} \| 
&\leq& c_6 \exp[(d-1)(1-\theta)\delta t_n + d\,\theta\,t_n] .
\end{eqnarray}
\end{prop}

\begin{proof}
For any $M\in\scrS_d$ as in (\ref{Iwasawa}) we have, for all $0<a_d\leq 1$,
\begin{equation}
\| M \| \ll a_1 = (a_2\cdots a_d)^{-1} \ll a_d^{-(d-1)},
\end{equation}
and
\begin{equation}
\| M^{-1} \| \ll a_d^{-1} .
\end{equation}
Combine this with Lemmas \ref{onetwo} and \ref{prime} to obtain the bounds  
\begin{equation}
\| {M^{(n)}}^{-1} \| \ll C_2 \; \delta(M^{(n)})^{-1}
=  C_2 \;\delta(M^{(0)} E^{t_n})^{-1}
< C_2 {C'}^{-1} \exp(\theta t_n)
\end{equation}
and
\begin{multline}
\| M^{(n)} \| \ll C_2^{d-1} \;\delta(M^{(n)})^{-(d-1)}
=  C_2^{d-1}\; \delta(M^{(0)} E^{t_{n}})^{-(d-1)} \\
< C_2^{d-1} {C'}^{-(d-1)} \exp[(d-1)\theta t_{n}]. 
\end{multline}
The remaining estimates follow immediately from 
(\ref{M1}), (\ref{M2})  and the equations
\begin{equation}
P^{(n)} = M^{(n)} E^{-t_n} {M^{(0)}}^{-1}  , \qquad
T^{(n)} = M^{(n)} E^{-\delta t_n} {M^{(n-1)}}^{-1} .
\end{equation}
\end{proof}

\begin{prop}\label{bound on gamma n}
Choose $M=M^{(0)}$ as in {\rm (\ref{Malf})},
and suppose $\vecalf$ satisfies condition {\rm (\ref{dio2})}.
Then there is a constant $c_7>0$ such that for all $n\in\Nn\cup\{0\}$,
\begin{equation}\label{gamma1}
c_7
\exp\bigg[-\theta \bigg(\frac{d^2}{1-\theta}-(d-1)\bigg) t_n\bigg] 
\leq 
|\gamma^{(n)}|
\leq 
c_1
\exp[(d-1)\theta t_{n}] 
\end{equation}
with $c_1$ as in {\rm (\ref{M1})}.
\end{prop}

\begin{proof}
The upper bound for $|\gamma^{(n)}|$ follows from 
(\ref{M1}), since $\gamma^{(n)}= (M^{(n)})_{dd}$ and hence
$|\gamma^{(n)}|\leq \| M^{(n)} \|$.

 From (\ref{gamman}) and the Diophantine condition (\ref{dio2}) we have
\begin{equation}\label{gamman3}
|\gamma^{(n)}| = \gamma
\exp[(d-1) t_n]
|\trans\vecp^{(n)}_{21}\vecalf + p^{(n)}_{22})|
> C \exp[(d-1) t_n] |\vecp^{(n)}_{21}|^{-(d-1)(1+\epsilon)} .
\end{equation}
Since
\begin{equation}
1+\epsilon=\frac{d-1+\theta}{(d-1)(1-\theta)}
\end{equation}
and
\begin{equation}
|\vecp^{(n)}_{21}| \leq \| P^{(n)} \| 
\end{equation}
the proposition follows from the estimate (\ref{P1}).
\end{proof}

\subsection{Hyperbolicity of the transfer matrices}\label{hyp}

Let
\begin{equation}
\vecomega^{(n)}_\perp = \{ \vecxi \in\Rr^d : \,\vecxi\cdot\vecomega^{(n)}=0\}
\end{equation} 
be the orthogonal complement of the vector 
\begin{equation}
\vecomega^{(n)} = \begin{pmatrix} \vecalf^{(n)} \\ 1 \end{pmatrix} \in\Rr^d .
\end{equation}

\begin{lem}\label{hype}
For all $\vecxi\in\vecomega_\perp^{(n-1)}$, $n\in\Nn$,
\begin{equation}
\trans {T^{(n)}}^{-1} \vecxi =  \exp(-\delta t_n) 
\trans\big(M^{(n-1)} {M^{(n)}}^{-1}\big)  \vecxi
\end{equation}
\end{lem}

\begin{proof}
This follows directly from the relation
\begin{equation}
E^{\delta t_n} \trans M^{(n-1)} \vecxi
= \exp(-\delta t_n) 
\begin{pmatrix} \trans A^{(n-1)} 
\vecxi' \\ 0 \end{pmatrix}
=\exp(-\delta t_n) \trans M^{(n-1)} \vecxi
\end{equation}
where $\vecxi'\in\Rr^{d-1}$ comprises the first $d-1$ components of $\vecxi$.
\end{proof}

\begin{prop} \label{hypprop}
Choose $M=M^{(0)}$ as in {\rm (\ref{Malf})},
and suppose $\vecalf$ satisfies condition {\rm (\ref{dio2})}.
Then there is a constant $\Lambda>0$ such that 
for all $\vecxi\in\vecomega_\perp^{(n-1)}$, $n\in\Nn$,
\begin{equation}
\| \trans {T^{(n)}}^{-1} \vecxi \| 
\leq \frac12\Lambda  
\exp(-\varphi_n) \|\vecxi\| 
\end{equation}
with
\begin{equation}\label{phidef}
\varphi_n=(1-\theta)\delta t_n-d\,\theta\, t_{n-1}.
\end{equation}
\end{prop}

\begin{proof}
 From Lemma \ref{hype},
\begin{equation}
\|\trans {T^{(n)}}^{-1} \vecxi\| \leq  \exp(-\delta t_n) 
\| M^{(n-1)} \|\; \|{M^{(n)}}^{-1}\|\;  \|\vecxi\| ,
\end{equation}
and the proposition follows from the bounds (\ref{M1}), (\ref{M2}).
\end{proof}

Given any positive sequence $\varphi_0,\varphi_1,\ldots$, the values $t_n$ that solve eq.~\eqref{phidef} with $t_0=0$ are
\begin{equation}
t_n=\frac{1}{1-\theta}\sum_{j=1}^n (1 +\beta)^{n-j} \varphi_j.
\end{equation}
where 
\begin{equation}\label{def beta}
\beta=\frac{d\theta}{1-\theta}.
\end{equation}
E.g., for constant $\varphi_n=\varphi>0$, we have
\begin{equation} \label{tn}
t_n=
\begin{cases}
n \varphi & (\theta=0) \\
\frac{\varphi}{d\theta} [(1+\beta)^n -1] & (0<\theta<1).
\end{cases}
\end{equation}

\subsection{The resonance cone}

As we shall see,  a crucial step in our renormalization scheme
is to eliminate all far-from-resonance modes in the Fourier series,
i.e., all modes labeled by integer vectors outside the cone
\begin{equation}\label{def resonance cone Kn}
K^{(n)} = \{ \vecxi \in\Rr^d \colon 
|\vecxi\cdot\vecomega^{(n)} |\leq \sigma_n \|\vecxi\| \}
\end{equation}
for a given $\sigma_n>0$.

\begin{lem}\label{lemma resonance cone}
Choose $M=M^{(0)}$ as in {\rm (\ref{Malf})},
and suppose $\vecalf$ satisfies condition {\rm (\ref{dio2})}.
Then 
\begin{equation}\label{formula An}
\sup_{\vecxi\in K^{(n-1)}-\{0\}}
\frac{\| \trans {T^{(n)}}^{-1} \vecxi \| }
{\|\vecxi\|}
\leq
\left[\frac\Lambda2 + c_6 \sigma_{n-1}
\e^{d\,\delta t_n} \right]
\exp\big[-(1-\theta)\delta t_n +d\,\theta\, t_{n-1}\big],
\end{equation}
for all $n\in\Nn$.
\end{lem}

\begin{proof}
We write $\vecxi=\vecxi_1 + \vecxi_2$,
where
\begin{equation}
\vecxi_1 = \frac{\vecxi\cdot\vecomega^{(n-1)}}{ \| \vecomega^{(n-1)}\|^2}
\,\vecomega^{(n-1)} , \qquad
\vecxi_2 \in \vecomega^{(n-1)}_\perp .
\end{equation}
Firstly,
\begin{equation}
\| \trans {T^{(n)}}^{-1} \vecxi_1 \|
\leq \| \trans {T^{(n)}}^{-1}\|\,\| \vecxi_1 \|
= \| {T^{(n)}}^{-1} \| \, \frac{| \vecxi\cdot\vecomega^{(n-1)} |}
{\|\vecomega^{(n-1)}\|}
\leq \sigma_{n-1} \| {T^{(n)}}^{-1} \| \, \| \vecxi \| 
\end{equation}
since $\vecxi\in K^{(n-1)}$ and $\|\vecomega^{(n-1)}\|
=\|(\trans\vecalf^{(n-1)},1)\|\geq 1$. Hence in view of (\ref{T2})
\begin{equation}
\| \trans {T^{(n)}}^{-1} \vecxi_1 \|
\leq c_6 \sigma_{n-1} \exp[(d-1)(1-\theta)\delta t_n + d\,\theta\,t_n]
\|\vecxi\|.
\end{equation}
Secondly, from Proposition \ref{hypprop} we infer
\begin{equation}
\| \trans {T^{(n)}}^{-1} \vecxi_2 \| \leq 
\frac12\Lambda  
\exp\big[-(1-\theta)\delta t_n +d\,\theta\, t_{n-1}\big] \|\vecxi\| .
\end{equation}
This proves \eqref{formula An}.
\end{proof}

\begin{remark}
Note that if the $t_n$ are chosen as in {\rm (\ref{tn})}, and 
\begin{equation}\label{siggy}
\sigma_{n-1} \leq \frac12c_6^{-1} \Lambda \exp(-d\, \delta t_n),
\end{equation}
then
\begin{equation}\label{L2}
\| \trans {T^{(n)}}^{-1} \vecxi \| 
\leq \Lambda \exp(-\varphi) \|  \vecxi \| 
\end{equation}
for all $\vecxi\in K^{(n-1)}$, $n\in\Nn$ and $\varphi>0$.
\end{remark}


\section{Renormalization of vector fields}
\label{Section: Renormalization of vector fields}

\subsection{Definitions}

The transformation of a vector field $X$ on a manifold $M$ by a diffeomorphism $\psi\colon
M\to M$ is given by the {\it pull-back} of $X$ under $\psi$:
$$
\psi^*X=(D\psi)^{-1}X\circ \psi.
$$

As the tangent bundle of the $d$-torus is trivial,
$T\Tt^d\simeq\Tt^d\times\Rr^d$, we identify the
set of vector fields on $\Tt^d$ with the set of functions from $\Tt^d$
to $\Rr^d$, that can be regarded as maps of $\Rr^d$ by lifting to the
universal cover.
We will make use of the analyticity to extend to the complex domain,
so we will deal with complex analytic functions.
We will also be considering an extra variable related to a parameter.

\begin{remark}\label{remark analyticity}
We will be using maps between Banach spaces over $\Cc$ with a notion of analyticity stated as follows (cf. e.g. \cite{Hille}):
a map $F$ defined on a domain is analytic if it is locally bounded and G\^ateux differentiable.
If it is analytic on a domain, it is continuous and Fr\'echet differentiable.
Moreover, we have a convergence theorem which is going to be used later on.
Let $\{F_k\}$ be a sequence of functions analytic and uniformly locally bounded on a domain $D$.
If $\lim_{k\to+\infty}F_k=F$ on $D$, then $F$ is analytic on $D$.
\end{remark}

Let $\rho,a,b>0$, $r=(a,b)$ and consider the domain
\begin{equation}
D_\rho\times B_r,
\end{equation}
where $D_\rho =\{\vecx\in\Cc^{d} \colon \|\im \vecx\| < \rho/2\pi\}$ for the norm $\|{\vecu}\|=\sum_i|u_i|$ on $\Cc^d$, and
\begin{equation}
B_r  = 
\left\{\vecy=(y_1,\dots,y_d)\in\Cc^d\colon 
\sum_{i=1}^{d-1} |y_i| < a
\text{ and }
|y_d| < b 
\right\}.
\end{equation}

Take complex analytic functions $f\colon D_\rho\times B_r \to\Cc^{d}$ that are
$\Zz^d$-periodic on the first coordinate and on the form of the Fourier series 
\begin{equation}\label{F series vf}
f(\vecx,\vecy)=\sum\limits_{\veck\in\Zz^d}f_\veck(\vecy)\e^{2\pi \i \veck\cdot \vecx}.
\end{equation}
Its coefficients are analytic functions $f_\veck\colon B_{r}\to\Cc^d$ with a continuous extension to the closure $\overline{B_r}$, endowed with the sup-norm:
$$
\|f_\veck\|_{r}=\sup_{\vecy\in B_r}\|f_\veck(\vecy)\|.
$$
The Banach spaces $\A_{\rho,r}$ and
$\A'_{\rho,r}$ are the subspaces of such functions such that the respective norms 
\begin{eqnarray*}
\|f\|_{\rho,r} 
&=&
\sum\limits_{\veck\in\Zz^d} \|f_\veck\|_{r} \, \e^{ \rho \|\veck\|},
\\
\|f\|'_{\rho,r} 
&=&
\sum\limits_{\veck\in\Zz^d} \left(1+2\pi\|\veck\|\right)\|f_\veck\|_{r} \,  \e^{\rho \|\veck\|}
\end{eqnarray*}
are finite.
Also, write the constant Fourier mode of $f\in\A_{\rho,r}$ through the projection
\begin{equation}
\Ee f(\vecy)=\int_{\Tt^d}f(\vecx,\vecy)d\vecx=f_0(\vecy)
\end{equation}
into the projected space denoted by $\Ee\A_{r}$.
The norm of its derivative $Df_0$ is given by the operator norm $\|Df_0\|_r=\sup_{\|g\|_r=1}\|Df_0\,g\|_r$.

Some of the properties of the above spaces are of easy verification.
For instance, given any $f,g\in\A'_{\rho,r}$ we have:
\begin{itemize}
\item
$\|f(\vecx,\vecy)\|\leq\|f\|_{\rho,r}\leq\|f\|'_{\rho,r}$ where $(\vecx,\vecy)\in D_\rho\times B_r$,
\item
$\|f\|_{\rho-\delta,r}\leq\|f\|_{\rho,r}$ with $\delta<\rho$.
\end{itemize}

In order to setup notations write, according to section \ref{section:Multidimensional continued fractions}, $\vecomega^{(0)}=\vecomega\in\Rr^d-\{0\}$, $\lambda_0=1$ and, for $n\in\Nn$,
\begin{equation} \label{eq def omega n}
\vecomega^{(n)}  
= {\gamma^{(n)}}^{-1} M^{(n)} 
\left(\begin{smallmatrix}0\\ \vdots \\ 0 \\1\end{smallmatrix}\right)
= \lambda_nP^{(n)}\vecomega
= \eta_nT^{(n)}\vecomega^{(n-1)},
\end{equation} 
where 
\begin{equation} \label{eq def eta n lambda n}
\lambda_{n} =
\frac \gamma {\gamma^{(n)}} \e^{(d-1)t_{n}}
\quad\text{and}\quad
\eta_n = 
\frac{\lambda_n}{\lambda_{n-1}}.
\end{equation}

In the following, we will be interested in equilibria-free vector fields
 with a ``twist'' along the parameter direction.
By rescaling this direction we will find the right parameter which guarantees the conjugacy to a linear flow.
For a fixed $n\in\Nn\cup\{0\}$, we will be studying vector fields of the form
\begin{equation}\label{form of vfs X}
X(\vecx,\vecy)=X^0_{n}(\vecy)+f(\vecx,\vecy),
\qquad
(\vecx,\vecy)\in D_\rho\times B_r,
\end{equation}
where $f\in \A_{\rho,r}$ and
\begin{equation}
X^0_{n}(\vecy)=\vecomega^{(n)}+{\gamma^{(n)}}^{-1} M^{(n)}\vecy.
\end{equation} 
(We drop the second coordinate of the vector field because it will always be equal to zero -- there is no dynamics along the parameter direction.)
The linear transformation on $\vecy$ deforms the set $B_r$ along the directions of the columns of $M^{(n)}$ (see \eqref{Malf2}). 
In particular, its $d$th column corresponds to $\vecomega^{(n)}$.

For the space of the above vector fields we use the same notation $\A_{\rho,r}$ and the same norm $\|\cdot\|_{\rho,r}$ without ambiguity.


\subsection{Resonance modes}

Given $\sigma_n>0$ we define the {\em far
from resonance} Fourier modes $f_\veck$ as in \eqref{F series vf} with respect to $\vecomega^{(n)}$ to be the ones whose indices $\veck$ are in the cone 
\begin{equation}
I_n^-=
\{\veck\in\Zz^d\colon |\veck\cdot\vecomega^{(n)}| > \sigma_n\|\veck\|\}.
\end{equation}
Similarly, the {\em resonant} modes correspond to the cone 
\begin{equation}
I_n^+=\Zz^d-I^-_n.
\end{equation}

It is also useful to define the projections
$\Ii_n^+$ and $\Ii_n^-$ on $\A_{\rho,r}$ and $\A'_{\rho,r}$ by restricting the Fourier modes to $I_n^+$
and $I_n^-$, respectively.
The identity operator is $\Ii=\Ii_n^++\Ii_n^-$.

Moreover, take
\begin{equation}\label{defn of An}
A_n=\sup_{\veck\in I^+_n-\{0\}}
\frac{\|\trans{T^{(n+1)}}^{-1}\veck\|}{\|\veck\|}.
\end{equation}

A useful property of the above cones is included in the Lemma below.

\begin{lemma}\label{lemma res cone with y, vf}
If $\veck\in I_n^-$ and $\vecy\in B_{r_n}$ with $r_n=(a_n,b_n)$,
\begin{equation}\label{cdns on an and bn cone}
a_n \leq \sigma_n \left(\frac12-b_n\right)\,|\gamma^{(n)}|\,\|M^{(n)}\|^{-1}
\quad\text{and}\quad
b_n < \frac12,
\end{equation}
then
\begin{equation}\label{bdd res cone with y}
\left|
\veck\cdot X_n^0(\vecy)
\right| 
> \frac{\sigma_n}2 \|\veck\|.
\end{equation}
\end{lemma}

\begin{proof}
For every $\vecy\in B_{r_n}$ and $\veck\in I_n^-$,
\begin{equation}
\begin{split}
|\veck\cdot(\omega^{(n)}+{\gamma^{(n)}}^{-1} M^{(n)}\vecy)| &=
|(1+y_d)\veck\cdot\vecomega^{(n)} + {\gamma^{(n)}}^{-1}\veck\cdot M^{(n)}(y_1,\dots,y_{d-1},0)|
\\
& >
(1-b_n)\sigma_n\|\veck\|
-a_n\, |\gamma^{(n)}|^{-1} \|M^{(n)}\| \,\|\veck\|.
\end{split}
\end{equation}
Our choice of $a_n$ yields \eqref{bdd res cone with y}.
\end{proof}


\subsection{Basis change, time rescaling and reparametrization}

The fundamental step of the renormalization is a transformation of the
domain of definition of our vector fields.
This is done by a linear change of basis (coming essentially from the multidimensional continued fraction expansion of $\vecomega$ -- see section \ref{section:Multidimensional continued fractions}),
a linear rescaling of time because the orbits take longer to cross the new torus, and a change of variables for the parameter $\vecy$ in order to deal with the zero mode of the perturbation.

Let $\rho_{n-1},a_{n-1},b_{n-1}>0$, $r_{n-1}=(a_{n-1},b_{n-1})$ and consider a vector field 
\begin{equation}
X(\vecx,\vecy)=X_{n-1}^0(\vecy)+f(\vecx,\vecy),
\qquad
(\vecx,\vecy)\in D_{\rho_{n-1}}\times B_{r_{n-1}},
\end{equation}
with $f\in\A_{\rho_{n-1},r_{n-1}}$.
We are interested in the following coordinate and time linear changes:
\begin{equation}\label{coord x and time change}
\vecx\mapsto {T^{(n)}}^{-1}\vecx,
\qquad
t\mapsto \eta_n t.
\end{equation}
Notice that negative time rescalings are possible, meaning that we are inverting the direction of time. 
In addition to \eqref{coord x and time change} we will use a transformation on $\vecy$, a map $\vecy\mapsto \Phi_n(X)(\vecy)$ depending on $X$ in a way to be defined later.

Therefore, consider the transformation
\begin{equation}\label{def Ln}
L_n(\vecx,\vecy)=({T^{(n)}}^{-1}\vecx, \Phi_n(X)(\vecy)),
\qquad
(\vecx,\vecy)\in\Cc^{2d},
\end{equation}
that determines a vector field in the new coordinates as the image of the map
$$
X\mapsto\LL_n(X)=\eta_n L_n^*X.
$$
That is, for $(\vecx,\vecy)\in L_n^{-1} D_{\rho_{n-1}}\times B_{r_{n-1}}$,
\begin{equation}
\begin{split}
\LL_n(X)(\vecx,\vecy)
=& \, \eta_n T^{(n)}
[\vecomega^{(n-1)}+{\gamma^{(n-1)}}^{-1}M^{(n-1)}\Phi_n(X)(\vecy) + f_0\circ \Phi_n(X)(\vecy)]
\\
&
+\eta_n T^{(n)}(f-f_0) \circ L_n(\vecx,\vecy).
\end{split}
\end{equation}
In order to eliminate the $\veck=0$ mode of the perturbation of $X$ in the new coordinates and to normalise the linear term in $\vecy$ to ${\gamma^{(n)}}^{-1}M^{(n)}\vecy$, using the definitions of $T^{(n)}$ and $\eta_n$ we choose
\begin{equation}\label{formula for Mn and bn}
\Phi_n(X)\colon \vecy
\mapsto 
\left(\id+\gamma^{(n-1)}{M^{(n-1)}}^{-1} f_0\right)^{-1} 
(\e^{-d\delta t_n}y_1,\dots,\e^{-d\delta t_n}y_{d-1},y_d),
\end{equation}
if possible.
Hence,
\begin{equation}\label{formula resulting vf Ln}
\LL_n(X)(\vecx,\vecy)
=X_n^0(\vecx,\vecy) 
+\widetilde\LL_n(f-f_0)(\vecx,\vecy),
\end{equation}
where
\begin{equation}
\widetilde\LL_n \colon  f\mapsto \eta_n T^{(n)}f\circ L_n.
\end{equation}

Denote by $\Delta_\mu$ the set of $X\in\A_{\rho_{n-1},r_{n-1}}$ such that $\|f_0\|_{r_{n-1}}<\mu$.

\begin{lemma}\label{lemma b and M}
Let $r_n=(a_n,b_n)$ and $\mu_{n-1}>0$
such that
\begin{equation}\label{control on r}
\begin{split}
a_n & \leq 
\e^{d\delta t_n} 
\left[a_{n-1}-\left(1+|\gamma^{(n-1)}|\,\|{M^{(n-1)}}^{-1}\|\right)\mu_{n-1}\right]
\\
b_n & \leq 
b_{n-1}-
\left(1+|\gamma^{(n-1)}|\,\|{M^{(n-1)}}^{-1}\|\right)\mu_{n-1}.
\end{split}
\end{equation}
There exist an analytic map $\Phi_n\colon\Delta_{\mu_{n-1}}\to \Diff(B_{r_n},\Cc^d)$ such that, for each $X\in \Delta_{\mu_{n-1}}$, $\Phi_n(X)$ is given by \eqref{formula for Mn and bn} and
\begin{equation}\label{estimate on ball acted by M and b}
\Phi_n(X)(B_{r_n}) \subset B_{r_{n-1}}.
\end{equation}
In case $f_0$ is real-analytic, $\Phi_n(X)|{\Rr^d}$ is also real-valued. 
\end{lemma}

\begin{proof}
For $X\in \A_{\rho_{n-1},r_{n-1}}$ with $\|f_0\|_{r_{n-1}}<\mu_{n-1}$ and $\widehat\delta_{n-1}=(\delta_{n-1},\delta_{n-1})$ with
\begin{equation}
\delta_{n-1}=\mu_{n-1}|\gamma^{(n-1)}|\,\|{M^{(n-1)}}^{-1}\|,
\end{equation}
 we have by the Cauchy estimate
$$
\|Df_0\|_{r_{n-1}-\widehat\delta_{n-1}} \leq 
\frac{\|f_0\|_{r_{n-1}}}{\delta_{n-1}}  
< \frac1{|\gamma^{(n-1)}|\,\|{M^{(n-1)}}^{-1}\|}.
$$
So, $F=\id+\gamma^{(n-1)}{M^{(n-1)}}^{-1}f_0$ is a diffeomorphism on $B_{r_{n-1}-\widehat\delta_{n-1}}$.
Now, if $R_1<a_{n-1}-\delta_{n-1}-\mu_{n-1}$, $R_2<b_{n-1}-\delta_{n-1}-\mu_{n-1}$ and $R=(R_1,R_2)$, we have $B_R\subset F(B_{r_{n-1}-\widehat\delta_{n-1}})$ and $F^{-1}(B_R)\subset B_{r_{n-1}-\widehat\delta_{n-1}}$.
Therefore, $\Phi_n(X)$ as given by \eqref{formula for Mn and bn} is a diffeomorphism on $B_{r_n}$ by choosing $R=(\e^{-d\delta t_n}a_n,b_n)$, and thus we get \eqref{estimate on ball acted by M and b}.
In addition, $X\mapsto \Phi_n(X)$ is analytic from its dependence on $f_0$. 
When restricted to a real domain for a real-analytic $f_0$, $\Phi_n(X)$ is also real-analytic.
\end{proof}

Let the translation $R_\vecz$ on $\Cc^{2d}$ be defined for
$\vecz\in\Cc^d$ and given by 
\begin{equation}\label{translation Rz}
R_\vecz\colon (\vecx,\vecy)\mapsto (\vecx+\vecz,\vecy).
\end{equation}
Notice that we have the following ``commutative'' relation:
\begin{equation}\label{commutation property Rz and Ln}
L_n^*R_\vecz^*=R_{{T^{(n)}}\vecz}^*L_n^*,
\qquad
\vecz\in\Cc^d.
\end{equation}
This also follows from the fact that $\Phi_n$ is unchanged by the introduction of the translation $R_\vecz$.

\subsection{Analyticity improvement}

\begin{lemma}\label{proposition TT}
If $\delta>0$ and
\begin{equation}\label{hyp on rho n' for vf}
\rho_n'\leq\frac{\rho_{n-1}}{A_{n-1}}-\delta,
\end{equation}
then $\widetilde\LL_n$ as a map from $(\Ii^+_{n-1}-\Ee)\A_{\rho_{n-1},r_{n-1}}\cap \Delta_{\mu_{n-1}}$ into $(\Ii-\Ee)\A'_{\rho_n',r_n}$
is continuous and compact with
\begin{equation}\label{bdd tilde LL n}
\|\widetilde\LL_n\|\leq
|\eta_{n}|\, \|{T^{(n)}} \|\,
\left(1+\frac{2\pi}{\delta}\right).
\end{equation}
\end{lemma}

\begin{remark}
This result means that every vector field in $\Ii_{n-1}^+\A_{\rho_{n-1},r_{n-1}}\cap \Delta_{\mu_{n-1}}$, i.e. a function on $D_{\rho_{n-1}}\times B_{r_{n-1}}$ into $\Cc^d$, has an analytic extension to 
${T^{(n)}}^{-1}D_{\rho_n'}\times B_{r_{n-1}}$.
\end{remark}

\begin{proof}
Let $f\in(\Ii^+_{n-1}-\Ee)\A_{\rho_{n-1},r_{n-1}}\cap \Delta_{\mu_{n-1}}$.
Then,
\begin{equation}
\|f\circ L_n\|'_{\rho_n',r_n}
\leq 
\sum_{\veck\in I^+_{n-1}-\{0\}} 
\left(1+2\pi\|\trans{T^{(n)}}^{-1}\,\veck\|\right)
\|f_\veck\circ M_n\|_{r_n}
\e^{(\rho_n'-\delta+\delta)\|\trans{T^{(n)}}^{-1} \veck\|}.
\end{equation}
By using the relation $\xi \e^{-\delta\,\xi}\leq \delta^{-1}$ with $\xi\geq0$, \eqref{defn of An} and \eqref{estimate on ball acted by M and b}, we get
\begin{equation}
\begin{split}
\|f\circ L_n\|'_{\rho_n',r_n}
\leq &
\left(1+2\pi/\delta\right)
\sum_{I^+_{n-1}-\{0\}} 
\|f_\veck\|_{r_{n-1}}
\e^{A_{n-1}(\rho_n'+\delta)\|\veck\|}
\\
\leq &
\left(1+2\pi/\delta\right)
\|f\|_{\rho_{n-1},r_{n-1}}.
\end{split}
\end{equation}
Finally, $\|\widetilde\LL_n f\|'_{\rho_n',r_n}\leq |\eta_{n}|\,\|{T^{(n)}} \| \,\|f\circ L_n\|'_{\rho_n',r_n}$.

The above for $D_{\rho_n'}\times B_{r_n}$ is also valid for $D_{\zeta}\times B_{r_n}$, $\zeta>\rho_n'$ but satisfying a similar inequality to \eqref{hyp on rho n' for vf}.
Therefore, $\widetilde\LL_n=\II\circ\JJ$, where
$\JJ\colon(\Ii^+_{n-1}-\Ee)\A_{\rho_{n-1},r_{n-1}}\to\A'_{\zeta,r_n}$ is bounded as $\widetilde\LL_n$, and the inclusion map
$\II\colon\A'_{\zeta,r_n}\to\A'_{\rho_n',r_n}$ is compact.
\end{proof}

For $0<\rho_n''\leq \rho'_n$, consider the inclusion
\begin{equation}
\II_n\colon \A'_{\rho_n',r_n}\to \A'_{\rho_n'',r_n}
\end{equation}
by restricting $X\in \A'_{\rho_n',r_n}$ to the smaller domain $D_{\rho_n''}\times B_{r_n}$.
When restricted to non-constant modes, its norm can be estimated as follows.

\begin{lemma}\label{lemma cutoff}
If $\phi_n \geq 1$ and 
\begin{equation}\label{hyp rho''n vf}
0<\rho''_n \leq \rho'_n - \log(\phi_n),
\end{equation}
then
\begin{equation}
\|\II_n (\Ii-\Ee)\| \leq \phi_n^{-1}.
\end{equation}
\end{lemma}

\begin{proof}
For $f\in(\Ii-\Ee)\A'_{\rho_n',r_n}$, we have
\begin{equation}
\|\II_n(f)\|'_{\rho_n'',r_n} \leq
\sum_{\veck\not=0}(1+2\pi\|\veck\|)\|f_\veck\|_{r_n}\e^{\rho'_n\|\veck\|}
\phi_n^{-\|\veck\|}
 \leq
\phi_n^{-1} \|f\|'_{\rho_n',r_n}.
\end{equation}
\end{proof}


\subsection{Elimination of far from resonance modes}

The theorem below (to be proven in Section \ref{section proof of main theorem}) states the
existence of a nonlinear change of coordinates $U$, isotopic to the
identity, that cancels the $I_n^-$ modes of any
$X$ as in (\ref{form of vfs X}) with sufficiently small $f$.
We are eliminating only the far from resonance modes,
this way avoiding the complications usually related to small
divisors.
We remark that the ``parameter'' direction $\vecy$ is not affected by this
change of coordinates.

For given $\rho_n,r_n,\varepsilon,\nu>0$, denote by $\VV_\varepsilon$ the open ball in $\A'_{\rho_n+\nu,r_n}$ centred at $X_n^0$ with radius $\varepsilon$.

\begin{theorem}\label{main theorem1}
Let $r_n$ be as in \eqref{cdns on an and bn cone}, $\sigma_n<\|\vecomega^{(n)}\|$ and
\begin{equation}\label{formula epsilon}
\varepsilon_n=
\frac{\sigma_n}{42}
\min\left\{
\frac{\nu}{4\pi},
\frac{\sigma_n}{72\|\vecomega^{(n)}\|}
\right\}.
\end{equation}
For all $X\in\VV_{\varepsilon_n}$ there exists an isotopy 
\begin{equation}
\begin{split}
U_{t}\colon D_{\rho_n}\times B_{r_n} & \to D_{\rho_n+\nu}\times B_{r_n},
\\
(\vecx,\vecy) & \mapsto (\vecx+u_t(\vecx,\vecy),\vecy),
\end{split}
\end{equation}
of analytic diffeomorphisms with $u_t$ in $\A'_{\rho_n,r_n}$, $t\in[0,1]$,
satisfying 
\begin{equation}\label{equation homotopy method}
\Ii_n^-U_{t}^*X=(1-t)\,\Ii_n^-X,
\qquad
U_{0}=\id.
\end{equation}
This defines the maps 
\begin{equation}
\begin{split}
\fU_t\colon & \VV_{\varepsilon_n} \to \A'_{\rho_n,r_n} \\
& X\mapsto \id+u_t
\end{split}
\end{equation}
and
\begin{equation}
\begin{split}
\UU_t\colon & \VV_{\varepsilon_n} \to  \Ii^+\A_{\rho_n,r_n}\oplus(1-t)\Ii_n^-\A'_{\rho_n+\nu,r_n} \\
& X\mapsto U_t^*X 
\end{split}
\end{equation}
which are analytic, and 
satisfy the inequalities
\begin{equation}\label{estimate U around X0}
\begin{split}
\|\fU_t(X)-\id\|'_{\rho_n,r_n}
\leq & 
\frac{42t}{\sigma_n} \|\Ii_n^-f\|_{\rho_n,r_n}\\
\|\UU_t(X)-X_n^0\|_{\rho_n,r_n}
\leq &
(3-t)
\|f\|'_{\rho_n+\nu,r_n}.
\end{split}
\end{equation}
If $X$ is real-analytic, then $\fU_t(X)(\Rr^{2d})\subset\Rr^{2d}$.
\end{theorem}

\begin{remark}
Further on we will be using the above result for $t=1$. So that all far from resonance modes are eliminated.
\end{remark}

Recall the definition of the translation $R_\vecz$ in \eqref{translation Rz}.

\begin{lemma}\label{theorem 2 elim terms}
In the conditions of Theorem {\rm \ref{main theorem1}},
if $\vecx\in \Rr^d$ and $X\in \VV_{\varepsilon_n}$, then
\begin{equation}\label{relation tilde U and U}
\fU_t(X\circ R_\vecx)=R_\vecx^{-1}\circ \fU_t(X)\circ R_\vecx
\end{equation}
on $\DD_{\rho_n,r_n}$.
\end{lemma}

\begin{proof}
Notice that $R_\vecx (D_{\rho_n}\times B_{r_n})=D_{\rho_n}\times B_{r_n}$.
If $U_t=\fU_t(X)$ is a solution of the homotopy equation \eqref{equation homotopy method} on $D_{\rho_n}\times B_{r_n}$, then 
$\tilde U_t=R_\vecx^{-1}\circ \fU_t(X)\circ R_\vecx$
solves the same equation for $\tilde X=X\circ R_\vecx$, i.e. 
$\Ii_n^-\tilde X\circ \tilde U_t=(1-t)\Ii_n^-\tilde X$, on $D_{\rho_n}\times B_{r_n}$.
\end{proof}

\subsection {Trivial limit of renormalization}\label{The Limit Set of the Renormalization}

Let a sequence of ``widths'' $0<\sigma_n<1$ of the resonance cones $I_n^+$ be given.
The {\em $n$th step renormalization} operator is thus
$$
\RR_n=\UU_n \circ \II_n \circ \LL_n \circ\RR_{n-1}
\quad\text{and}\quad
\RR_0=\UU_0,
$$
where $\UU_n$ is the full elimination of the
modes in $I^-_n$ as in Theorem \ref{main theorem1} (for $t=1$).
Notice that
$\RR_n(X^0+\vecv)=X^0_n$,
for every $\vecv\in\Cc^d$.
 From the previous sections the map $\RR_n$ on its domain is analytic.
Also, in case a vector field $X$ is real-analytic, the same is true for
$\RR_n(X)$.

Fix the constants $\nu$ and $\delta$ as in Theorem \ref{main theorem1} and Lemma \ref{proposition TT}, respectively, and choose $0<\lambda<1$.
Take
\begin{equation}\label{formula theta n vf}
\Theta_n =
\min\left\{\varepsilon_n, 
\frac{\lambda^n \sigma_n^2}{\prod_{i=1}^n\|{T^{(i)}}^{-1}\|^2},
\lambda^n\frac{
\frac{\sigma_n |\gamma^{(n)}|}{\|M^{(n)}\|}
- \frac{\sigma_{n+1}|\gamma^{(n+1)}|}{\e^{d\delta t_{n+1}}\|M^{(n+1)}\|}}
{1+|\gamma^{(n)}|\, \|{M^{(n)}}^{-1}\|}
\right\}
\end{equation}
by assuming that the sequence of times $t_n$ guarantees that $\Theta_n>0$.
Now, write
\begin{equation}
B_n=\prod_{i=0}^n A_i.
\end{equation}
with $A_i$ given by \eqref{defn of An}.
By recalling the inequalities \eqref{hyp on rho n' for vf} and \eqref{hyp rho''n vf} we choose, for a given $\rho_0>0$,
\begin{equation}\label{formula rho n vf}
\rho_n = 
\frac1{B_{n-1}}
\left[ \rho_0 - 
\sum_{i=0}^{n-1} B_i \log\left(\phi_{i+1}\right)
-(\delta+\nu)\sum_{i=0}^{n-1} B_i
\right],
\end{equation}
where
\begin{equation}
\phi_n=
\max\left\{
2|\eta_n|\, \|T^{(n)}\| (1+2\pi\delta^{-1})\frac{\Theta_{n-1}}{\Theta_n},
1\right\} 
\geq1
\end{equation}
is to be used in Lemma \ref{lemma cutoff}.

Define the following function for every $\vecomega\in\Rr^d$ associated to the choice of $\sigma_n$:
\begin{equation}\label{def B omega}
\BB(\vecomega)=\sum_{i=0}^{+\infty} B_i \log\left(\phi_{i+1}\right) + (\delta+\nu)\sum_{i=0}^{+\infty} B_i.
\end{equation}

The convergence of the renormalization scheme now
follows directly from our construction.

\begin{theorem}\label{convergence for Nn}\label{corollary Rn}
Suppose that
\begin{equation}\label{cdn on the frequency}
\BB(\vecomega)<+\infty
\end{equation}
and $\rho>\BB(\vecomega)+\nu$.
There is $K,b>0$ and $r_n=(a_n,b_n)$ with $a_n>0$ and $b_n>b>0$, such that if $X$ is in a sufficiently small open ball around $X^0$ in $\A_{\rho,r_0}$, then 
\begin{itemize}
\item[(i)]
$X$ is in the domain of $\RR_n$ and
\begin{equation}\label{bound on LLn less than epsilon}
\|\RR_n(X)-\RR_n(X^0)\|_{\rho_n,r_n}
\leq
K\Theta_n \|X-X^0\|_{\rho,r_0},
\quad
n\in\Nn\cup\{0\},
\end{equation}
\item[(ii)]
for each $|s|<b$ there exists in $B_{r_{n-1}}\subset\Cc^d$ the limits
\begin{equation}\label{def an sum}
p_n^s(X) =\lim_{m\to+\infty} \Phi_n(\RR_{n-1}(X))\dots \Phi_m(\RR_{m-1}(X))
(0,\dots,0,s) 
\end{equation}
and
\begin{equation}\label{estimate p-s}
\lim_{n\to+\infty}\|p_n^s(X)-(0,\dots,0,s)\|=0,
\end{equation}
\item[(iii)]
the map $X\mapsto p_n(X)$ is analytic and takes any real-analytic $X$ into an analytic curve $s\mapsto p_n^s(X)$ in $\Rr^d$.
\end{itemize}
\end{theorem}

\begin{proof}
Let $\xi>0$ and $\rho_0=\rho-\nu-\xi>0$ such that $\rho_0 >\BB(\vecomega)$.
Hence, by \eqref{formula rho n vf}, we have $R>0$ satisfying $\rho_n>R B_{n-1}^{-1}$ for all $n\in\Nn$.

Denote by $c$ the radius of an open ball in $\A_{\rho,r_0}$ centred at $X^0$ and containing $X$.
If $c\leq\varepsilon_0$ we can use Theorem \ref{main theorem1} to obtain $\RR_0(X)\in\Ii_0^+\A_{\rho_0,r_0}$ with $r_0=(a_0,b_0)$ satisfying \eqref{cdns on an and bn cone} and
$$
\|\RR_0(X)-\RR_0(X^0)\|_{\rho_0,r_0}\leq
2 \|X-X^0\|'_{\rho+\xi,r_0}
\leq 2\xi^{-1}\|X-X^0\|_{\rho,r_0}.
$$
Let $K=2(\xi\Theta_0)^{-1}$ and assume that $c\leq K^{-1}\min\{b_0(1-\lambda), \frac12-b_0\}$.
So, \eqref{bound on LLn less than epsilon} holds for $n=0$.

Now, with $n\in\Nn$ we choose the following $r_n$:
\begin{equation}
a_n  =
\sigma_n\left(\frac12-b_0\right)\frac{|\gamma^{(n)}|}{\|M^{(n)}\|}
\quad\text{and}\quad
b_n =
b_0 - cK \sum_{i=0}^{n-1}\lambda^i,
\end{equation}
so that $1/2>b_n>b=b_0-cK(1-\lambda)^{-1}$.
The inequalities in \eqref{cdns on an and bn cone} follow immediately.
Moreover, \eqref{control on r} is also satisfied with $\mu_{n-1}=cK\Theta_{n-1}$ because
\begin{equation}
\begin{split}
a_{n-1} - \e^{-d\delta t_n} a_n  
& \geq
(\frac12-b_0)(1+|\gamma^{(n-1)}|\,\|{M^{(n-1)}}^{-1}\|)\Theta_{n-1}
\\
& \geq
(1+|\gamma^{(n-1)}|\,\|{M^{(n-1)}}^{-1}\|)cK\Theta_{n-1},
\\
b_{n-1}-b_n & = cK \lambda^{n-1}
\\
& \geq
(1+|\gamma^{(n-1)}|\,\|{M^{(n-1)}}^{-1}\|)cK\Theta_{n-1}.
\end{split}
\end{equation}

Suppose that $X_{n-1}=\RR_{n-1}(X)\in\Ii^+_{n-1}\A_{\rho_{n-1},r_{n-1}}$ and
$$
\|X_{n-1}-X_{n-1}^0\|_{\rho_{n-1},r_{n-1}}\leq
K \Theta_{n-1} \|X-X^0\|_{\rho,r}.
$$
Since \eqref{control on r} holds, Lemmas \ref{lemma b and M} and \ref{proposition TT} are valid and, together with \eqref{formula resulting vf Ln} and Lemma \ref{lemma cutoff}, can be used to estimate $\II_n\circ\LL_n(X_{n-1})$:
\begin{equation}
\begin{split}
\|\II_n\circ \LL_n(X_{n-1})-X_n^0\|'_{\rho''_n,r_n}
&\leq 
|\eta_n|\, \|T^{(n)}\| (1+2\pi\delta^{-1})
\phi_n^{-1} K \Theta_{n-1} \|X-X^0\|_{\rho,r_0}
\\
&=
\frac12 K \Theta_n \|X-X^0\|_{\rho,r_0}.
\end{split}
\end{equation}
This vector field is inside the domain of $\UU_n$ as \eqref{cdns on an and bn cone} and $\frac12 c\,K\Theta_n<\varepsilon_n$ are satisfied.
Thus \eqref{bound on LLn less than epsilon} follows from \eqref{estimate U around X0}.

Denote by $f_0^{(n)}$ the constant mode of the perturbation term of $X_n$.
By Lemma \ref{lemma b and M}, $\Phi_n(X_{n-1})\colon B_{r_n}\to B_{r_{n-1}}$ is given by 
$$
\vecy\mapsto (\id+g_n)\diag(\e^{-d\delta t_n},\dots,\e^{-d\delta t_n},1)\vecy,
$$
where
\begin{equation}
g_n=\left(\id+\gamma^{(n-1)}{M^{(n-1)}}^{-1}f_0^{(n-1)}\right)^{-1}-\id
\end{equation}
is defined on $B_{r'_n}$ with $r'_n=(\e^{-d\delta t_n} a_n,b_n)$.
So, for $\vecz\in B_{r'_n}$ there is $\vecxi\in B_{r'_n}$ such that 
\begin{equation}
\begin{split}
g_n(\vecz) & = [I+\gamma^{(n-1)}{M^{(n-1)}}^{-1}Df_0^{(n-1)}(\vecxi)]^{-1}[\vecz-\gamma^{(n-1)}{M^{(n-1)}}^{-1}f_0^{(n-1)}(0)]-\vecz
\\
& = -[I+\gamma^{(n-1)}{M^{(n-1)}}^{-1}Df_0^{(n-1)}(\vecxi)]^{-1}\gamma^{(n-1)}{M^{(n-1)}}^{-1}[Df_0^{(n-1)}(\vecxi)\,\vecz+f_0^{(n-1)}(0)]
\end{split}
\end{equation}
and
\begin{equation}\label{bdd on g n}
\begin{split}
\|g_n\|_{r'_n} 
&\leq 
\frac{|\gamma^{(n-1)}|\,\|{M^{(n-1)}}^{-1}\|}{1-|\gamma^{(n-1)}|\|{M^{(n-1)}}^{-1}\|\,\|Df_0^{(n-1)}\|_{r'_n}} 
\left(\|r'_n\|\, \|Df_0^{(n-1)}\|_{r'_n} 
+\|f_0^{(n-1)}\|_{r'_n}\right).
\end{split}
\end{equation}

The choice of $r_n$ means that
\begin{equation}
\min\{
a_{n-1}-\e^{-d\delta t_n} a_n ,b_{n-1}-b_n\}
\gg
\min\left\{
\frac{\sigma_{n-1}|\gamma^{(n-1)}|}{\|M^{(n-1)}\|}, \lambda^{n-1}
\right\}.
\end{equation}
By using \eqref{formula theta n vf} and the Cauchy estimate,
\begin{equation}
\|Df_0^{(n-1)}\|_{r'_n}\leq
\frac{\|f_0^{(n-1)}\|_{r_{n-1}}}{\min\{a_{n-1}-\e^{-d\delta t_n}a_n,b_{n-1}-b_n\}}
\ll
\frac{\lambda^{n-1}}{|\gamma^{(n-1)}|\,\|{M^{(n-1)}}^{-1}\|}
\end{equation}
Thus, 
\begin{equation}\label{up est g n}
\|g_n\|_{r'_n}\ll\lambda^{n-1}.
\end{equation}

Writing $\vecy_s=(0,\dots,0,s)$, by induction we have
\begin{equation}
\Phi_n(X_{n-1})\dots \Phi_m(X_{m-1})(\vecy_s) = 
\vecy_s+ \sum_{i=n}^m 
\diag(\e^{-d(t_{i-1}-t_{n-1})},\dots,\e^{-d(t_{i-1}-t_{n-1})} ,1)
g_i(\vecxi_i),
\end{equation}
for some $\vecxi_k\in B_{r'_k}$.
Therefore, from \eqref{up est g n}, there exists $p_n^s(X)\in\Cc^d$ unless $X$ is real which clearly gives $p_n^s(X)\in\Rr^d$.
In addition,
\begin{equation}
\|p_n^s(X)-\vecy_s\|
\leq 
\sum_{i=n}^{+\infty} \|g_i\|_{r'_i}
\ll
\frac{\lambda^{n-1}}{1-\lambda}.
\end{equation}

The maps $X\mapsto p_n^s(X)$ are analytic since the convergence is uniform.
Lemma \ref{lemma b and M} gives us the nested sequence $\Phi_n(X_{n-1})(B_{r_n})\subset B_{r_{n-1}}$.
So, as  $\vecy_s\in \cap_{i\in\Nn}B_{r_i}$, it follows that $p_n^s(X)\in B_{r_{n-1}}$.
\end{proof}

\begin{remark}\label{remark on small analyt r}
The above can be generalised for a small analyticity radius $\rho$ by considering a sufficiently large $N$ and applying the above theorem to $\widetilde X=\UU_N\LL_N\dots\UU_1\LL_1\UU_0(X)$, where $X$ is close enough to $X^0$.
We recover the large strip case since $\rho_N$ is of the order of $B_{N-1}^{-1}$. 
It remains to check that $\rho_N>\BB(\vecomega^{(N)})+\nu$.
This follows from the fact that $\BB(\vecomega^{(N)})=B_{N-1}^{-1}[\BB(\vecomega)-\BB_N(\vecomega)]$ where $\BB_N(\vecomega)$ is the sum of the first $N$ terms of $\BB(\vecomega)$ so that $\BB_N(\vecomega)\to \BB(\vecomega)$ as $N\to+\infty$.
\end{remark}

\begin{lemma}\label{dioph in BC}
If $\vecomega=\left(\begin{smallmatrix}\vecalf\\1\end{smallmatrix}\right)$ in $\Rr^d$ is diophantine, i.e. $\vecalf$ satisfies \eqref{dio2} with exponent $\epsilon$ (related to $\theta$ by \eqref{relation theta epsilon} and to $\beta$ by \eqref{def beta}), then \eqref{cdn on the frequency} is verified.
\end{lemma}

\begin{proof}
Let us set $\delta t_n = \xi t_{n-1}, \, \sigma_n = \exp (-c\delta t_n), \, n\geq 1$,
where positive constants $\xi,c$ will be chosen later. 
Obviously,
$t_n=(1+\xi)t_{n-1}=[(1+\xi)/\xi]\delta t_n$ and $\delta t_n= (1+\xi)\delta t_{n-1}$. 
We shall assume that
\begin{equation}
\label{1cond}
c<d(1+\xi),
\end{equation}
so that $\sigma_{n-1}\exp(d\delta t_n) = \exp (-c\delta t_{n-1} + d \delta t_n)=
\exp[(d-c/(1+\xi))\delta t_n]$ is much larger than $\Lambda$ given by Proposition \ref{hypprop}. 
Hence, using \eqref{formula An} we have
\begin{equation}
\label{AnK}
A_{n-1} \ll \exp\left[\left(-\frac{c}{1+\xi} + d-(1-\theta) + \frac{d\theta}{\xi}\right)\delta t_n\right].
\end{equation}

We next estimate $\|\vecomega^{(n)}\|$ and $\varepsilon_n$. 
It follows from \eqref{eq def omega n} that $\|\vecomega^{(n)}\|\ll\|M^{(n)}\|\,|{\gamma^{(n)}}^{-1}|$. 
Thus, using \eqref{M1}, \eqref{gamma1} we have
\begin{equation}\label{omegaK}
\|\vecomega^{(n)}\|
\ll
\exp\left(\frac{\theta}{1-\theta}d^2t_n\right)=
\exp\left(\frac{\theta}{1-\theta}d^2\frac{1+\xi}{\xi}\delta t_n\right).
\end{equation}
Since $\|\vecomega^{(n)}\|\geq 1$ one gets from \eqref{formula epsilon} that $\varepsilon_n \sim \sigma_n^2/\|\vecomega^{(n)}\|$
which together with \eqref{omegaK} implies
\begin{equation}\label{epsilonK}
\exp\left[\left(-2c - \frac{\theta}{1-\theta}d^2\frac{1+\xi}{\xi}\right)\delta t_n\right]
\ll \varepsilon_n \ll 
\exp(-2c \delta t_n).
\end{equation}
Here $X\sim Y$ means that there exist two positive constants $C_1,C_2>0$ such
that $C_1 Y<X<C_2 Y$.
Using again \eqref{M1}, \eqref{gamma1} we get 
\begin{equation}
\frac{\sigma_n|\gamma^{(n)}|}{\|M^{(n)}\|}\gg \exp \left[\left(-c - \frac{\theta}{1-\theta}d^2\frac{1+\xi}{\xi}\right)\delta t_n\right].
\end{equation}
Also, since $\|M^{(n+1)}\| \geq \gamma^{(n+1)}$, 
\begin{equation}
\frac{\sigma_{n+1}|\gamma^{(n+1)}|}{\|M^{(n+1)}\|\exp(d\delta t_{n+1})}\ll \exp [-(c+d)(1+\xi)\delta t_n].
\end{equation}
We shall assume that $c$ and $\xi$ are chosen in such a way that
\begin{equation}\label{2cond*}
-c - \frac{\theta}{1-\theta}d^2\frac{1+\xi}{\xi}> - (c+d)(1+\xi),
\end{equation}
so that
\begin{equation}\label{sgm3K}
\frac{\sigma_n|\gamma^{(n)}|}{\|M^{(n)}\|}-
\frac{\sigma_{n+1}|\gamma^{(n+1)}|}{\|M^{(n+1)}\|\exp(dt_{n+1})} 
\gg 
\frac{\sigma_n|\gamma^{(n)}|}{\|M^{(n)}\|}.
\end{equation}
Inequality \eqref{2cond*} is equivalent to the following condition
\begin{equation}\label{2cond}
c > \frac{\theta}{1-\theta}\frac{1+\xi}{\xi^2}d^2 - \frac{1+\xi}{\xi}d.
\end{equation}

Finally, we want $A_n$ to be small and, hence, require the exponent in \eqref{AnK} to be negative
\begin{equation}\label{3cond}
-\frac{c}{1+\xi} + d-(1-\theta) + \frac{d\theta}{\xi}<0.
\end{equation}

Suppose that conditions \eqref{1cond}, \eqref{2cond}, \eqref{3cond} are satisfied.
It follows immediately from the estimates above and \eqref{eq def eta n lambda n}, \eqref{M1}, \eqref{M2}, \eqref{T2}, \eqref {gamma1} that
\begin{equation}\label{log}
|\log\Theta_n|, |\log\Theta_{n-1}|, \log\|T^{(n)}\|, |\log|\eta_n|| 
\ll 
\delta t_n.
\end{equation}
At the same time
\begin{equation}\label{An1K}
B_n=\prod_{i=0}^n A_i\ll C^n\exp(-\alpha t_{n+1}),
\end{equation}
where 
\begin{equation}
\label{alphaK}
\alpha = \frac{c}{1+\xi} - d+(1-\theta) - \frac{d\theta}{\xi} >0.
\end{equation}
Since $B_n$ decays exponentially with $t_n$ and $\log \phi_n$ grows at
most linearly the series \eqref{cdn on the frequency} converges.
To finish the proof it is enough to show that conditions \eqref{1cond}, \eqref{2cond}, \eqref{3cond} can be satisfied. 
Indeed, since $0<\theta <1$ we can choose $\xi$ so large
that $1-\theta - d\theta/\xi>0$ and
\begin{equation}
\label{2cond**} 
\frac{\theta}{1-\theta}\frac{1+\xi}{\xi^2}d^2 - \frac{1+\xi}{\xi}d<0.
\end{equation}
It is easy to see that all three inequalities \eqref{1cond}, \eqref{2cond}, \eqref{3cond} are
satisfied if $(1+\xi) (d-\beta) < c < (1+\xi)d$, where $\beta =1-\theta - d\theta/\xi>0$.
\end{proof}

\subsection{Analytic conjugacy to linear flow}

As a consequence of Theorem \ref{corollary Rn}, we obtain an analytic
conjugacy between a vector field and the linear flow, 
thus proving Theorem \ref{theorem: main. intro}.
In the following we always assume to be in the conditions of 
Section \ref{The Limit Set of the Renormalization}.

Let $r=r_0$ and
\begin{equation}
\Delta=\{X\in\A_{\rho,r}\colon \|X-X^0\|_{\rho,r}<c\}
\end{equation}
inside the domain of $\RR_n$ for all $n\in\Nn\cup\{0\}$.
By taking $X\in\Delta$, we denote $X_n=\RR_n(X)\in\Ii_n^+\A_{\rho_n,r_n}$ so that
\begin{equation}\label{formula R n X}
X_n= \lambda_n\, (U_0 \circ L_1\circ U_1 \cdots L_n\circ U_n )^*(X),
\end{equation}
where $U_k=\fU_k(\II_k\LL_k(X_{k-1}))$ is given by Theorem \ref{main theorem1} for $t=1$ at the $k$th step and $L_k$ is the linear rescaling as in \eqref{def Ln} for $X_{k-1}$.

Denote by $V_n$ the coordinate change 
\begin{equation}
V_n\colon (\vecx,\vecy)\mapsto ({P^{(n)}}^{-1}\vecx, \Phi_1(X_0)\dots \Phi_n(X_{n-1})(\vecy))
\end{equation}
and set $V_0=\id$.
Thus, $L_n=V_{n-1}^{-1}\circ V_n$ and
\begin{equation}\label{formula X n with Vs}
X_n= \lambda_n\, (V_n\circ U_n)^*(V_{n-1} \circ U_{n-1}\circ V_{n-1}^{-1})^* \cdots (V_1\circ U_1\circ V_1^{-1})^*U_0^*(X).
\end{equation}
In particular, the $\vecy$-coordinate is only transformed by the second component of $V_n$.

Notice that if $X_n=X_n^0$ for some $n\in\Nn$, 
$$
\vecy=\Phi_1(X_0)\dots \Phi_n(X_{n-1})
(0,\dots,0,s)
\in\Cc^d,
$$
with $|s|<b$,
corresponds to the parameter for which $X$ is conjugated to $(1+s)\,\vecomega^{(n)}$.
The parameter value for the general case $X_n-X_n^0\to 0$ as $n\to+\infty$ is $p^s(X)=p_1^s(X)$.

\begin{lemma}
There is an open ball $B$ about $X^0$ in $\Delta$ such that we can find a sequence $R_n>0$ satisfying $R_{-1}=\rho$,
\begin{equation}\label{bounds on Rn for vf}
R_n+2\pi 42 K \Theta_n^{1/2} \|X-X^0\|_{\rho,r} \leq R_{n-1} \leq 
\frac{\rho_{n-1}}{\|P^{(n-1)}\|},
\quad
X\in B,
\end{equation}
and
\begin{equation}\label{control on Rn ell for vf}
\lim_{n\to+\infty}
R_n^{-1}\Theta_n^{1/2} = 0.
\end{equation}
\end{lemma}

\begin{proof}
Let $\rho_*=\min \rho_n$.
It is enough to check that $\Theta_n^{1/2}\ll \lambda^n \rho_*\prod_{i=1}^n\|T^{(i)}\|^{-1}$ with $0<\lambda<1$ and taking $R_n=c \lambda^{-n}\Theta_n^{1/2}$ for some positive constant $c$.
This immediately implies \eqref{control on Rn ell for vf} and \eqref{bounds on Rn for vf} by considering a small enough upper bound for $\|H-H^0\|_{\rho,r}$.
\end{proof}

Let $\Diff_{per}(D_\zeta,\Cc^d)$, $\zeta>0$, be the Banach space of $\Zz^d$-periodic diffeomorphisms $g\colon D_\zeta\to\Cc^d$ with finite norm $\|g\|_\zeta=\sum_\veck\|g_\veck\|\e^{\zeta\|\veck\|}$, where $g_\veck\in\Cc^d$ are the coefficients of the Fourier representation. 
It is simple to check that $\|g\circ P^{(n)} \|_{R_n} \leq \|g\|_{\rho_n}$.

Denote by $u_n$ the analytic function
\begin{equation}
\begin{split}
u_n\colon \Delta & \to \Diff_{per}(D_{\rho_n},\Cc^d)
\\
X & \mapsto \fU_n(\II_n\LL_n(X_{n-1}))(\cdot,p_{n+1}^s(X)).
\end{split}
\end{equation}
As $p_{n+1}^s(X)\in B_{r_n}$, $D_{\rho_n}\times\{\vecy=p_{n+1}^s(X)\}$ is inside the domain $\DD_{\rho_n,r_n}$ of $\fU_n(\II_n\LL_n(X_{n-1}))$ given in Theorem \ref{main theorem1}.
Now, for each $X$, define the isotopic to the identity diffeomorphism
\begin{equation}
W_n(X)=
{P^{(n)}}^{-1}\circ u_n(X) \circ P^{(n)},
\end{equation}
on ${P^{(n)}}^{-1}D_{\rho_n}$.
If $X$ is real-analytic, then $W_n(X)(\Rr^d)\subset\Rr^d$, since this property holds for $u_n(X)$.
We also have $W_n(X^0)=\id$.

\begin{lemma}
For all $n\in\Nn\cup\{0\}$,
$W_n\colon B\to\Diff_{per}(D_{R_n},\Cc^d)$ is analytic satisfying
$W_n(X)\colon D_{R_n}\to D_{R_{n-1}}$ and
\begin{equation}\label{norm of Wn-I in Drho n+1}
\|W_n(X)-\id\|_{R_n}
\leq 42K\Theta_n^{1/2} \|X-X^0\|_{\rho,r},
\quad
X\in B.
\end{equation}
\end{lemma}

\begin{proof}
For any $X\in\Delta$, in view of \eqref{estimate U around X0} we get
\begin{equation*}
\begin{split}
\|W_n(X)-\id\|_{R_n}
& =
\| {P^{(n)}}^{-1}\circ[u_n(X) -\id ]\circ P^{(n)}   \|_{R_n}
\\
&\leq
\frac{42}{\sigma_{n}} \|{P^{(n)}}^{-1}\|\, \|\II_n\LL_n(X_{n-1})-X_{n}^0\|_{\rho_n,r_n}.
\end{split}
\end{equation*}
We can bound the above by \eqref{norm of Wn-I in Drho n+1}.

Now, for $\vecx\in D_{R_n}$ and $X\in B\subset\Delta$,
\begin{equation*}
\begin{split}
\|\im W_n(X)(\vecx)\| 
& \leq 
\|\im(W_n(X)(\vecx)-\vecx)\| + \|\im \vecx\|
\\
&<
\|W_n(X)-\id\|_{R_n}+R_n/2\pi
\leq
R_{n-1}/2\pi.
\end{split}
\end{equation*}
So we have $W_n(X)\colon D_{R_n}\to D_{R_{n-1}}$ 
and $W_n(X)\in\Diff_{per}(D_{R_n},\Cc^d)$.
 From the properties of $\fU_n$, $W_n$ is analytic as a map from $B$
into $\Diff_{per}(D_{R_n},\Cc^d)$.
\end{proof}

Consider the analytic map $H_n\colon B\to\Diff_{per}(D_{R_n},\Cc^d)$ defined by
the coordinate transformation $H_n(X)\colon D_{R_n}\to D_{\rho_0}$ as
\begin{equation}\label{def H n}
H_n(X)=W_0(X)\circ\dots\circ W_n(X).
\end{equation}

\begin{lemma}\label{lemma Hn-Hn-1 vf}
There exists $c>0$ such that for $X\in B$ and $n\in\Nn$,
\begin{equation}
\|H_n(X)-H_{n-1}(X)\|_{R_n}
\leq c \Theta_n^{1/2} \|X-X^0\|_{\rho,r}.
\end{equation}
\end{lemma}

\begin{proof}
For each $k=0,\dots, n-1$, consider the transformations
\begin{equation*}
\begin{split}
G_k(z,X)= &
 (W_k(X)-\id)\circ(\id+G_{k+1}(z,X))+G_{k+1}(z,X),
\\
G_n(z,X)= &
z(W_n(X)-\id),
\end{split}
\end{equation*}
with $(z,X)\in  \{z\in\Cc\colon |z| < 1+ d_n \} \times B$, where we have $c'>0$ such that
$$
d_n=\frac{c'}{\Theta_n^{1/2}\|X-X^0\|_{\rho,r}}-1>0.
$$
If the image of $D_{R_n}$ under $\id+G_{k+1}(z,X)$ is inside the domain of $W_k(X)$, or simply
$$
\|G_{k+1}(z,X)\|_{R_n}\leq (R_k-R_n)/2\pi,
$$
then $G_k$ is well-defined as an analytic map into $\Diff_{per}(D_{R_n},\Cc^d)$, and
$$
\|G_k(z,X)\|_{R_n} \leq 
\|W_k(X)-\id\|_{R_k} + \|G_{k+1}(z,X)\|_{R_n}.
$$
An inductive scheme shows that 
\begin{equation*}
\begin{split}
\|G_n(z,X)\|_{R_n} 
\leq &
(R_{n-1}-R_n)/2\pi,
\\
\|G_k(z,X)\|_{R_n} 
\leq &
\sum_{i=k}^{n-1}
\|W_i(X)-\id\|_{R_i}
+
|z| \, \|W_n(X)-\id\|_{R_n}
\\
\leq &
(R_{k-1}-R_{n})/2\pi.
\end{split}
\end{equation*}

By Cauchy's formula
\begin{equation*}
\begin{split}
\|H_n(X)-H_{n-1}(X)\|_{R_{n}}
&=
\|G_0(1,X)-G_0(0,X)\|_{R_{n}}
\\
&=
\norm{\frac1{2\pi i} \oint_{|z|=1+d_n/2}\frac{G_0(z,X)}{z(z-1)}dz}_{R_{n}},
\end{split}
\end{equation*}
and
\begin{equation*}
\begin{split}
\|H_n(X)-H_{n-1}(X)\|_{R_{n}}
&\leq 
\frac2{d_n}\sup_{|z|=1+d_n/2}\|G_0(z,X)\|_{R_{n}}
\\
 &\ll
\Theta_n^{1/2} \|X-X^0\|_{\rho,r}.
\end{split}
\end{equation*}
\end{proof}

Consider $C^1_{per}(\Rr^d,\Cc^d)$ to be the Banach space of the $\Zz^d$-periodic $C^1$ functions between $\Rr^d$ and $\Cc^d$ with norm
\begin{equation}
\|f\|_{C^1} = \max_{k\leq1}\max_{\vecx\in\Rr^d} \|D^kf(\vecx)\|.
\end{equation}

\begin{lemma}
There exists $C>0$, an open ball $B'\subset B$ about $X^0$ and an analytic map $H\colon B'\to \Diff_{per}(\Rr^d,\Cc^d)$ such that for $X\in B'$, $H(X)=\lim_{n\to+\infty}H_n(X)$ and
\begin{equation}
\|H(X)-\id\|_{C^1}  \leq C \|X-X^0\|_{\rho,r}.
\end{equation}
If $X\in B'$ is real-analytic, then $H(X)\in \Diff_{per}(\Rr^d,\Rr^d)$.
\end{lemma}

\begin{proof}
As the domains $D_{R_n}$ are shrinking, we consider the
restrictions of $W_n(X)$ and $H_n(X)$ to $\Rr^d$, and estimate their
$C^1$ norms from the respective norms in $\Diff_{per}(D_{R_n},\Cc^d)$. 
More precisely, for any $X\in B$, making use of Lemma \ref{lemma Hn-Hn-1 vf},
\begin{equation}\label{estimate var H n}
\begin{split}
\|H_n(X)-H_{n-1}(X)\|_{C^1} 
&\leq 
\max_{k\leq1}\sup_{\vecx\in D_{R_n/2}} \|D^k [H_n(X)(\vecx)-H_{n-1}(X)(\vecx)]\| 
\\
&\leq
\frac{4}{R_n}  \|H_n(X)-H_{n-1}(X)\|_{R_n},
\end{split}
\end{equation}
which goes to zero by \eqref{control on Rn ell for vf}.
Notice that here we have used Cauchy's estimate $\|D^1g\|_\zeta\leq (2\pi /\e^{} \delta) \|g\|_{\zeta+\delta}$ with $\zeta,\delta>0$.

Therefore, it is shown the existence of the limit
$H_n(X)\to H(X)$ as $n\to+\infty$, in the Banach space 
$C_{per}^1(\Rr^d,\Cc^d)$.
Moreover, $\|H(X)-\id\|_{C^1} \ll \|X-X^0\|_{\rho,r}$.
The convergence of $H_n$ is uniform in $B$ so $H$ is analytic.
As the space of close to identity diffeomorphisms is closed for the $C^1$ norm, $H(X)$ is a diffeomorphism for any $X$ sufficiently close to $X^0$, i.e. $X\in B'$.
The fact that, for real-analytic $X$, $H(X)$ takes real values for real arguments, follows from the same property of each $W_n(X)$.
\end{proof}

To simplify notation, write $\pi_\vecy X=X(\cdot,\vecy)$.

\begin{lemma}
For every real-analytic $X\in B'$ and $|s|<b$, $[H(X)]^*(\pi_{p^s(X)}X)=(1+s)\,\vecomega$ on $\Rr^d$.
\end{lemma}

\begin{proof}
For each $n\in\Nn$ the definition of $H_n(X)$ and \eqref{formula X n with Vs} imply that
\begin{equation}\label{equation Hn and RRn}
H_n(X)^*(\pi_{p^s(X)}X)=
\lambda_n^{-1} \pi_{ p^s(X)} {V_n^{-1}}^* (X_n).
\end{equation}
The r.h.s. can be written as
\begin{multline}
\lambda_{n}^{-1} {P^{(n)}}^{-1}  
[\vecomega^{(n)} + {\gamma^{(n)}}^{-1}M^{(n)} \Phi_n(X_{n-1})^{-1}\cdots \Phi_1(X_0)^{-1} p^s(X)]
+ \lambda_{n}^{-1} \pi_{ p^s(X)} {V_n^{-1}}^* (X_n-X_n^0)
= \\
=(1+s)\,\vecomega 
+ \lambda_{n}^{-1} {P^{(n)}}^{-1} {\gamma^{(n)}}^{-1}M^{(n)}  p_{n+1}^s(X)
-s\,\vecomega
+ \lambda_{n}^{-1} \pi_{p^s(X)} {V_n^{-1}}^* (X_n-X_n^0).
\end{multline}
Its terms can be estimated, for $\vecx\in \Rr^d$, by
\begin{equation}
\|\lambda_{n}^{-1}\pi_{p^s(X)} {V_n^{-1}}^* (X_n-X_n^0)(\vecx)\|
\leq
|\lambda_{n}^{-1}|\,\|{P^{(n)}}^{-1}\|\,\|X_n-X_n^0\|_{\rho_n,r_n}
\ll  \Theta_n^{1/2},
\end{equation}
and 
\begin{equation}
\begin{split}
\frac{1}{|\lambda_n\gamma^{(n)}|}
\|{P^{(n)}}^{-1} M^{(n)} [p_{n+1}^s(X)-(0,\dots,0,s)]\| 
&=
\frac{\|M^{(0)}\|}{|\gamma|\e^{(d-1)t_n}}\|E^{t_n}[p_{n+1}^s(X)-(0,\dots,0,s)]\| 
\\
&\ll
\|p_{n+1}^s(X)-(0,\dots,0,s)\|
\end{split}
\end{equation}
which is controlled by \eqref{estimate p-s}.
Consequently, the limit of \eqref{equation Hn and RRn} as $n\to+\infty$ is $(1+s)\,\vecomega$.
Using the convergence of $H_n$ we complete the proof.
\end{proof}

\begin{lemma}\label{proposition: comm H and R}
If $X\in B'$ and $\vecx\in\Rr^d$, then
\begin{equation}\label{eq comm rel for H}
H(X\circ R_\vecx)=\widehat R_\vecx^{-1}\circ H(X)\circ \widehat R_\vecx,
\end{equation}
where $\widehat R_\vecx\colon \vecz\mapsto \vecz+\vecx$ is a translation on $\Cc^d$.
\end{lemma}

\begin{proof}
For each $n\in\Nn$, \eqref{relation tilde U and U} 
and \eqref{commutation property Rz and Ln}
yield that $\UU_n(X\circ R_\vecx)=\UU_n(X)\circ R_\vecx$ and
$\LL_{n}(X\circ R_\vecx)=\LL_{n}(X)\circ R_{T^{(n)}\vecx}$.
This implies immediately that
\begin{equation}\label{relation Rn and Rz}
\RR_n(X\circ R_\vecx)=\RR_n(X)\circ R_{{P^{(n)}} \vecx}.
\end{equation}
Next, from a simple adaptation of \eqref{relation tilde U and U} and the formula
$\widehat R_{P^{(n)}\vecz}=P^{(n)} \widehat R_\vecz {P^{(n)}}^{-1}$ for $\vecz\in\Cc^d$, we get
\begin{equation}
\begin{split}
W_n(X\circ R_\vecx) =&
{P^{(n)}}^{-1}\circ\fU_{ a_n}(\LL_n\RR_{n-1}(X\circ R_\vecx))\circ {P^{(n)}}
\\
=&
\widehat R_\vecx^{-1}\circ W_n(X)\circ \widehat R_\vecx.
\end{split}
\end{equation}
Thus, $H_n(X\circ R_\vecx)=\widehat R_\vecx^{-1}\circ H_n(X)\circ \widehat R_\vecx$.
The convergence of $H_n$ implies \eqref{eq comm rel for H}.
\end{proof}

\begin{theorem}\label{lemma: existence of h infty diff}
If $\vecv\in\vf^\vecomega(\Tt^d)$ is sufficiently close to
$\vecomega$, then there exists an analytic curve $s\mapsto\vecp^s\in\Rr^d$ for $|s|<b$, and
$h\in\Diff^\omega(\Tt^d)$ homotopic to the identity such that 
\begin{equation}\label{equivalence vf and omega}
h^*(\vecv+\vecp^s)=(1+s)\vecomega.
\end{equation}
The maps $\vecv\mapsto \vecp$ and $\vecv\mapsto h$ are analytic.
\end{theorem}

\begin{proof}
The lift $\widetilde\vecv$ to $\Rr^d$ of $\vecv$ is assumed to have an analytic
extension in $D_\rho$.
Consider the real-analytic vector field
$$
Y(\vecx,\vecy)=\widetilde\vecv(\vecx)+\vecy
$$
in $\A_{\rho,r}$.
Suppose that $\vecv$ is close enough to $\vecomega$ such that
$Y\in B'$ and $Y\circ R_\vecz\in B'$ for some $\eta>0$ and $\vecz\in D_\eta$.
Then, the parameter $\vecp^s=p^s(Y)\in\Rr^d$ and the $C^1$-diffeomorphism $h=H(Y)\bmod1$ verify \eqref{equivalence vf and omega}.

We now want to extend $h$ analytically to a complex neighbourhood of its
domain.
Take $\widetilde h(\vecz)=\vecz+ H(Y\circ R_\vecz)(0)$, $\vecz\in D_\eta$.
The maps $\vecz\mapsto Y\circ R_\vecz$ and $X\mapsto H(X)$ are analytic and $C^1_{per}(\Rr^d,\Cc^d)\ni g\mapsto g(0)$ is bounded. 
As $\widetilde h$ involves their composition, it is analytic on the domain 
$D_\eta$ and $\Zz^d$-periodic.
 From \eqref{eq comm rel for H}, for
any $\vecx\in\Rr^d$, we have
\begin{equation}
\begin{split}
\widetilde h(\vecx)\bmod1 
& = 
(\vecx+\widehat R_\vecx^{-1}\circ H(Y)\circ \widehat R_\vecx(0))\bmod1 \\
& = 
(\vecx+H(Y)(\vecx)-\vecx)\bmod1 \\
& = 
h(\vecx).
\end{split}
\end{equation}
The extension of $h$ is a complex analytic diffeomorphism, thus $h$ is a real-analytic diffeomorphism.
\end{proof}

\section{Renormalization of Hamiltonian flows}
\label{Section: Renormalization of Hamiltonian flows}

\subsection{Preliminaries}

Consider the symplectic manifold $T^*\Tt^d$ with respect to the canonical symplectic form $\sum_{i=1}^ddy_i\wedge dx_i$.
As the cotangent bundle of $\Tt^d$ is trivial,
$T^*\Tt^d\simeq\Tt^d\times\Rr^d$, we identify 
functions on $T^*\Tt^d$ with functions on $\Tt^d\times\Rr^d$.
By lifting to the universal cover, we consider functions from $\Rr^{2d}$ into $\Rr$ and extend them to the complex domain.

Let $\Omega$ be a neighbourhood of $\Rr^d\times\{0\}$ in $\Cc^{2d}$.
A Hamiltonian is a complex analytic function $H\colon \Omega\to\Cc$,
$\Zz^d$-periodic on the first coordinate, written on
the form of a Taylor-Fourier series 
\begin{equation}
H(\vecx,\vecy)=\sum\limits_{(\veck,\vecnu)\in I}H_{\veck,\vecnu}\vecy^\vecnu\e^{2\pi \i \veck\cdot \vecx},
\qquad
(\vecx,\vecy)\in\Omega,
\end{equation}
where $I=\Zz^d\times(\Nn\cup\{0\})^d$, $H_{\veck,\vecnu}\in\Cc$ and $\vecy^\vecnu=y_1^{\nu_1}\dots y_d^{\nu_d}$.

Let the positive real numbers $\rho$ and $r$ be given in order to determine the domain
\begin{equation}\label{def domain DD}
\DD_{\rho,r}=D_\rho\times B_r,
\end{equation}
where 
\begin{equation}
\begin{split}
D_\rho & =\{\vecx\in\Cc^{d} \colon \|\im \vecx\| < \rho/2\pi\}
\text{ and } 
\\
B_r & =\{\vecy\in\Cc^d\colon \|\vecy\| < r\},
\end{split}
\end{equation} 
for the norm $\|{\vecu}\|=\sum_{i=1}^d|u_i|$ on $\Cc^d$.
Consider the Banach space $\A_{\rho,r}$ of Hamiltonians defined on $\Omega=\DD_{\rho,r}$, which extend continuously to the boundary and with finite norm 
\begin{equation}\label{def norm for H}
\|H\|_{\rho,r} 
=\sum\limits_{(\veck,\vecnu)\in I} |H_{\veck,\vecnu}| \, r^{\|\vecnu\|} \, \e^{ \rho \|\veck\|}.
\end{equation}
Similarly, take a norm on the product space $\A_{\rho,r}^{2d}=\A_{\rho,r}\times\dots\times\A_{\rho,r}$ given by $\|(H_1,\dots,H_{2d})\|_{\rho,r}=\sum_{i=1}^{2d}\|H_i\|_{\rho,r}$.
Using this we have the Banach space $\A'_{\rho,r}$ of the Hamiltonians $H\in\A_{\rho,r}$ with finite norm
$$
\|H\|'_{\rho,r} 
=\|H\|_{\rho,r} 
+\|\nabla H\|_{\rho,r}.
$$
A useful property is the Cauchy estimate: for any $\delta>0$ we have
\begin{equation}\label{Cauchy estimate H}
\begin{split}
\|\partial_iH\|_{\rho,r} & \leq \frac{2\pi}{\delta} \|H\|_{\rho+\delta,r},
\quad H\in\A_{\rho+\delta,r},
\quad 1\leq i\leq d,
\\
\|\partial_jH\|_{\rho,r} & \leq \frac{1}{\delta} \|H\|_{\rho,r+\delta},
\quad H\in\A_{\rho,r+\delta},
\quad d+1\leq j\leq 2d,
\end{split}
\end{equation}
where $\partial_k$ denotes the partial derivative with respect to the $k$th argument.
In particular
\begin{equation}
\|H\|'_{\rho,r} \leq \left(1+\frac{2\pi+1}{\delta}\right)\|H\|_{\rho+\delta,r+\delta}.
\end{equation}

The constant Fourier modes are written by the projections
\begin{equation}
\Ee_0F=\int_{\Tt^d}F(\vecx,0)d\vecx=F_{0,0}
\quad\text{and}\quad
\Ee F(\vecy)=\int_{\Tt^d}F(\vecx,\vecy)d\vecx=\sum_{\nu}F_{0,\nu}\vecy^\nu.
\end{equation}
The space where $\Ee F$ lies is denoted by $\Ee \A_r$ and the natural induced norm is $\|\cdot\|_r$.
Similarly, we define $\Ee\A'_r$ with norm $\|\cdot\|'_r$.
Recall Remark \ref{remark analyticity} about the notion of analyticity of maps over the above spaces.

In the next sections we will be studying Hamiltonians close enough to $H^0$ as given by \eqref{def H0 intro} for a given choice of $\vecomega\in\Rr^d-\{0\}$ and an invertible symmetric $d\times d$ matrix $Q$.
The norm of matrices is given by $\|Q\|=\max_{j=1\dots d}\sum_{i=1}^d|Q_{i,j}|$, where $Q_{i,j}$ are the entries of the $d\times d$ matrix $Q$.


\subsection{Change of basis and rescaling}\label{section: rescalings for H}

The following transformations leave invariant the dynamics of the flow generated by the Hamiltonian, producing an equivalent system.
They consist of 
\begin{itemize}
\item
an affine linear symplectic transformation of the phase space, 
\begin{equation}\label{defn Ln for H}
L_n\colon (\vecx,\vecy)
\mapsto 
({T^{(n)}}^{-1}\vecx, \trans T^{(n)}\vecy +\vecb_n),
\qquad 
(\vecx,\vecy)\in\Cc^{2d},
\end{equation}
for some $\vecb_n\in\Cc^d$,
\item
a linear time (energy) change, 
\begin{equation} \label{energy rescaling}
H
\mapsto 
\eta_n H,
\end{equation}
where $\eta_n$ is as in \eqref{eq def eta n lambda n},
\item
a linear action rescaling,
\begin{equation} \label{action rescaling}
H
\mapsto 
\frac1{\mu_n}H(\cdot, \mu_n \cdot)
\end{equation}
with a choice of $\mu_n>0$ to be specified later on,
\item
and the (trivial) elimination of the constant term
\begin{equation} \label{elim const term}
H \mapsto 
(\Ii-\Ee_0)H.
\end{equation}
\end{itemize}

Recall the definition of the sequence $\vecomega^{(n)}$ given by \eqref{eq def omega n}.
For $n\in\Nn$ and $\rho_{n-1},r>0$, consider Hamiltonians of the form
\begin{equation}
H(\vecx,\vecy)= H_{n-1}^0(\vecy) + F(\vecx,\vecy),
\qquad
(\vecx,\vecy)\in \DD_{\rho_{n-1},r},
\end{equation}
where
\begin{equation}
H_{n-1}^0(\vecy)= \vecomega^{(n-1)}\cdot\vecy +\frac12\trans\vecy  Q_{n-1}\vecy
\end{equation}
and $H_0^0=H^0$, with $Q_{n-1}$ being a $d\times d$ symmetric matrix.
By performing the transformations above, we get a new Hamiltonian that is the image of the map
$$
H\mapsto\LL_n(H)= (\Ii-\Ee_0)\frac{\eta_n}{\mu_n} H\circ L_n(\cdot,\mu_n\cdot).
$$
In order to simplify notation, we write
\begin{equation}
\Phi_n(\vecy)=\mu_n\trans T^{(n)}\vecy +\vecb_n.
\end{equation}
So, for any $(\vecx,\vecy)\in L_n^{-1}\DD_{\rho_{n-1},r}$,
\begin{multline}\label{formula Ln of H}
\LL_n(H)(\vecx,\vecy)
= \\ =
(\Ii-\Ee_0)\frac{\eta_n}{\mu_n} \left[
\vecomega^{(n-1)}\cdot \Phi_n(\vecy)
+ \frac12 \trans \Phi_n(\vecy) Q^{(n-1)} \Phi_n(\vecy)
+ F\circ L_n(\vecx,\mu_n \vecy) \right].
\end{multline}
By decomposing $F=(\Ii-\Ee)F+F_0$ and using the Taylor expansion of $F_0$:
\begin{equation}\label{Taylor expansion F0}
F_0\circ \Phi_n(\vecy) =
F_0(\vecb_n)+\mu_n \trans\nabla F_0(\vecb_n)\trans T^{(n)}\vecy
+
\frac{\mu_n^2}2 \trans\vecy T^{(n)} D^2F_0(\vecb_n)\trans T^{(n)}\vecy
+\Upsilon_n(\vecy),
\end{equation}
with $\Upsilon_n(\vecy)=\OO(\|\vecy\|^3)$, we get
\begin{equation}\label{eqn for LL n}
\begin{split}
\LL_n(H)(\vecx,\vecy) = &
\vecomega^{(n)}\cdot\vecy
+\eta_n 
\left[\trans\vecb_nQ_{n-1} + \trans\nabla F_0(\vecb_n)\right]
\trans T^{(n)}\vecy
\\
&
+\frac{\eta_n\mu_n}2 
\trans\vecy T^{(n)} 
\left[Q_{n-1} +D^2F_0(\vecb_n)\right]
\trans T^{(n)}\vecy
\\
&
+ \frac{\eta_n}{\mu_n} \Upsilon_n(\vecy)
+ \frac{\eta_n}{\mu_n} (\Ii-\Ee)F\circ L_n(\vecx,\mu_n \vecy).
\end{split}
\end{equation}
To ``normalize'' the linear terms in $\vecy$ of $\Ee\LL_n(H)$ equal to $\omega^{(n)}\cdot\vecy$, we choose $\vecb_n$ inside the domain of $\nabla F_0$ such that
\begin{equation}\label{cdn b_n for H}
Q_{n-1} \vecb_n + \nabla F_0(\vecb_n)=0.
\end{equation}
Hence,
\begin{equation}\label{final LLn for H}
\LL_n(H)(\vecx,\vecy)
= H_n^0(\vecx,\vecy) 
+ \widehat\LL_n(F_0)(\vecy)
+\widetilde\LL_n(F-F_0)(\vecx,\vecy).
\end{equation}
Here we have introduced the operators 
\begin{equation}
\widehat\LL_n\colon F_0\mapsto \frac{\eta_n}{\mu_n} \Upsilon_n
\end{equation}
and
\begin{equation}\label{def tilde Ln H}
\widetilde\LL_n\colon (\Ii-\Ee)F 
\mapsto 
\frac{\eta_n}{\mu_n} (\Ii-\Ee)F \circ L_n(\cdot,\mu_n \cdot)
\end{equation}
defined in $\Ee\A_r$ and $(\Ii-\Ee)\A_{\rho_{n-1},r}$, respectively.
Moreover, we have used the symmetric $d\times d$ matrix
\begin{equation}\label{def Q n}
Q_n = \eta_n\mu_n T^{(n)}
\left[Q_{n-1} +D^2F_0(\vecb_n)\right]
\trans T^{(n)}.
\end{equation}

Denote by $\Delta_\gamma$ the set of $H\in\A_{\rho_{n-1},r}$ such that $\|F_0\|_{\rho_{n-1},r}<\gamma$.

\begin{lemma}\label{prop exist b n for H}
If $\det(Q_{n-1})\not=0$ and
\begin{equation}\label{1st cdn epsilon H}
\gamma_n=
\frac{r^2}{16\|Q_{n-1}^{-1}\|},
\end{equation}
there is $b_n\in C^1(\Delta_{\gamma_n},\Cc^d)$ such that, for all $H\in \Delta_{\gamma_n}$, $\vecb_n=b_n(H)$ satisfies \eqref{cdn b_n for H} and
\begin{equation}\label{bd b n H}
\|b_n(H)\|<(2/r)\|Q_{n-1}^{-1}\|\,\|F_0\|_r
<\frac r8.
\end{equation}
Moreover, $\det(Q_n)\not=0$ and
\begin{equation}\label{growth of inv Q}
\|Q_{n}^{-1}\| \leq
\frac{\|{T^{(n)}}^{-1}\|\, \|\trans{T^{(n)}}^{-1}\|}
{\mu_n|\eta_n|
(\|Q_{n-1}^{-1}\|^{-1}- \frac{16}{r^2} \|F_0\|_r)}.
\end{equation}
In the case $F_0$ is real-analytic and $Q_{n-1}$ is real, then $b_n(H)\in\Rr^d$ and $Q_n$ is also real.
\end{lemma}

\begin{proof}
Consider the differentiable function $\FF(H,\vecb)=\vecb+Q_{n-1}^{-1}\nabla F_0(\vecb)$ defined on $\Delta_{\gamma_n}\times B_{r/2}$.
Notice that $\FF(H_{n-1}^0,0)=0$ and the derivative with respect to the second argument, 
$$
D_2\FF(H,\vecb)=I+Q_{n-1}^{-1}D^2 F_0(\vecb), 
\qquad(H,\vecb)\in \Delta_{\gamma_n}\times B_{r/2}, 
$$
admits a bounded inverse because
\begin{equation}\label{bdd 2nd der F 0}
\begin{split}
\|D^2 F_0\|_{r/2} 
& =\max_{d+1\leq j\leq 2d} \|\partial_j\nabla F_0\|_{r/2}
\\
&\leq (4/r) \|\nabla F_0\|_{3r/4}
\\
&\leq (16/r^2) \|F_0\|_r
\\
&< \|Q_{n-1}^{-1}\|^{-1}
\end{split}
\end{equation}
by the Cauchy estimate.
The implicit function theorem implies the existence of a $C^1$ function $b\colon H\mapsto b(H)$ in a neighbourhood of $H_{n-1}^0$ such that 
$$
\FF(H,b(H))=b(H)+ Q_{n-1}^{-1}\nabla F_0(b(H))=0,
$$ 
i.e. a solution of \eqref{cdn b_n for H}. 
Notice that for any $H\in\Delta_{\gamma_n}$ the operator $\id-\FF(H,\cdot)$ is a contraction with a unique fixed point $\vecb$.
Hence the domain of the $C^1$ function $H\mapsto b(H)=\vecb$ is extendable to $\Delta_{\gamma_n}$ and thus \eqref{bd b n H}.
Assuming $F_0$ to be real-analytic and $Q_{n-1}$ with real entries, the same argument is still valid when considering $B_{r/2}\cap\Rr^d$.
Thus, $b(H)$ is real and $Q_n$ is a real symmetric matrix.

 From \eqref{bdd 2nd der F 0},
$$
\|Q_{n-1}^{-1}D^2F_0(b_n(H))\|<1,
\quad
H\in\Delta_{\gamma_n}.
$$ 
Hence, $A=Q_{n-1}[I+Q_{n-1}^{-1}D^2F_0(b_n(H))]$ is invertible.
Moreover, 
\begin{equation}
\|A^{-1}\|\leq 1/(\|Q_{n-1}^{-1}\|^{-1}-\|D^2F_0\|_{r/2}).
\end{equation}
Now, $Q_{n}^{-1}=(\eta_n\mu_n)^{-1}\trans {T^{(n)}}^{-1}A^{-1}{T^{(n)}}^{-1}$, thus \eqref{growth of inv Q}.
\end{proof}

\begin{lemma}\label{prop hat L for H}
If $r<r'$ and
\begin{equation}\label{cdn mu n H}
\mu_n < 
\frac{r}
{4r'\|\trans T^{(n)}\|},
\end{equation}
then $\widehat\LL_n\colon\Ee\A_{r}\cap\Delta_{\gamma_n}\to\Ee\A'_{r'}$ and
\begin{equation}\label{bdd hat LL n for H}
\|\widehat\LL_n\| 
\leq
\mu_n^2
|\eta_n|
\left(1+\frac1{2r'}\right) 
\frac{(4r'\,\|\trans T^{(n)}\|)^3}
{r^2(r-4r'\mu_n\,\|\trans T^{(n)}\|)}.
\end{equation}
\end{lemma}

\begin{proof}
Let $H\in\Delta_{\gamma_n}$, $R=\frac{r}{4r'\mu_n \|\trans T^{(n)}\|}>1$ and the map
\begin{equation}
\begin{split}
f\colon\{z\in\Cc\colon |z|\leq R\} & \to \Cc^d \\
z & \mapsto F_0(z\mu_n\trans T^{(n)}\vecy+b_n(H)).
\end{split}
\end{equation}
Hence $\Upsilon_n$ as in \eqref{Taylor expansion F0} can be written as
$$
f(1)-f(0)-Df(0)-\frac12 D^2f(0)=
\frac1{2\pi i}\oint_{|z|=R}\frac{f(z)}{z^3(z-1)}dz.
$$
Therefore,
\begin{eqnarray*}
\|\Upsilon_n\|'_{r'}
&=&
\frac1{2\pi}
\left
\|\oint_{|z|=R}\frac{f(z)}{z^3(z-1)}dz
\right\|'_{r'} 
\\
&\leq&
\frac{1}{R^2(R-1)}
\sup_{|z|=R}\|F_0(z\mu_n \trans T^{(n)}\cdot+b_n(H))\|'_{r'}.
\end{eqnarray*}
For $\|\vecy\|<r'$, in view of \eqref {bd b n H},
$$
\sup_{|z|=R}
\|z\mu_n \trans T^{(n)}\vecy+b_n(H)\| \leq 
R \mu_n \|\trans T^{(n)}\| \,r'
+\|b_n(H)\|
<r/2,
$$
and
\begin{equation}
\begin{split}
\sup_{|z|=R}
\|F_0(z\mu_n \trans T^{(n)}\cdot+b_n(H))\|'_{r'}
&\leq
\|F_0\|_{r/2}
+R \mu_n \, \|\trans T^{(n)}\| \,  \|\nabla F_0\|_{r/2} 
\\
&\leq
\|F_0\|_{r/2}
+ \frac1{2r'}\|F_0\|_{r}
\leq 
\left(1+\frac1{2r'}\right)
\|F_0\|_r.
\end{split}
\end{equation}
Thus, $\|\Upsilon_n\|'_{r'} \leq (1+1/2r')[R^2(R-1)]^{-1} \|F_0\|_r$ and
$$
\|\widehat\LL_n(F_0)\|'_{r'} =
\frac{|\eta_n|}{\mu_n}
\|\Upsilon_n\|'_{r'}
\leq 
\frac{|\eta_n|}{\mu_n}
\left(1+\frac1{2r'}\right) 
\frac{(4r'|\mu_n|\,\|\trans T^{(n)}\|)^3}
{r^2(r-4r'|\mu_n|\,\|\trans T^{(n)}\|)}
\|F_0\|_r.
$$
\end{proof}

Finally, we have the relation for each $\vecz\in\Cc^d$:
\begin{equation}\label{comm Rz Ln for H}
R_\vecz\circ L_n=L_n\circ R_{{T^{(n)}}\vecz},
\end{equation}
because $\Ee H\circ R_\vecz=\Ee H$.


\subsection{Far from resonance modes}

Given $\sigma_n,\tau_n>0$, we call {\em far from resonance} modes with respect to $\vecomega^{(n)}$ the Taylor-Fourier modes with indices in 
\begin{equation}\label{def I- for H}
I_n^-
=\left\{ (\veck,\vecnu)\in I\colon 
|\vecomega^{(n)}\cdot \veck| > \sigma_n \|\veck\| , 
\|\vecnu\| < \tau_n \|\veck\|
\right\}.
\end{equation}
The {\em resonant modes} are in $I_n^+=I- I_n^-$.
We also have the projections
$\Ii_n^+$ and $\Ii_n^-$ over the spaces of Hamiltonians by restricting the Taylor-Fourier modes to $I_n^+$ and $I_n^-$, respectively.
The identity operator is $\Ii=\Ii_n^++\Ii_n^-$.

Furthermore, consider the same sequence of positive numbers $A_n$ given in \eqref{defn of An}.


\subsection{Analyticity improvement}

\begin{lemma}\label{lemma tilde Ln for H}
If $\delta>0$,
\begin{equation}\label{cdns on rho n and tau n for H}
\rho_n'\leq\frac{\rho_{n-1}}{A_{n-1}} -\delta
\quad\text{and}\quad
\tau_n \geq \frac{2}{\log2}(\rho'_n+\delta)\|\trans{T^{(n)}}^{-1}\|,
\end{equation}
then $\widetilde\LL_n$ as a map from $(\Ii^+_{n-1}-\Ee)\A_{\rho_{n-1},r}\cap\Delta_{\gamma_n}$ to $(\Ii-\Ee)\A'_{\rho'_n,r'}$ is continuous and compact with
\begin{equation}\label{bd tilde Ln for H}
\|\widetilde\LL_n\|
\leq
 \left(1+\frac{2\pi}{\delta}+\frac{r}{2{r'}^2\log2}\right) 
\frac{|\eta_n|}{\mu_n}.
\end{equation}
\end{lemma}

\begin{proof}
Let $F\in(\Ii^+_{n-1}-\Ee)\A_{\rho_{n-1},r}\cap\Delta_{\gamma_n}$,
\begin{equation}
E = \{(0,\vecnu)\colon \vecnu \in (\Nn\times\{0\})^d\}
\quad\text{and}\quad
J_n =\{\veck\in\Zz^d \colon |\veck\cdot\vecomega^{(n)}|\leq\sigma_n\|\veck\|\}.
\end{equation}
Using Lemma \ref{prop exist b n for H} and \eqref{cdn mu n H} we have
\begin{equation}
\psi_n =\mu_n \|\trans T^{(n)}\|\,r' + \|b_n(H)\|
\leq 
\frac r4 +\frac 2r \|Q_{n-1}^{-1}\|\,\|F_0\|_r
< \frac  r2.
\end{equation}

We want to find an upper bound on
\begin{multline}\label{formula norm F o L}
\|F\circ L_n(\cdot,\mu_n \cdot)\|'_{\rho'_n,r'}
 \\ \leq
\sum_{I_{n-1}^+-E} 
\left(1+2\pi\|\trans{T^{(n)}}^{-1} \veck\|
+\mu_n \|\trans T^{(n)}\|\,\|\vecnu\|/r'\right)
|F_{\veck,\vecnu}|
\psi_n^{\|\vecnu\|}
\e^{\rho'_n\|\trans{T^{(n)}}^{-1} \veck\|}
\\ \leq
\sum_{I_{n-1}^+-E} 
\left(1+\frac{2\pi}{\delta}\e^{\delta\|\trans{T^{(n)}}^{-1}\veck\|}
+\frac{r}{4{r'}^2\xi_n}\,\e^{\xi_n\|\vecnu\|}\right)
|F_{\veck,\vecnu}|
\psi_n^{\|\vecnu\|}
\e^{\rho'_n\|\trans{T^{(n)}}^{-1} \veck\|},
\end{multline}
where we have used the inequality $\zeta \e^{-\delta\,\zeta}\leq \delta^{-1}$ with $\zeta\geq0$ and again a choice of $\mu_n$ verifying \eqref{cdn mu n H}.
Here $\xi_n=\frac12\log(r/\psi_n)>\frac12\log2$.

Consider separately the two cases corresponding to the definition of the resonance cone $I_{n-1}^+$.

We deal first with the modes corresponding to $\veck\in J_{n-1}-\{0\}$.
By \eqref{defn of An} and \eqref{cdns on rho n and tau n for H} each of these modes in \eqref{formula norm F o L} is bounded from above by
\begin{equation}
\left(1+\frac{2\pi}{\delta}+\frac{r}{2{r'}^2\log2}\right) 
r^{\|\vecnu\|} \e^{\rho_{n-1}\|\veck\|}.
\end{equation}

Now, consider $\|\vecnu\|\geq\tau_n\|\veck\|$ with $\veck\not=0$, so that
\begin{equation}
\begin{split}
\|\trans{T^{(n)}}^{-1} \veck\|
&\leq 
\tau_n^{-1}\|\trans{T^{(n)}}^{-1}\|\,\|\vecnu\|.
\end{split}
\end{equation}
These modes in \eqref{formula norm F o L} are estimated by
\begin{equation}
\left(1+\frac{2\pi}{\delta}
+\frac{r}{4{r'}^2\xi_n} \, \e^{\xi_n\|\vecnu\|}\right) 
\left(  r\, 
\e^{-2\xi_n+(\rho'_n+\delta)\|\trans{T^{(n)}}^{-1}\| / \tau_n}
\right)^{\|\vecnu\|}
\leq
\left(1+\frac{2\pi}{\delta}+\frac{r}{2{r'}^2\log2}\right) 
r^{\|\vecnu\|},
\end{equation}
where we have used \eqref{cdns on rho n and tau n for H}.

Finally, we get
$$
\|F \circ L_n(\cdot,\mu_n \cdot)\|'_{\rho'_n,r'}
\leq \left(1+\frac{2\pi}{\delta}+\frac{r}{2{r'}^2\log2}\right)  \|F\|_{\rho_{n-1},r}, 
$$
and \eqref{bd tilde Ln for H} follows from \eqref{def tilde Ln H}.

The compacticity of the operator follows in the same way as in the proof of Lemma \ref{proposition TT}.
\end{proof}

Let $0<\rho''_n\leq \rho'_n$ and the inclusion
\begin{equation}
\II_n\colon \A'_{\rho'_n,r'}\to \A'_{\rho''_n,r'},
\quad
H\mapsto H|\DD_{\rho''_n,r'}.
\end{equation}
The norm of the $\veck\not=0$ modes can be improved by the application of such inclusion.
That is done in a similiar way as in Lemma \ref{lemma cutoff}, therefore we omit here the proof and just repeat the result for convenience.

\begin{lemma}\label{lemma cutoff for H}
If $\phi_n \geq 1$ and 
\begin{equation}\label{hyp rho''n H}
0<\rho''_n \leq \rho'_n - \log(\phi_n),
\end{equation}
then
\begin{equation}
\|\II_n (\Ii-\Ee)\| \leq \phi_n^{-1}.
\end{equation}
\end{lemma}


\subsection{Elimination of far from resonance modes}

The theorem below states the
existence of a symplectomorphism isotopic to the identity that cancels the far from resonance modes of a Hamiltonian close to $H_n^0$.

For given $\rho_n,r',\varepsilon,\nu>0$, denote by $\VV_\varepsilon$ the open ball in $\A'_{\rho_n+\nu,r'}$ centred at $H_n^0$ with radius $\varepsilon$.

\begin{theorem}\label{theorem existence of g}
Let $0<r<r'$, $\sigma_n>2r'\|Q_n\|$ and
\begin{equation}\label{formula epsilon for H}
\varepsilon_n=
\frac{\sigma_n^2 \min\left\{1,\frac\nu{2\pi},r'-r\right\}^2}
{48\|H_n^0\|_{r'} (2\pi+1)^2(1+2\pi+\frac{\tau_n+1}{r'})^2}.
\end{equation}
For every $H\in\VV_{\varepsilon_n}$ there exist an analytic symplectomorphism
$g\colon \DD_{\rho_n,r}\to \DD_{\rho_n+\nu,r'}$ in $\A_{\rho_n,r}^{2d}$
satisfying 
\begin{equation}\label{equation homotopy method for H}
\Ii_n^- H\circ g=0.
\end{equation}
This defines the maps 
\begin{equation}
\begin{split}
\fG\colon \VV_{\varepsilon_n} & \to  \A_{\rho_n,r}^{2d}
\\
H & \mapsto  g
\end{split}
\end{equation}
and 
\begin{equation}
\begin{split}
\UU\colon \VV_{\varepsilon_n} & \to \Ii_n^+\A_{\rho_n,r}
\\
H  & \mapsto H\circ g,
\end{split}
\end{equation}
which are analytic, and verify
\begin{equation}\label{estimate around omega y}
\begin{split}
\|\fG(H)-\id\|'_{\rho_n,r}
\leq & 
\frac{1}{\varepsilon_n} \|\Ii_n^-H\|_{\rho_n,r}
\\
\|\UU(H)-H_n^0\|_{\rho_n,r}
\leq &
\left(1+\sqrt\frac{12\|H_n^0\|_{r'}}{\varepsilon_n}\right)
\|H-H_n^0\|'_{\rho_n+\nu,r'}.
\end{split}
\end{equation}
If $H$ is real-analytic, then $\fG(H)(\Rr^{2d})\subset \Rr^{2d}$.
\end{theorem}

We prove this theorem in Section \ref{elim modes Ham}.

Recall the definition of the translation $R_\vecz$ on $\Cc^{2d}$ given by \eqref{translation Rz}.

\begin{lemma}\label{theorem comm g R}
In the conditions of Theorem {\rm \ref{theorem existence of g}}, 
if $\vecx\in\Rr^d$ and $H\in\VV_{\epsilon_n}$, then
\begin{equation}
\fG(H\circ R_\vecx)=R_\vecx^{-1}\circ \fG(H)\circ R_\vecx
\end{equation}
on $\DD_{\rho_n,r}$.
\end{lemma}

\begin{proof}
If $g=\fG(H)$ is a solution of \eqref{equation homotopy method for H} in $\DD_{\rho_n,r}$, then 
$\widetilde g=R_\vecx^{-1}\circ \fG(H)\circ R_\vecx$
solves the same equation for $\widetilde H=H\circ R_\vecx$, i.e. 
$\Ii^-\widetilde H\circ \widetilde g=0$ in $\DD_{\rho_n,r}$.
\end{proof}

\subsection {Convergence of renormalization}
\label{section:Convergence of renormalization for H}

For a resonance cone $I^+_n$ and $\mu_n>0$, the {\em $n$th step renormalization} operator is defined to be
$$
\RR_n=\UU_n \circ\II_n\circ \LL_n\circ\RR_{n-1}
\quad\text{and}\quad
\RR_0=\UU_0,
$$
where $\UU_n$ is as in Theorem \ref{theorem existence of g}, cancelling all the $I_n^-$ modes.
Notice that if $H^+(\vecy)= H^0(\vecy)+\vecv\cdot\vecy$ with $H^0$ non-degenerate, $\RR_n(H^+)=H^0_n$ for every $\vecv\in\Cc^d$.
This means that the renormalizations eliminate the direction corresponding to linear terms in $\vecy$.
 From the previous sections the map $\RR_n$ on its domain of validity is analytic by construction.
In addition, whenever a Hamiltonian $H$ is real-analytic, the same is true for $\RR_n(H)$.

Let $r'>r>0$, $\rho_0>0$ and fix a sequence $\sigma_n$ such that $\sigma_0>2r'\|Q\|$.
To complete the specification of the resonant modes and of $\varepsilon_n$ in Theorem \ref{theorem existence of g}, take
\begin{equation}
\tau_n  =
\frac{2\rho_{0}\|\trans{T^{(n)}}^{-1}\|}{ B_{n-1}\log2}
\end{equation}
according to Lemma \ref{lemma tilde Ln for H}.
Consider also the constants $\nu$ and $\delta$ as they appear in Theorem \ref{theorem existence of g} and Lemma \ref{lemma tilde Ln for H}, respectively.
For $n\in\Nn$, we define the non-increasing sequence
\begin{multline}\label{def Theta n H}
\Theta_n  =
\min\left\{
\Theta_{n-1},
\frac{\sigma_n^2}{(4r'\|Q\|)^2}
\prod_{i=1}^n \frac{2^6\zeta_i^2}{\|T^{(i)}\|^2\|\trans T^{(i)}\|^2},
\right.
\\
\left.
\frac{\varepsilon_n^3}{\prod_{i=1}^n\|{T^{(i)}}^{-1}\|^3},
\prod_{i=1}^n \frac{\min\{|\eta_i|^{-3},|\eta_i|\}}
{2^{10}\zeta_i^3 \|{T^{(i)}}^{-1}\|^2\|\trans{T^{(i)}}^{-1}\|^6}
\right\}
\leq 1,
\end{multline}
with $\Theta_0=1$.
The sequence $\rho_n$ is as defined in \eqref{formula rho n vf} with
\begin{equation}\label{def of phi mu}
\begin{split}
\phi_n & =
\max\left\{1,
2\left(1+\frac{2\pi}{\delta}+\frac{r}{2{r'}^2\log2}\right)
\left(1+\sqrt\frac{12\|H^0_n\|_{r'}}{\varepsilon_n}\right)
\frac{|\eta_n|\Theta_{n-1}}{\mu_n\Theta_n}
\right\}
\geq1,
\\
\mu_n & =
\left(
\frac{\Theta_n}{2^8
\zeta_n
\max\{1,|\eta_n|\}\Theta_{n-1}}\right)^{1/2}
\leq1.
\end{split}
\end{equation}
Here,
$$
\zeta_i=
\left(1+\sqrt\frac{12\|H^0_n\|_{r'}}{\varepsilon_n}\right)
\left(1+\frac1{2r'}\right)
\left(\frac{r'}r\right)^3
\|\trans T^{(i)}\|^3
>1.
$$
Notice that $\phi_n$ is our choice for Lemma \ref{lemma cutoff for H}.

Define the function $\vecomega\mapsto \bar\BB(\vecomega)$ as in \eqref{def B omega} but using the present choice of $\phi_n$.

\begin{theorem}\label{convergence of Rn for H}
Suppose that $\det(Q)\not=0$,
\begin{equation}\label{cdn freq for H}
\bar\BB(\vecomega) < +\infty,
\end{equation}
and $\rho>\bar\BB(\vecomega)+\nu$.
There exists $c,K>0$ such that if $H\in\A_{\rho,r'}$ and $\|H-H^0\|_{\rho,r'}<c$,
then $H$ is in the domain of $\RR_n$ and
\begin{equation}\label{bound on LLn less than epsilon for H}
\|\RR_n(H)-\RR_n(H^0)\|_{\rho_n,r} \leq
K \Theta_n \|H-H^0\|_{\rho,r'},
\quad
n\in\Nn\cup\{0\}.
\end{equation}
\end{theorem}

\begin{proof}
Let $\xi>0$ and $\rho_0=\rho-\nu-\xi>0$ such that $\rho_0 >\bar\BB(\vecomega)$.
Hence, by the definition \eqref{formula rho n vf} of $\rho_n$, there is $R>0$ satisfying $\rho_n>R B_{n-1}^{-1}$ for all $n\in\Nn$.

If $c\leq\varepsilon_0$ we use Theorem \ref{theorem existence of g} to get $\RR_0(H)\in\Ii_0^+\A_{\rho_0,r}$ with 
$$
\|\RR_0(H)-\RR_0(H^0)\|_{\rho_0,r}\leq
K \Theta_0 \|H-H^0\|_{\rho,r'}
$$
for some $K>0$.

Now, suppose that $H_{n-1}=\RR_{n-1}(H)\in\Ii^+_{n-1}\A_{\rho_{n-1},r}$, $n\in\Nn$, and 
\begin{equation}\label{ineq conv of ren H}
\begin{split}
\|H_{n-1}-H_{n-1}^0\|_{\rho_{n-1},r}
& \leq
K \Theta_{n-1} \|H-H^0\|_{\rho,r'},
\\
\|Q_{n-1}\|
&\leq
\|Q\|
\prod_{i=1}^{n-1} \frac32 \mu_{i}|\eta_{i}| \,\|T^{(i)}\|\,\|\trans T^{(i)}\|,
\\
\|Q_{n-1}^{-1}\|
&\leq
\|Q^{-1}\|
\prod_{i=1}^{n-1}2\mu_{i}^{-1}|\eta_{i}|^{-1} \|{T^{(i)}}^{-1}\|\, \|\trans{T^{(i)}}^{-1}\|.
\end{split}
\end{equation}
So, for $c$ small enough, we get
\begin{equation}\label{up bd Qn-1}
\|Q_{n-1}^{-1}\| 
\ll \frac{\Theta_{n-1}^{1/2}}{\Theta_{n-1}} 
\prod_{i=1}^{n-1} \frac{2^5\zeta_i^{1/2}\|{T^{(i)}}^{-1}\|\,\|\trans{T^{(i)}}^{-1}\|}{|\eta_i|^{1/2}}
\leq \frac{r^2}{32cK\Theta_{n-1}}.
\end{equation}
Thus, Lemma \ref{prop exist b n for H} is valid and as a consequence $\|b_n(H_{n-1})\| < r/8$.

After performing the operators $\LL_n$ and $\II_n$, we want to estimate the norm of the resulting Hamiltonians.
The constant and non-constant Fourier modes are dealt separately in
\begin{equation}
\II_n\LL_n(H)=
H_n^0 + 
\widehat\LL_n(\Ee H_{n-1}- H_{n-1}^0) +
\II_n\widetilde\LL_n((\Ii-\Ee)H_{n-1}).
\end{equation}
For the former we use Lemma \ref{prop hat L for H} and for the latter Lemmas \ref{lemma tilde Ln for H} and \ref{lemma cutoff for H}.
That is, the definition of $\mu_n$ implies that $\mu_n\leq\frac{r}{8r'\|\trans T^{(n)}\|}$ and
\begin{equation}
\begin{split}
\|\widehat\LL_n(\Ee H_{n-1}- H_{n-1}^0)\|'_{r'} & \leq 
2^7K\left(1+\frac{1}{2r'}\right) \left(\frac{r'}{r}\right)^3 
\mu_n^2|\eta_n|\,\|\trans T^{(n)}\|^3 \Theta_{n-1} \|H-H^0\|_{\rho,r'}
\\
& \leq
\frac{K}{2(1+\sqrt{12\|H^0_n\|_{r'}/\varepsilon_n})}\Theta_n \|H-H^0\|_{\rho,r'}.
\end{split}
\end{equation}
Furthermore, $\phi_n$ yields
\begin{equation}
\begin{split}
\|\II_n\widetilde\LL_n((\Ii-\Ee)H_{n-1})\|'_{\rho''_n,r'}
&\leq 
K
\left( 1+ \frac{2\pi}{\delta}+\frac{r}{2{r'}^2\log2}
\right)
\mu_n^{-1}\phi_n^{-1}|\eta_n|
\Theta_{n-1} \|H-H^0\|_{\rho,r'}
\\
& \leq
\frac{K}{2(1+\sqrt{12\|H^0_n\|_{r'}/\varepsilon_n})} \Theta_n \|H-H^0\|_{\rho,r'}.
\end{split}
\end{equation}

Moreover, assuming $c$ to be small enough, we obtain from \eqref{def Q n}, $\|Q_{n-1}\|^{-1}\leq \|Q_{n-1}^{-1}\|$ and \eqref{up bd Qn-1} that
\begin{equation}
\begin{split}
\|Q_n\| & \leq 
\mu_n |\eta_n|\, \|T^{(n)}\|\,\|\trans T^{(n)}\|
\|Q_{n-1}\|(1+ 16r^{-2}cK\Theta_{n-1}\|Q_{n-1}\|^{-1})
\\
& \leq
\|Q\|
\prod_{i=1}^{n} \frac32 \mu_i|\eta_i| \, \|T^{(i)}\|\,\|\trans T^{(i)}\|
\leq \frac{\sigma_n}{4r'}.
\end{split}
\end{equation}
By using \eqref{growth of inv Q} and again \eqref{up bd Qn-1},
\begin{equation}
\begin{split}
\|Q_n^{-1}\| & \leq 
\frac{\|{T^{(n)}}^{-1}\|\, \|\trans{T^{(n)}}^{-1}\| \|Q_{n-1}^{-1}\|} 
{\mu_n |\eta_n|
(1 -  16 r^{-2} cK\Theta_{n-1}\|Q_{n-1}^{-1}\|)}
\\
&\leq
\|Q^{-1}\|
\prod_{i=1}^{n}2\mu_{i}^{-1}|\eta_{i}|^{-1}\|{T^{(i)}}^{-1}\|\, \|\trans{T^{(i)}}^{-1}\|.
\end{split}
\end{equation}

The Hamiltonian $\II_n\LL_n(H_{n-1})$ is inside the domain of $\UU_n$ since for $c$ small enough $\frac12 c\,K\Theta_n<\varepsilon_n$ and 
$\|Q_n\|<\sigma_n/(2r')$.
The result follows from \eqref{estimate around omega y}.
\end{proof}

Cf. Remark \ref{remark on small analyt r} on how to generalise the above for a small analyticity radius $\rho$.

\begin{lemma}
If $\vecomega=\left(\begin{smallmatrix}\vecalf\\1\end{smallmatrix}\right)\in\Rr^d$ is diophantine, then \eqref{cdn freq for H} is verified.
\end{lemma}

\begin{proof}
The proof follows the same lines as for Lemma \ref{dioph in BC}, using the same choices of $\sigma_n$ and $t_n$.
In fact, to show \eqref{cdn freq for H} it is only necessary to check that the series $\sum B_n|\log|\eta_{n+1}||$, $\sum B_n\log\|T^{(n+1)}\|$, $\sum B_n\log \|\vecomega^{(n+1)}\|$ and $\sum B_n|\log\sigma_{n+1}|$ converge.
This is already done in the proof of the lemma.
\end{proof}

\subsection{Construction of the invariant torus}

In the following we will always assume to be in the conditions of 
section \ref{section:Convergence of renormalization for H}.
We use Theorem \ref{convergence of Rn for H} to determine the existence of an $\vecomega$-invariant torus for the flow of analytic Hamiltonians $H$ close enough to $H^0$ (Theorem \ref{theorem: main. intro for H}).
This follows from the construction of an analytic conjugacy between the linear flow on $\Tt^d$ of rotation vector $\vecomega$ and an orbit of $H$.
We will use the notations
$$
\lambda_n=\prod_{i=1}^n\eta_i
\quad\text{and}\quad
\chi_n =\prod_{i=1}^n\mu_i.
$$

Let the set $\Delta$ be given by
\begin{equation}
\Delta=\{H\in \A_{\rho,r'}\colon \|H-H^0\|_{\rho,r'}<c\},
\end{equation}
which is contained in the domain of $\RR_n$ for all $n\in\Nn\cup\{0\}$.
Given $H\in\Delta$, denote by $H_n=\RR_n(H)\in\Ii_n^+\A_{\rho_n,r}$.
It is simple to check that
\begin{equation}\label{formula for Hn}
\begin{split}
H_n &=
\frac{\lambda_n}{\chi_n}
[(\Ii-\Ee_0)
H\circ g_0\circ L_1^\mu \circ g_1 \circ\cdots\circ L_n^\mu] 
\circ g_n
\\
& = 
\frac{\lambda_n}{\chi_n}
((\Ii-\Ee_0)
H\circ g_0 \circ 
[\PP_1(H)\circ g_1\circ \PP_1(H)^{-1}] \circ 
\\
&
\cdots\circ
[\PP_{n-1}(H)\circ g_{n-1}\circ \PP_{n-1}(H)^{-1}]\circ \PP_n(H))
\circ g_n.
\end{split}
\end{equation}
Here, $g_k=\fG_k(\LL_k(H_{k-1}))$ is given by Theorem \ref{theorem existence of g} at the $k$th step and
\begin{equation}
L_k^\mu\colon(\vecx,\vecy)\mapsto L_k(\vecx,\mu_k\vecy),
\end{equation}
with $L_k$ as in \eqref{defn Ln for H} obtained for $H_{k-1}$.
In addition, the affine symplectic map $\PP_n(H)=L_1^\mu\cdots L_n^\mu$ is
\begin{equation}
\PP_n(H)\colon
(\vecx,\vecy)\mapsto
\left({P^{(n)}}^{-1}\vecx ,
\Phi_1(H)\dots \Phi_n(H_{n-1})(\vecy)\right)
\end{equation}
and $\PP_0(H)=\id$,
where $\Phi_k(H_{k-1})(\vecy)=\mu_k\trans T^{(k)}\vecy+b_k(H_{k-1})$.
Notice that
\begin{equation}
\Phi_1(H)\dots \Phi_n(H_{n-1}) (\vecy)=\chi_n \trans P^{(n)} \vecy + v_n(H),
\end{equation}
with 
$$
v_n(H)=b_1(H)+\sum_{i=2}^{n} \chi_{i-1}\trans P^{(i-1)} b_{i}(H_{i-1}).
$$

Let 
\begin{equation}\label{def an sum for H}
\begin{split}
a_n(H) &= 
\lim_{m\to+\infty} \Phi_n(H_{n-1})\dots \Phi_m(H_{m-1})(0)
\\
&= 
b_n(H_{n-1}) + \sum_{i=n+1}^{+\infty} \mu_n\dots\mu_{i-1}
\trans T^{(n)} \dots \trans T^{(i-1)} b_i(H_{i-1})
\end{split}
\end{equation}
if the limit exists.
If that is the case, 
\begin{equation}
a(H)= a_1(H) =\lim_{n\to+\infty} v_n(H)
\end{equation}
and 
\begin{equation}
a(H)-v_n(H)= \chi_{n} \trans P^{(n)} a_{n+1}(H).
\end{equation}

\begin{lemma}
For $H\in\Delta$ there is $a_n(H)\in B_{r/2}\subset\Cc^d$ and
the map $a_n\colon H\mapsto a_n(H)$ from $\Delta$ into $B_{r/2}$ is analytic taking any real-analytic $H$ into $\Rr^d$.
\end{lemma}

\begin{proof}
 From Lemma \ref{prop exist b n for H} we obtain $\|b_k(H_{k-1})\|\ll 1$ for any $k\in\Nn$.
Thus, by the definition of $\mu_n$, there is $0<\lambda<1$ such that
\begin{equation}
\mu_n\dots\mu_{i-1}
\|\trans T^{(n)} \dots \trans T^{(i-1)} b_i(H_{i-1})\|
\ll \lambda^{i-n},
\end{equation}
where $1\leq n\leq i-1$.
Hence, \eqref{def an sum for H} converges and each
$a_n(H)$ is well-defined in $\Cc^d$, unless $H$ is real which gives $ a_n(H)\in\Rr^d$.
The maps $H\mapsto  a_n(H)$ are analytic since the convergence is uniform.
\end{proof}

\begin{lemma}
There is an open ball $B$ centred at $H^0$ in $\Delta$ such that we can find sequences $R_n,r_n>0$ satisfying $R_{-1}=\rho$, $r_{-1}=r'$,
\begin{gather}\label{cdn on Rn for H}
R_n+ 2\pi K \Theta_n^{2/3} \|H-H^0\|_{\rho,r'}
\leq
R_{n-1}
\leq
\frac{\rho_{n-1}}{\|P^{(n-1)}\|},
\\ \label{cdn on rn for H}
r_n+ K \Theta_n^{2/3} \|H-H^0\|_{\rho,r'}
\leq
r_{n-1}\leq
\frac{\chi_{n-1} r}{2\|\trans{P^{(n-1)}}^{-1}\|},
\end{gather}
$H\in B$, and
\begin{equation}\label{lim Rn ell Theta n for H}
\lim_{n\to+\infty} 
R_n^{-1} \Theta_n^{2/3} = 0.
\end{equation}
\end{lemma}

\begin{proof}
Let $\rho_*=\min \rho_n$.
Since $\chi_n$ is decreasing, it is enough to check that 
$$
\Theta_n^{2/3}\ll 
\min\left\{
\lambda^n \rho_*\prod_{i=1}^n\|T^{(i)}\|^{-1},
\chi_n\prod_{i=1}^n\|\trans{T^{(i)}}^{-1}\|^{-1}
\right\}
$$ 
for some $0<\lambda<1$ by taking $R_n=c_1 \lambda^{-n}\Theta_n^{2/3}$ and $r_n=c_2\Theta_n^{2/3}$ with small constants $c_1,c_2>0$.
Thus, the inequalities \eqref{cdn on Rn for H} and \eqref{cdn on rn for H} hold whenever we take a sufficiently small bound on $\|H-H^0\|_{\rho,r}$.
The limit \eqref{lim Rn ell Theta n for H} is now immediate.
\end{proof}

For each $H\in\Delta$, consider the isotopic to the identity analytic symplectomorphism
\begin{equation}
W_n(H) = \PP_n(H)\circ\fG_n(\LL_n(H_{n-1}))\circ \PP_n(H)^{-1}
\end{equation}
on $ \PP_n(H) \DD_{\rho_n,r}$.
In particular, $W_n(H^0)=\id$.
Notice that for $H$ real-analytic, 
$$
W_n(H)(\Rr^{2d})\subset\Rr^{2d}.
$$

For a given $H\in\Delta$, define the norm $\|X\|_n=\|X\circ V_{a(H)}\|_{R_n,r_n}$, whenever $X\circ V_{a(H)}\in\A_{R_n,r_n}^{2d}$, where we have introduced the vertical translation 
\begin{equation}
V_\vecz\colon(\vecx,\vecy)\mapsto (\vecx,\vecy+\vecz),
\end{equation}
for any $\vecz\in\Cc^d$.

\begin{lemma} 
$W_n$ is analytic on $B$, and satisfies
$$
W_n(H)\colon V_{a(H)}(\DD_{\rho_n,r_n})\to V_{a(H)}(\DD_{\rho_{n-1},r_{n-1}})
$$ 
and
\begin{equation}\label{Wn-id for H}
\|W_n(H)-\id\|_n \leq K' \Theta_n^{2/3} \|H-H^0\|_{\rho,r},
\quad
H\in B,
\end{equation}
for some constant $K'>0$.
\end{lemma}

\begin{proof}
For $H\in\Delta$ and $(\vecx,\vecy)\in \DD_{R_n,r_n}$,
\begin{equation}
\begin{split}
\|\im P^{(n)}\vecx\| 
&<
\|P^{(n)}\| R_n/2\pi \leq \rho_n/2\pi,
\\
\|\Phi_n^{-1}(H_{n-1})\dots \Phi_1^{-1}(H)(\vecy+a(H))\| 
& =
\|\chi_n^{-1} \trans {P^{(n)}}^{-1} ( \vecy +a(H) - v_n(H))\|
\\
& \leq
\chi_n^{-1} \|\trans {P^{(n)}}^{-1}\| r_n + \|a_{n+1}(H)\|
< r.
\end{split}
\end{equation}
Therefore, $\PP_n(H)^{-1} \circ V_{a(H)} (\DD_{R_n,r_n}) \subset \DD_{\rho_n,r}$.
Moreover, using \eqref{estimate around omega y}, 
\begin{equation}
\begin{split}
\|W_n(H)-\id\|_n 
& =
\| 
\widehat\PP_n(H)\circ [\fG_n(\II_n\LL_n(H_{n-1}))-\id] \circ \PP_n(H)^{-1} \circ V_{a(H)}
\|_{R_n,r_n}
\\
&\leq
\varepsilon_n^{-1}
\|\widehat\PP_n(H)\| \, 
\|\II_n\LL_n(H_{n-1})-H_n^0\|_{\rho_n,r'}
\\
&\ll
\Theta_n^{2/3},\|H-H^0\|_{\rho,r},
\end{split}
\end{equation}
where $\widehat\PP_n(H)$ corresponds to the linear part $(\vecx,\vecy)\mapsto ({P^{(n)}}^{-1}\vecx, \chi_n\trans P^{(n)}\vecy)$ of $\PP_n(H)$ which has norm bounded by $\|\widehat\PP_n(H)\| \leq \|{P^{(n)}}^{-1}\|+\chi_n\|\trans P^{(n)}\|$.

Now, for $(\vecx,\vecy)\in \DD_{R_n,r_n}$ and $H\in B$,
\begin{equation*}
\begin{split}
\|\pi_1 \im W_n(H)\circ V_{a(H)}(\vecx,\vecy)\| 
& \leq 
\|\im(\pi_1 W_n(H)\circ V_{a(H)}(\vecx,\vecy)-\vecx)\| + \|\im \vecx\|
\\
&<
\|W_n(H)-\id\|_n+R_n/2\pi
<
R_{n-1}/2\pi,
\\
\|\pi_2 W_n(H)\circ V_{a(H)}(\vecx,\vecy)-a(H)\| 
& \leq 
\|\pi_2 W_n(H)\circ V_{a(H)}(\vecx,\vecy)-\vecy-a(H)\|+\|\vecy\|
\\
&<
\|W_n(H)-\id\|_n+r_n
<
r_{n-1}.
\end{split}
\end{equation*}
So, $W_n(H)\colon V_{a(H)}(\DD_{R_n,r_n})\to V_{a(H)}(\DD_{R_{n-1},r_{n-1}})$.
\end{proof}

Consider the analytic map $\Gamma_n$ on $B$ satisfying
$\Gamma_n(H)\colon V_{a(H)}(\DD_{R_n,r_n})\to  V_{a(H)}(\DD_{\rho,r'})$,
\begin{equation}
\Gamma_n(H) = W_0(H)\circ \cdots \circ W_n(H).
\end{equation}
We then rewrite \eqref{formula for Hn} as
\begin{equation}\label{relation H and Hn}
H\circ \Gamma_n(H) = \frac{\chi_n}{\lambda_n} H_n\circ \PP_n(H)^{-1} + E(H),
\end{equation}
where $E(H)$ represents a constant (irrelevant) term.
Since each $W_n(H)$ is symplectic, thus $\Gamma_n(H)$ is symplectic and $H\circ\Gamma_n(H)$ is canonically equivalent to the Hamiltonian $H_n$.
In particular, if $H_n=H_n^0$ for some $n$, there is an $\vecomega$-invariant torus in the phase space of $H_n$.
We are interested in the general case, $H_n-H_n^0\to0$ as $n\to+\infty$.

By a simple adaptation of Lemma \ref{lemma Hn-Hn-1 vf}, we have the following result for $\Gamma_n$.

\begin{lemma}\label{lemma Gamma n-Gamma n-1}
There is $c>0$ such that for $H\in B$
$$
\|\Gamma_n(H)-\Gamma_{n-1}(H)\|_n \leq
c \Theta_n^{2/3} \|H-H^0\|_{\rho,r'}.
$$
\end{lemma}

Consider the Banach space
$C^1_{per}(\Rr^d,\Cc^{2d})$ of $C^1$ functions $\Zz^d$-periodic, endowed with the norm
$$
\|f\|_{C^1} = \max_{k\leq1}\max_{\vecx\in\Rr^d} \|D^kf(\vecx)\|.
$$

\begin{lemma}
There exist $C>0$, an open ball $B'\subset B$ centred at $H^0$ and an analytic map $\Gamma\colon B'\to \Diff_{per}(\Rr^d,\Cc^{2d})$ such that, for every $H\in B'$, $\Gamma(H)=\lim_{n\to+\infty}\Gamma_n(H)(\cdot,a(H))$ and 
\begin{equation}\label{estimate Gamma id for H}
\|\Gamma(H)-(\id,a(H))\|_{C^1} \leq C \|H-H^0\|_{\rho,r'}.
\end{equation}
If $H\in B'$ is real-analytic, then $\Gamma(H)\in \Diff_{per}(\Rr^d,\Rr^{2d})$.
\end{lemma}

\begin{proof}
For each $H\in B$, by the first inequality in \eqref{Cauchy estimate H},
\begin{equation}\label{estimate var Gamma n}
\begin{split}
\|[\Gamma_n(H)-\Gamma_{n-1}(H)](\cdot,a(H))\|_{C^1} 
&\leq 
\max_{k\leq1}\sup_{\vecx\in D_{\rho_n/2}} 
\|D^k [\Gamma_n(H)(\vecx,a(H))-\Gamma_{n-1}(H)(\vecx,a(H))]\| 
\\
&\leq
\frac{4\pi}{R_n}  
\|\Gamma_n(H)-\Gamma_{n-1}(H)\|_n
\end{split}
\end{equation}
which is estimated using \eqref{lim Rn ell Theta n for H}.
Hence, $\Gamma_n(H)(\cdot,a(H))$ converges
in the Banach space $C_{per}^1(\Rr^d,\Cc^{2d})$, and \eqref{estimate Gamma id for H} holds.
The convergence of $\Gamma_n$ is uniform in $B$, thus $\Gamma$ is analytic.
If $H$ is sufficiently close to $H^0$, $\Gamma(H)$ is in fact a diffeomorphism as the space of close to identity diffeomorphisms is closed for the $C^1$ norm.
Finally, for $H$ real-analytic we have $\Gamma(H)(\Rr^d)\subset \Rr^{2d}$ in view of the similar property for each $W_n(H)$.
\end{proof}

The Hamiltonian vector field of $H$ is denoted by $X_H=\Jj \nabla H$, where $\Jj\colon(x,y)\mapsto (y,-x)$.

\begin{lemma}
For $H\in B'$, we have on $\Rr^d$
\begin{equation}\label{eq Gamma n for H}
X_H\circ\Gamma(H) = D (\Gamma(H))\,\vecomega.
\end{equation}
\end{lemma}

\begin{proof}
Since $\Gamma_n(H)$ is a symplectomorphism, we have for $\vecx\in\Rr^d$,
\begin{equation}
\begin{split}
Y_n(\vecx) &=
X_H\circ\Gamma_n(H)\circ V_{a(H)}(\vecx,0)
-D(\Gamma_n(H))\circ V_{a(H)}(\vecx,0) \, X_{H^0}(\vecx,0)
\\
& =
[D(\Gamma_n(H))\circ V_{a(H)} \, 
X_{H\circ\Gamma_n(H)\circ V_{a(H)}-H^0}](\vecx,0).
\end{split}
\end{equation}
Hence,
\begin{equation}\label{rel n conjugacy}
\|Y_n (\vecx)\|
\leq 
\|D(\Gamma_n(H))(\vecx,a(H))\| \,
\|\nabla [H\circ\Gamma_n(H)\circ V_{a(H)}-H^0](\vecx,0)\|.
\end{equation}
We can now estimate the above norm by recalling \eqref{relation H and Hn} in
\begin{equation}
\begin{split}
\nabla [H\circ\Gamma_n(H)\circ V_{a(H)}-H^0]
= &
\frac{\chi_n}{\lambda_n}
\nabla[(H_n-H_n^0)\circ\PP_n(H)^{-1}\circ V_{a(H)} 
\\
& 
+H_n^0\circ\PP_n(H)^{-1}\circ V_{a(H)}-H_n^0\circ\PP_n(H_0)^{-1}]
\end{split}
\end{equation}
and
\begin{multline}
\frac{\chi_n}{\lambda_n}
\nabla[H_n^0\circ\PP_n(H)^{-1}\circ V_{a(H)}-H_n^0\circ\PP_n(H_0)^{-1}]
=\\=
\trans[a(H)-v_n(H)]
\left[Q+ 
\sum_{i=1}^{n-1} \frac{1}{\lambda_i\chi_i}{P^{(i)}}^{-1} D^2F_0^{(i)}(b_{i+1}(H_i)) \trans{P^{(i)}}^{-1}
\right].
\end{multline}
Notice that by induction we get
\begin{equation}
Q_n=\lambda_n\chi_n P^{(n)}Q\trans P^{(n)}
+\sum_{i=0}^{n-1} \frac{\lambda_n\chi_n}{\lambda_i\chi_i}
P^{(n)}{P^{(i)}}^{-1}D^2F_0^{(i)}(b_{i+1}(H_i))\trans{P^{(i)}}^{-1}\trans P^{(n)}.
\end{equation}
Since $\sum_{i=1}^{n-1} (\chi_i|\lambda_i|)^{-1} \|{P^{(i)}}^{-1}\| \, \|\trans{P^{(i)}}^{-1}\| \Theta_i \ll 1$,
\begin{equation}
\frac{\chi_n}{|\lambda_n|}
\|\nabla[(H_n-H_n^0)\circ\PP_n(H)^{-1}\circ V_{a(H)}](\vecx,0)\|
 \ll
\frac{1}{|\lambda_n|}\|\trans {P^{(n)}}^{-1}\|\,\|H_n-H_n^0\|_{\rho_n,r}
\end{equation}
and
\begin{equation}
\begin{split}
\|a(H)-v_n(H)\|
&\leq
\chi_n\|\trans P^{(n)}\|\,\|a_{n+1}(H)\|
\\
&\ll
\chi_n\|\trans P^{(n)}\|\,\Theta_n
\end{split}
\end{equation}
we have that
\begin{equation}
\|\nabla [H\circ\Gamma_n(H)\circ V_{a(H)}-H^0](\vecx,0)\|
\ll
\Theta_n^{2/3}.
\end{equation}
Finally, from the convergence of $\Gamma_n$ and
\begin{equation}
\|D\Gamma_n(H)(\vecx,a(H))\|
\ll
\frac{1}{R_n} \|\Gamma_n(H)\|_n \ll \frac{1}{R_n},
\end{equation}
we find that $\|Y_n(\vecx)\|$ converges uniformly to $0$ as $n\to+\infty$.
\end{proof}

\begin{lemma}\label{proposition: comm H and R for H}
If $H\in B'$ and $\vecx\in\Rr^d$, then
\begin{equation}\label{comm Gamma Rz for H}
\Gamma(H\circ R_\vecx)=R_\vecx^{-1}\circ \Gamma(H)\circ \widehat R_\vecx
\end{equation}
where $\widehat R_\vecx\colon \vecz\mapsto \vecz+\vecx$ is a translation on $\Cc^d$.
\end{lemma}

\begin{proof}
For each $n\in\Nn$, \eqref{comm Rz Ln for H} implies that $\PP_n(H\circ R_z)=\PP_n(H)$ and we know that $\PP_n(H)\circ R_{P^{(n)}\vecz}^{-1} = R_\vecz^{-1}\circ\PP_n(H)$, $\vecz\in\Cc^d$.
So, from Lemma \ref{theorem comm g R},
\begin{equation}
\begin{split}
W_n(H\circ R_\vecx) &=\PP_n(H)\circ\fG_n(\LL_n\RR_{n-1}(H\circ R_\vecx))\circ 
\PP_n(H)^{-1}
\\
&=
R_\vecx^{-1}\circ W_n(H)\circ R_\vecx.
\end{split}
\end{equation}
Thus, $\Gamma_n(H\circ R_\vecx)=R_\vecx^{-1}\circ \Gamma_n(H)\circ R_\vecx$ and \eqref{comm Gamma Rz for H} follows from the convergence of $\Gamma_n$.
\end{proof}

The flow generated by $X_H$ is denoted by $\phi_H^t$ taken at time $t\geq0$.
Hence, 
$$
\phi_{H^0}^t|_{\Tt^d\times\{0\}}=\widehat R_{\vecomega t}.
$$
We prove below the existence of an invariant torus $\TTT$ for $H$, i.e. an analytic conjugacy between $\phi_H^t|_\TTT$ and $\widehat R_{\vecomega t}$.

\begin{theorem}
Let $D\subset\Rr^d$ be an open ball about the origin.
If $H\in C^\vecomega(\Tt^d\times D)$ is sufficiently close to
$H^0$, then there exist $\veca\in\Rr^d$ and
a $C^\omega$-diffeomorphism $\gamma\colon\Tt^d\to \Tt^d\times D$ such that 
\begin{equation}\label{eq conj to inv torus}
\phi_H^t\circ \gamma=\gamma\circ\widehat R_{\vecomega t},
\quad
t\geq0,
\end{equation}
and $\TTT=\gamma(\Tt^d)\simeq\Tt^d$ is a submanifold homotopic to $\{\vecy=\veca\}$.
Furthermore, the maps $H\mapsto\veca$ and $H\mapsto\gamma$ are analytic.
\end{theorem}

\begin{proof}
The lift $\widetilde H$ to $\Rr^d\times D$ of $H$ is assumed to have a unique analytic
extension to $\DD_{\rho,r'}$.
Consider the real-analytic Hamiltonian
$G=\widetilde H\in \A_{\rho,r'}$.
Suppose that $G$ is close enough to $H^0$ such that
$G\in B'$ and $G\circ R_\vecz\in B'$ for $\eta>0$ and $\vecz\in D_\eta$.
Then, $\veca=a(G)$ and $\gamma=\Gamma(G)|_{[0,1)^d}$, $C^1$ and homotopic to $(\id,\veca)$, verifies \eqref{eq conj to inv torus}.
This follows from \eqref{eq Gamma n for H} and the equivalent equation
$$
\frac d{dt}\Big|_{t=0}(\phi_H^t\circ\gamma)=\frac d{dt}\Big|_{t=0}(\gamma\circ\widehat R_{\vecomega t}),
$$
which we integrate for initial condition $\phi_H^0=\widehat R_0=\id$.

We now want to extend analytically $\gamma$ to a complex neighbourhood of its
domain.
Take $\widetilde \gamma(\vecz)=R_\vecz\circ \Gamma(G\circ R_\vecz)(0)$,
$\vecz\in D_\eta$.
The maps $\vecz\mapsto G\circ R_\vecz$ and $H\mapsto\Gamma(H)$ are analytic and
$C^1_{per}(\Rr^d,\Cc^{2d})\ni g\mapsto g(0)$ is bounded.
As $\widetilde\gamma\colon D_\eta\to\Cc^{2d}$ involves their composition, it is analytic and $\Zz^d$-periodic.
 From \eqref{comm Gamma Rz for H}, for any $\vecx\in\Rr^d$, we have
$$
\widetilde\gamma(\vecx) =\Gamma(G)\circ \widehat R_\vecx(0) =\Gamma(G)(\vecx)=\gamma(\vecx).
$$

Finally, since $\Gamma$ is analytic, the same is true for the map $H\mapsto\gamma$.
\end{proof}

\begin{remark}
A quasiperiodic invariant torus $\TTT$ is always Lagrangian. 
If in addition $\TTT$ is homotopic to $\{\vecy=\veca\}$, it is the graph of a function $\psi\in C^\omega(\Tt^d,D)$ (see \cite{Herman6}).
\end{remark}

\begin{appendix}
\section{Elimination of modes}

\subsection{Homotopy method for vector fields}\label{section proof of main theorem}

In this section we prove Theorem \ref{main theorem1}
using the homotopy method (cf. \cite{jld}).
As $n$ is fixed, we will drop it from
our notations. 
In addition we write $\rho'=\rho_n$ and $\rho=\rho_n+\nu$.
We will be using the symbol $D_\vecx$ for the derivative with respect
to $\vecx$.

Firstly, we include a technical lemma that will be used in the
following.

\begin{lemma}\label{lemma f-f*}
Let $f\in\A'_{\rho,r}$.
If $U=\id+(u,0)$ where $u\colon D_{\rho'}\times B_r\to D_{(\rho-\rho')/2}$ is in $\A_{\rho',r}$
and $\|u\|_{\rho',r}<(\rho-\rho')/4\pi$,
then
\begin{itemize}
\item\label{f*}
$\|f\circ U\|_{\rho',r}\leq \|f\|_{(\rho+\rho')/2,r}$,
\item\label{Df*}
$\|D_\vecx f\circ U\|\leq \|f\|'_{(\rho+\rho')/2,r}$,
\item\label{f*-f}
$\|f\circ U-f\|_{\rho',r}\leq \|f\|'_{(\rho+\rho')/2,r}\,\|u\|_{\rho',r}$,
\item\label{Df*-Df}
$\|D_\vecx f\circ U-D_\vecx f\|\leq \frac{4\pi}{\rho-\rho'}\|f\|'_{\rho,r}\,\|u\|_{\rho',r}$.
\end{itemize}
\end{lemma}

The proof of these inequalities is straightforward and thus will be
omitted. 
Now, assume that
$$
\delta = 
42\varepsilon/\sigma
< 1/2.
$$
For vector fields in the form $X=\vecomega+\pi_2+f$, where $\pi_2\colon(\vecx,\vecy)\mapsto \vecy$ is seen as a function in $\A'_{\rho,r}$, consider $f$ to be in the open ball in $\A'_{\rho,r}$ centred at the origin with radius $\varepsilon$.
The coordinate transformation $U$ is written as $U=\id+(u,0)$, with $u$ in
$$
\BB=\left\{u\in\Ii^-\A'_{\rho',r}\colon  u\colon D_{\rho'}\times B_r\to D_\rho,
\|u\|'_{\rho',r}<\delta\right\}.
$$
Notice that we have
\begin{eqnarray*}
\Ii^-U^*(X)
&=&
\Ii^-(DU)^{-1}(\vecomega+\pi_2+f\circ U,0) \\
&=&
(\Ii^-(I+D_\vecx u)^{-1}(\vecomega+\pi_2+f\circ U),0).
\end{eqnarray*}

 From now on the parameter $r$ is omitted whenever there is no ambiguity.
Define the operator
$F\colon\BB\to\Ii^-\A_{\rho'}$,
\begin{equation}\label{definition of F}
F(u)=\Ii^-(I+D_\vecx u)^{-1}(\vecomega+\pi_2+f\circ U).
\end{equation}
$F(u)$ takes real values for real arguments whenever $u$ has that property.
It is easy to see that
the derivative of $F$ at $u$ is the linear map from
$\Ii^-\A'_{\rho'}$ to $\Ii^-\A_{\rho'}$:
\begin{eqnarray}\label{derivative of F}
DF(u)\,h 
& = & 
\Ii^-(I+D_\vecx u)^{-1} [ D_\vecx f\circ U\,h 
\\\nonumber
& &
-D_\vecx h\,(I+D_\vecx u)^{-1}\left(\vecomega+\pi_2+f\circ U\right)].
\end{eqnarray}

We want to find a solution of
\begin{equation}\label{F(u)}
F(u_t)=(1-t)F(u_0),
\end{equation}
with $0\leq t\leq1$ and ``initial'' 
condition $u_0=0$.
Differentiating the above equation with respect to $t$, we get
\begin{equation}\label{homotopy trick eq}
DF(u_t)\frac{du_t}{dt}=-F(0).
\end{equation}

\begin{proposition}\label{DF(u) invert}
If $u\in\BB$, then $DF(u)^{-1}$
is a bounded linear operator from
$\Ii^-\A_{\rho'}$ to $\Ii^-\A'_{\rho'}$ and
$$
\|DF(u)^{-1}\| < {\delta}/{\varepsilon}.
$$
\end{proposition}

 From the above proposition (to be proved in Section \ref{proof of
prop}) we
integrate (\ref{homotopy trick eq}) with respect to $t$,
obtaining the integral equation:
\begin{equation}\label{u_l}
u_t=-\int_0^t DF(u_s)^{-1}\,F(0)\,ds.
\end{equation}
In order to check that $u_t\in\BB$ for any $0\leq t\leq1$, we
estimate its norm:
$$
\begin{array}{rl}
\|u_t\|'_{\rho'} 
\leq 
&
t\sup\limits_{v\in\BB}\|DF(v)^{-1}F(0)\|'_{\rho'}
\\
\leq&
t\sup\limits_{v\in\BB}\|DF(v)^{-1}\|\,\|\Ii^-f\|_{\rho'}
<t\delta\|f\|_{\rho'}/\varepsilon,
\end{array}
$$
so, $\|u_t\|'_{\rho'}<\delta$.
Therefore, the solution of (\ref{F(u)}) exists in $\BB$ and is given
by (\ref{u_l}).
Moreover, if $X$ is real-analytic, then $u_t$ takes real values for real arguments.

It is now easy to see that
$$
U_t^*X-X^0 =
\Ii^+\sum_{n\geq2}(-D(U_t-\id))^nX^0
+\Ii^+ U_t^*f + (1-t)\Ii^-f.
$$
So, using Lemma \ref{lemma f-f*},
\begin{eqnarray*}
\|U_t^*X-X^0\|_{\rho'}
&\leq&
\frac1{1-\|u_t\|'_{\rho'}}
(\|\vecomega\|\,{\|u_t\|'_{\rho'}}^2+\|f\|_\rho) + (1-t)\|f\|_{\rho'}
\\
&<&
\frac{1}{1-\delta}\left(
\delta^2\|\vecomega\|\|f\|_{\rho'}/\varepsilon^2+1
\right)\|f\|_{\rho}+ (1-t)\|f\|_{\rho'}
\\
&<&
\left[\frac1{1-\delta}\left(\frac{\delta^2\|\vecomega\|}\varepsilon+1\right)+1-t\right]
\|f\|'_\rho.
\end{eqnarray*}
Moreover,
$\|U_t^*X-X^0-\Ii^+f - (1-t) \Ii^-f\|_{\rho'}=\OO(\|f\|_{\rho}^2)$, hence
the derivative of $X\mapsto U_t^*X$ at $X^0$ is $\Ii-t\Ii^-$.


\subsubsection{Proof of Proposition \ref{DF(u) invert}} \label{proof of prop}

\begin{lemma}\label{DF(0) invert}
If $\|f\|'_\rho<\varepsilon<\sigma/6$, then 
$$
DF(0)^{-1}\colon\Ii^-\A_{\rho'}\to\Ii^-\A'_{\rho'}
$$ 
is continuous and
$$
\|DF(0)^{-1}\| < \frac3{\sigma-6\|f\|'_\rho}.
$$
\end{lemma}

\begin{proof}
 From (\ref{derivative of F}) one has 
$$
\begin{array}{rl}
DF(0)\,h
&
=\Ii^- ( \widehat f-D_\vecomega)\,h
\\
&=
-\left(
\Ii-\Ii^-\widehat
f\,D_\vecomega^{-1}
\right)
D_\vecomega\,h,
\end{array}
$$
where $\widehat f\,h=Df\,h-Dh\,f$ and $D_\vecomega\,h=D_\vecx h\,(\vecomega+\pi_2)$. 
Thus, the inverse of this operator, if it exists, is given by
$$
DF(0)^{-1}=-D_\vecomega^{-1}
\left(\Ii-\Ii^-\widehat
f\,D_\vecomega^{-1}\right)^{-1}.
$$

The inverse of $D_\vecomega$ is the linear map from
$\Ii^-\A_{\rho'}$ to $\Ii^-\A'_{\rho'}$:
$$
D_\vecomega^{-1}\,g(\vecx,\vecy)=\sum\limits_{\veck\in I^-}\frac{g_\veck(\vecy)}{2\pi
\i \veck\cdot X^0(\vecy)} \e^{2\pi \i\veck\cdot \vecx},
$$
and is well-defined since Lemma \ref{lemma res cone with y, vf} implies that 
$$
|\veck\cdot X^0(\vecy)| > \sigma\|\veck\|/2,
$$
with $\veck\in I^-$ and $\vecy\in B_r$.
So,
\begin{eqnarray*}
\|D_\vecomega^{-1}\,g\|'_{\rho'}
&=&
\sum\limits_{\veck\in I^-}\frac{1+2\pi\|\veck\|}{2\pi} 
\sup_{\vecy\in B_r}
\left\|\frac{g_\veck(\vecy)}{\veck\cdot X^0(\vecy)}\right\|\e^{\rho'\|\veck\|}
\\
&<&
\sum\limits_{\veck\in I^-}\frac{1+2\pi\|\veck\|}{\pi\sigma
\|\veck\|} \|g_\veck\|_r\e^{\rho'\|\veck\|} 
\\
&\leq&
\frac3\sigma\|g\|_{\rho'}.
\end{eqnarray*}
Hence, 
$\|D_\vecomega^{-1}\| < 3/\sigma$.
It is possible to bound from above the norm of $\widehat f\colon
\Ii^-\A'_{\rho'}\to\A_{\rho'}$ by $\|\widehat f\|\leq 2\|f\|'_{\rho'}$.
Therefore,
$$
\|\Ii^-\widehat f \,D_\vecomega^{-1}\| < \frac6\sigma\|f\|'_{\rho'}<1,
$$
and
$$
\left\|\left(\Ii-\Ii^-\widehat f\,D_\vecomega^{-1}\right)^{-1}\right\| < \frac\sigma{\sigma-6\|f\|'_{\rho'}}.
$$
The statement of the lemma is now immediate.
\end{proof}

\begin{lemma}\label{DF(u)-DF(0) bound}
Given $u\in\BB$, 
the linear operator 
$DF(u)-DF(0)$
mapping $\Ii^-\A'_{\rho'}$ into $\Ii^-\A_{\rho'}$, is bounded and
$$
\|DF(u)-DF(0)\| < \frac{\|u\|'_{\rho'}}{1-\|u\|'_{\rho'}}
\left[
\left(\frac{4\pi}{\rho-\rho'}+ \frac{4-2\|u\|'_{\rho'}}{1-\|u\|'_{\rho'}}\right) \|f\|'_\rho 
+\frac{2-\|u\|'_{\rho'}}{1-\|u\|'_{\rho'}}
\|\vecomega+\pi_2\|
\right].
$$
\end{lemma}

\begin{proof}
The formula (\ref{derivative of F}) gives
\begin{eqnarray*}
\left[DF(u)-DF(0)\right]\,h &=&
\Ii^-(I+D_\vecx u)^{-1} 
\left[D_\vecx f\circ U\,h-(I+D_\vecx u)D_\vecx f\,h   \right. 
\\
&& \left. 
-D_\vecx h\,(I+D_\vecx u)^{-1}(\vecomega+\pi_2+f)\circ U 
\right.
\\
&& \left.
+ (I+D_\vecx u)D_\vecx h\,(\vecomega+\pi_2+f)\right]
\\
&=&
\Ii^-(I+D_\vecx u)^{-1} \{A+B+C\},
\end{eqnarray*}
where
\begin{equation*}
\begin{split}
A &=
\left[D_\vecx f\circ U-D_\vecx f-D_\vecx u\,D_\vecx f\right]\,h
\\
B&=
D_\vecx u\,D_\vecx h\,(\vecomega+\pi_2+f)
\\
C&= 
-D_\vecx h\,(I+D_\vecx u)^{-1}\left[f\circ U-f-D_\vecx u\,(\vecomega+\pi_2+f)\right].
\end{split}
\end{equation*}
Using Lemma \ref{lemma f-f*},
\begin{eqnarray*}
\|A\|_{\rho'} &\leq&
\left( \frac{4\pi}{\rho-\rho'}\|f\|'_{\rho}\|u\|_{\rho'}+  \|f\|'_{\rho'}\|u\|'_{\rho'} \right)\|h\|_{\rho'},
\\
\|B\|_{\rho'} &\leq& 
\left(\|\vecomega+\pi_2\|+\|f\|_{\rho'}\right)
\|u\|'_{\rho'}\|h\|'_{\rho'},
\\
\|C\|_{\rho'} &\leq&
\frac{1}{1-\|u\|'_{\rho'}}   
\left[
\|f\|'_{(\rho+\rho')/2}\|u\|_{\rho'}+\|u\|'_{\rho'}
\left( \|\vecomega+\pi_2\|_r+\|f\|_{\rho'} \right) \right]
  \|h\|'_{\rho'}.
\end{eqnarray*}

\end{proof}

To conclude the proof of Proposition
\ref{DF(u) invert}, notice that
\begin{eqnarray*}
\|DF(u)^{-1}\| 
&\leq&
 \left( \|DF(0)^{-1}\|^{-1}-\|DF(u)-DF(0)\|\right)^{-1}
\\
&<& 
\left\{ 
\frac\sigma3-2\varepsilon-
\frac{\delta}{1-\delta}
\left[
\left(\frac{4\pi}{\rho-\rho'}+ \frac{4-2\delta}{1-\delta}\right)  \varepsilon
+\frac{2-\delta}{1-\delta}
\|\vecomega+\pi_2\|_r
\right]
\right\}^{-1}
\\
&<&
 \frac{\delta}{\varepsilon}.
\end{eqnarray*}
The last inequality is true if
\begin{equation*}\label{2nd bound on epsilon}
\varepsilon<  
\delta\left[
\frac\sigma3-\frac{2\delta}{(1-\delta)^2}\|\vecomega+\pi_2\|_r \right]
\left[ 1+2\delta+\frac{\delta^2}{1-\delta} \left( \frac{4\pi}{\rho-\rho'}+\frac{4-2\delta}{1-\delta}\right)
\right]^{-1}
\end{equation*}
with a positive numerator $N$ and denominator $D$ in the r.h.s.
This is so for our choices of $\varepsilon$ and $\delta<\frac12$, by observing
that 
\begin{equation}
\|\pi_2\|_r =\sup_{\vecy\in B_r} \|\vecy\|
\leq
a\,|\gamma^{(n)}|^{-1} \|M^{(n)}\| 
+ b \|\vecomega\| <\frac12 \|\vecomega\|,
\end{equation}
thus $\|\vecomega+\pi_2\|_r<\frac32\|\vecomega\|$ and
$$
\frac{2\delta}{(1-\delta)^2}\|\vecomega+\pi_2\|_r<
12\delta\|\vecomega\|
<\frac\sigma6.
$$
So, $N>\delta\sigma/6$, $D<7$, and finally $\varepsilon\leq
\frac{\sigma^2}{42 \|\vecomega\|} < \frac\sigma{42}<N/D$.

\subsection{Elimination of modes for Hamiltonians}
\label{elim modes Ham}

Here we present a proof of Theorem \ref{theorem existence of g}. 
It is similar to related methods appearing in e.g. \cite{Koch,Abad2}.
As we have fixed $n$, we will not include it in our notations.

Let $R=(R_1,R_2)$ and $R'=(R'_1,R'_2)$ be such that $R>R'>0$ componentwise.
We will be interested on the set $\GG_{R'}$ of analytic symplectomorphisms $g\colon\DD_{R'}\to\DD_{R}$ satisfying $g-\id\in \A_{R'}^{2d}$ and 
$$
\|g-\id\|_{R'}<\delta=\min\{(R_1-R'_1)/2\pi,R_2-R'_2\}.
$$
We use the notation $\{\cdot,\cdot\}$ for the usual Poisson bracket associated to $\Jj\colon(\vecx,\vecy)\mapsto (\vecy,-\vecx)$.
In the following $R-\delta$ stands for $R-\delta(1,1)$ and $\pi_2\colon(\vecx,\vecy)\mapsto\vecy$ is the projection on the second component.

\begin{lemma}\label{lemma estimates on H step elim}
Let $0<\xi\leq\frac12$.
If $G\in\A'_{R'}$ and $\|G\|'_{R'}<\xi\delta/(2\pi+1)$, then there is a unique analytic symplectomorphism $g\colon \DD_{R'-2\delta}\to \Cc^{2d}$ such that $\|g-\id\|_{R'-2\delta}<\xi\delta$ and
\begin{equation}
g=\id+\Jj\nabla G\circ\widehat g,
\end{equation}
where $\widehat g(\vecx,\vecy)=(\vecx,\pi_2 g(\vecx,\vecy))$, $(\vecx,\vecy)\in\DD_{R'-2\delta}$.
Moreover, for any $H\in\A_{R'}$
\begin{equation}\label{estimates for g H}
\begin{split}
\|H\circ g\|_{R'-2\delta} & \leq  \|H\|_{R'}\\
\|H\circ g-H\|_{R'-2\delta} & \leq 2\xi \|H\|_{R'}\\
\|H\circ g-H-\{H,G\}\|_{R'-2\delta} & \leq 2\xi^2 \|H\|_{R'}
\end{split}
\end{equation}
and the maps $G\mapsto g$ and $G\mapsto H\circ g$ are analytic.
\end{lemma}

\begin{proof}
Define the map $T\colon g\mapsto\id+\Jj\nabla G\circ\widehat g$ on the open ball $B$ in $\GG_{R'-2\delta}$ centred at the identity and with radius $\xi\delta$.
It is simple to check that $T(B)\subset B$, in particular $T(g)$ for $g\in B$ is symplectic.
We now show that $T$ is a contraction on $B$ and thus its unique fixed point is the map we are looking for.
In fact, whenever $g\in B$ we obtain
\begin{equation}
\begin{split}
\|DT(g)\| & \leq \|D\nabla G\circ \widehat g\|_{R'-2\delta} 
 \leq \|D\nabla G\|_{R'-\delta} \\
& \leq \frac{2\pi+1}\delta \|\nabla G\|_{R'} 
\leq \frac{2\pi+1}\delta \|G\|'_{R'} <\xi.
\end{split}
\end{equation}

For the estimates in \eqref{estimates for g H} (the first is now immediate) we introduce the differentiable function
\begin{equation}
\begin{split}
f\colon\{z\in\Cc\colon |z|<\zeta\} & \to \A_{R'} \\
z & \mapsto H\circ(\id+z\Jj\nabla G(\id+z(\widehat g-\id)))
\end{split}
\end{equation}
where $\zeta=1/\xi\geq2$.
Cauchy's integral formula yields that
\begin{equation}
\begin{split}
\|H\circ g-H\|_{R'-2\delta} & = \|f(1)-f(0)\|_{R'-2\delta} \\
& \leq \frac1{2\pi} \oint_{|z|=\zeta}\frac{\|f(z)\|_{R'-2\delta}}{|z(z-1)|}dz \\
& \leq \frac1{\zeta-1}\sup_{|z|=\zeta}\|f(z)\|_{R'-2\delta}
\leq 2\xi \|H\|_{R'}.
\end{split}
\end{equation}
and
\begin{equation}
\begin{split}
\|H\circ g-H-\{H,G\}\|_{R'-2\delta} & = \|f(1)-f(0)-f'(0)\|_{R'-2\delta} \\
& \leq \frac1{2\pi} \oint_{|z|=\zeta}\frac{\|f(z)\|_{R'-2\delta}}{|z^2(z-1)|}dz \\
& \leq \frac1{\zeta(\zeta-1)}\sup_{|z|=\zeta}\|f(z)\|_{R'-2\delta}
\leq
2\xi^2 \|H\|_{R'}.
\end{split}
\end{equation}
By the implicit function theorem the maps $G\mapsto g$ and $G\mapsto H\circ g$ are analytic.
\end{proof}

\begin{lemma}\label{lemma inv FF for Ham}
Let $\sigma> 2 R_2 \|Q\|$, $\varepsilon'>0$ and $H\in\A'_R$ such that
\begin{equation}
\|H-H^0\|_R < \varepsilon' \leq
\frac{\sigma\delta}{(2\pi+1)[1+2\pi+(\tau+1)/R_2]}.
\end{equation}
Then there is $G\in\Ii^-\A'_{R'}$ such that 
\begin{equation}\label{eq step elim Ham}
\Ii^-(H+\{H,G\})=0
\quad\text{and}\quad
\|G\|'_{R'}\leq  \frac\delta{(2\pi+1)\varepsilon'} \|\Ii^-H\|_{R'}.
\end{equation}
Moreover, the map $H\mapsto G$ is analytic.
\end{lemma}

\begin{proof}
Consider the linear operator associated to $H$:
\begin{equation}\label{def FF}
\FF\colon  \Ii^-\A'_{R'} \to \Ii^-\A_{R'},
\qquad
K\mapsto\Ii^-\{H,K\}.
\end{equation}
It is well-defined since
\begin{equation*}
\begin{split}
\|\Ii^-\{H,K\}\|_{R'} 
&\leq
\|\nabla H\|_{R'} \|\nabla K\|_{R'}
\\
&\leq
\|H\|'_{R'} \|K\|'_{R'}.
\end{split}
\end{equation*}
We will show that $\FF^{-1}\colon\Ii^-\A_{R'}\to\Ii^-\A'_{R'}$ is bounded and
\begin{equation}\label{norm of operator FF-1 for H}
\|\FF^{-1}\| 
< \frac{1}{\frac{\pi R_2\sigma}{(2\pi+1)R_2+\tau+1}-2\frac{2\pi+1}{\delta}\varepsilon'}
\leq\frac{\delta}{(2\pi+1)\varepsilon'}.
\end{equation}

We start by decomposing any Hamiltonian $H=H^0+F$ as
$$
H(\vecx,\vecy)=\sum_{\veck}H_\veck(\vecy)\, \e^{2\pi\i\veck\cdot\vecx}
\quad\text{with}\quad
H_\veck(\vecy)=\sum_\vecnu H_{\veck,\vecnu} \vecy^\vecnu.
$$
Write $D_0=\nabla_2 H^0\cdot\nabla_1$,
with $\nabla_1$ and $\nabla_2$ standing for the derivatives with respect to $\vecx$ and $\vecy$.
The definition of $\FF$ in \eqref{def FF} yields
\begin{equation*}
\FF(K)
=\Ii^- ( \widehat F-D_0)\,K
=
-\left(
\Ii-\Ii^-\widehat FD_0^{-1} 
\right)
D_0K,
\end{equation*}
where $\widehat F (K)=\{F,K\}$. 
If the inverse of $\FF$ exists is given by
\begin{equation}\label{inverse of FF}
\FF^{-1}=-D_0^{-1}
\left(\Ii-\Ii^-\widehat FD_0^{-1}\right)^{-1}.
\end{equation}
The map $D_0^{-1}\colon \Ii^-\A_{R'} \to \Ii^-\A'_{R'}$ is linear and given by
$$
D_0^{-1}\,W(\vecx,\vecy)=
\sum\limits_{\veck\in\Zz^d-\{0\}}
\frac{W_{\veck}(\vecy)}
{2\pi\i(\veck\cdot\nabla_2 H^0(\vecy))}
\e^{2\pi \i\veck\cdot \vecx},
\quad
W\in\Ii^-\A_{R'}.
$$

 For each $\veck\in I^-$, using \eqref{def I- for H} and $\|Q\|<\sigma/(2R_2)$ thus $|\veck\cdot Q\vecy/\veck\cdot\vecomega|<1/2$,
\begin{equation}
\frac{W_\veck(\vecy)}
{\veck\cdot\vecomega\left(1+\frac{\veck\cdot Q\vecy}{\veck\cdot\vecomega}\right)}
=
\frac{W_\veck(\vecy)}
{\veck\cdot\vecomega}
\sum_{n\geq 0}
\left(-\frac{\veck\cdot Q\vecy}{\veck\cdot\vecomega}\right)^n,
\end{equation}
we get the estimate
\begin{equation}
\begin{split}
\left\|\frac{W_{\veck}} {\veck\cdot\nabla_2 H^0}\right\|_{R_2} 
& \leq 
\sum_{n\geq0}\sum_\vecnu\frac{|W_{\veck,\vecnu}| \, R_2^{\|\vecnu\|} \|Q\|^nR_2^n}{\sigma^{n+1}\|\veck\|}
\\
& 
<
\sum_{n\geq0}\sum_\vecnu\frac{|W_{\veck,\vecnu}| \, R_2^{\|\vecnu\|}}{\sigma\|\veck\|}
\left(\frac12\right)^n
\\
&
=\frac{2}{\sigma\|\veck\|} \|W_\veck\|_{R_2}.
\end{split}
\end{equation}
Similarly, we find the bound
\begin{equation}
\left\|\frac{\nabla_2W_{\veck}}{\veck\cdot\nabla_2 H^0}\right\|_{R_2}
\leq
\sum_\vecnu \frac{2\|\vecnu\|\,|W_{\veck,\vecnu}|R_2^{\|\vecnu\|-1}}{\sigma\|\veck\|}
<
\frac{2\tau}{\sigma R_2}\|W_\veck\|_{R_2}.
\end{equation}
Finally,
\begin{equation}
\left\|\frac{W_{\veck}Q\veck}{(\veck\cdot\nabla_2 H^0)^2}\right\|_{R_2}
< \frac{2}{\sigma R_2\|\veck\|}\|W_\veck\|_{R_2}.
\end{equation}

It is now immediate to see that
\begin{equation*}
\|D_0^{-1}W\|_{R'} \leq
\frac2{2\pi\sigma} \|W\|_{R'},
\quad\text{and}\quad
\|\nabla_1(D_0^{-1}W)\|_{R'}
\leq
\frac2{\sigma} \|W\|_{R'}.
\end{equation*}
Moreover,
$$
\nabla_2
\left( \frac{W_\veck(\vecy)}
{\veck\cdot\nabla_2H^0(\vecy)}
\right)=
\frac{\nabla_2 W_\veck(\vecy)}{\veck\cdot\nabla_2H^0(\vecy)} 
-\frac{W_\veck(\vecy)Q\veck}
{(\veck\cdot\nabla_2H^0(\vecy))^2}
$$
which implies
\begin{equation}
\|\nabla_2(D_0^{-1}W)\|_{R'}
<
\frac{\tau+1}{\pi \sigma R_2} \|W\|_{R'}.
\end{equation}
Hence, 
$$
\|D_0^{-1}\| < \frac2\sigma 
\left(1+\frac1{2\pi}+\frac{\tau+1}{2\pi R_2}\right).
$$

As $\widehat F\colon \Ii^-\A'_{R'}\to\A_{R'}$ with $\|\widehat F\|\leq 2\,\|\nabla F\|_{R'}\leq 2\frac{2\pi+1}{\delta}\|F\|_R$ (by Cauchy's estimate),
$$
\|\Ii^-\widehat F \,D_0^{-1}\| < 
\frac4\sigma 
\left(1+\frac1{2\pi}+\frac{\tau+1}{2\pi R_2}\right)
\|\nabla F\|_{R'}<1,
$$
and
$$
\left\|\left(\Ii-\Ii^-\widehat F\,D_0^{-1}\right)^{-1}\right\| < 
\left[1-\frac4\sigma 
\left(1+\frac1{2\pi}+\frac{\tau+1}{2\pi R_2}\right)
\|\nabla F\|_{R'}
\right]^{-1}.
$$
Thus $\FF^{-1}$ exists given by \eqref{inverse of FF} and the estimate \eqref{norm of operator FF-1 for H} on its norm follows immediately.

A solution of \eqref{eq step elim Ham} is simply given by $G=\FF^{-1}(-\Ii^-H)$.
Therefore, $\|G\|'_{R'} \leq \|\FF^{-1}\|\,\|\Ii^-H\|_{R'}$.
\end{proof}

Consider the pairs $R=(\rho_n+\nu,r')$ and $R'=(\rho_n,r)$, $\sigma>2r'\|Q\|$ and $H_0=H$ as given in Theorem \ref{theorem existence of g}.
We are going to iterate the procedure indicated in the previous lemmas.
Let a sequence of Hamiltonians be given by
$$
H_k=H_{k-1}\circ g_k, 
\quad 
k\in\Nn, 
$$
where $G_k$ and $g_k$ are determined for $H_{k-1}$ by Lemmas \ref{lemma inv FF for Ham} and \ref{lemma estimates on H step elim}, respectively.
In addition, denote by
\begin{equation}
g^{(k)}=g_1\circ \dots\circ g_k
\end{equation}
the composition of all symplectomorphisms up to the $k$th-step so that $H_k=H\circ g^{(k)}$.
In order to determine the right domains of $H_k$, $G_k$ and $g_k$, define the sequences
\begin{equation}
\begin{split}
R_k & = R_{k-1}-4\delta_k=R-4\sum_{i=1}^k\delta_i,
\\
R'_k & =R_{k-1}-\delta_k,
\end{split}
\end{equation}
with $R_0 =R$ and
\begin{equation}
\delta_k= \frac1{2^k} \min\left\{1,\frac\nu{2\pi},r'-r\right\}
\leq \frac1{2^k}.
\end{equation}
From now on, assume that 
\begin{equation}
\varepsilon'=
\min\left\{
\frac12\|H^0\|_R,
\frac{\sigma\delta_1}{(2\pi+1)(1+2\pi+\frac{\tau+1}{r'})} 
\right\}.
\end{equation}

\begin{lemma}\label{estimates G and g}
If for every $k\in\Nn$, $\|\Ii^- H_{k-1}\|_{R_{k-1}} \leq \varepsilon'/2$ and 
$$
\|G_{k}\|'_{R'_k}<\frac{\delta_k}{(2\pi+1)\varepsilon'}\|\Ii^- H_{k-1}\|_{R_{k-1}}, 
$$
then $g_k(\DD_{R_k})\subset\DD_{R_{k-1}}$ and
\begin{equation}\label{final est on gkk}
\begin{split}
\|g^{(k)}-\id\|_{R_k} & \leq \sum_{i=1}^k \frac{\delta_i}{\varepsilon'} \|\Ii^- H_{i-1}\|_{R_{i-1}} \\
\|g^{(k)}-g^{(k-1)}\|_{R_{k}} & \leq \frac{3}{2\varepsilon'} \|\Ii^- H_{k-1}\|_{R_{k-1}}.
\end{split}
\end{equation}
\end{lemma}

\begin{proof}
Recall Lemma \ref{lemma estimates on H step elim} for $\xi=\|\Ii^-H_{k-1}\|_{R_{k-1}}/\varepsilon'$ and check that
$$
\|g_k-\id\|_{R_k}\leq \|g_k-\id\|_{R'_k-2\delta_k}< \frac{\delta_k}{\varepsilon'} \|\Ii^- H_{k-1}\|_{R_{k-1}}
$$
and $R_k+\delta_k< R_{k-1}$ componentwise.
Now,
\begin{equation}
\begin{split}
g^{(k)}-\id  & = 
\sum_{i=1}^{k}g_i\circ\cdots\circ g_k -\sum_{i=2}^{k}g_i\circ\cdots\circ g_k-\id \\
& =
\sum_{i=1}^{k-1}(g_i\circ\cdots\circ g_k-g_{i+1}\circ\cdots\circ g_k)+g_k-\id \\
& =
\sum_{i=1}^{k-1}(g_i-\id)\circ g_{i+1}\circ\cdots\circ g_k+g_k-\id.
\end{split}
\end{equation}
Thus,
\begin{equation}
\|g^{(k)}-\id\|_{R_k} \leq \sum_{i=1}^{k}\|g_i-\id\|_{R_i}
\leq \sum_{i=1}^k \frac{\delta_i}{\varepsilon'} \|\Ii^- H_{i-1}\|_{R_{i-1}}.
\end{equation}
Furthermore, as
\begin{equation}
g^{(k)}-g^{(k-1)} = 
(g^{(k-1)}-\id)\circ g_{k} -(g^{(k-1)}-\id)+(g_k-\id)
\end{equation}
we get
\begin{equation}
\begin{split}
\|g^{(k)}-g^{(k-1)}\|_{R_{k}} \leq &
(\|Dg^{(k-1)}-I\|_{R_{k}}+1)\,\|g_{k}-\id\|_{R_{k}}
\\
\leq &
\frac{\delta_k}{\varepsilon'}
\|\Ii^- H_{k-1}\|_{R_{k-1}}
\left(\frac{2\pi+1}{4\delta_k}
\sum_{i=1}^{k-1} \frac{\delta_{i}}{\varepsilon'}\|\Ii^- H_{i-1}\|_{R_{i-1}}+1\right)
\\
\leq &
\frac1{\varepsilon'}\left(1+\frac12\right) \|\Ii^- H_{k-1}\|_{R_{k-1}}.
\end{split}
\end{equation}
\end{proof}

Notice that since $\varepsilon'\leq \frac12\|H^ 0\|_R$, we have
\begin{equation}
\varepsilon'
\leq \|H^0\|_R-\varepsilon'\leq \|H\|_R 
\leq \|H^0\|_R +\varepsilon'
\end{equation}
and also
\begin{equation}
\frac12\|H^0\|_R \leq \|H\|_R \leq \frac32\|H^0\|_R.
\end{equation}

\begin{lemma}\label{estimates on Hk and gk}
For any $k\in\Nn$, if $\|\Ii^-H\|_R \leq {\varepsilon'}^2/(8\|H\|_R)$, then
\begin{eqnarray}\label{final est on Hkk 1}
\|\Ii^- H_k\|_{R_k} & \leq  &
\left(\frac{4\|H\|_R}{{\varepsilon'}^2}\right)^{2^k-1} \|\Ii^-H\|_R^{2^k}
\leq\frac{\varepsilon'}{2},
\\ \label{final est on Hkk 2}
\|H_{k}-H_{k-1}\|_{R_k} & \leq &
\frac{4 \|H\|_R}{\varepsilon'} \|\Ii^-H_{k-1}\|_{R_{k-1}},
\\\label{final est on Hkk 3}
\|H_k\|_{R_k} & \leq & 2\|H\|_R.
\end{eqnarray}
\end{lemma}

\begin{proof}
We will prove the above inequalities by induction.
The generating Hamiltonian $G_1$ given by Lemma \ref{lemma inv FF for Ham} and the symplectomorphism $g_1$ by Lemma \ref{lemma estimates on H step elim} satisfy $\|G_1\|'_{R'_1}\leq \delta_1\|\Ii^-H\|/[(2\pi+1)\varepsilon']$, $\|g_1-\id\|_{R'_1-2\delta_1}<\|\Ii^-H\|_R\delta_1/\varepsilon'$ and $\Ii^-H_{1} = \Ii^-H\circ g_{1} - \Ii^-(H+\{H,G_{1}\})$.
Hence,
\begin{equation}
\|\Ii^-H_{1}\|_{R_{1}}  \leq 
\|H\circ g_{1}-H_k-\{H,G_{1}\}\|_{R_{1}}  \leq
2\left(\frac{\|\Ii^-H\|_R}{\varepsilon'}\right)^2\|H\|_R.
\end{equation}
and
\begin{equation}
\|H_1-H\|_{R_1} \leq 
\|\nabla H\|_{R_1} \|g_1-\id\|_{R_1}
\leq \frac{2\pi+1}{4\varepsilon'} \|\Ii^-H\|_R \|H\|_R
\leq \frac{2}{\varepsilon'} \|\Ii^-H\|_R \|H\|_R.
\end{equation}
Thus, \eqref{final est on Hkk 1} and \eqref{final est on Hkk 2} are valid for $k=1$ and so is \eqref{final est on Hkk 3} because $\|H_1\|_{R_1}\leq\|H_1-H\|_{R_1}+\|H\|_R$.

Now, assume that the inequalities are true for $k$.
Under these conditions, Lemma \ref{lemma inv FF for Ham} guarantees the existence of $G_{k+1}$ so that
\begin{equation}
\|G_{k+1}\|'_{R_{k+1}} 
\leq 
\frac{\delta_{k+1}}{(2\pi+1)\varepsilon'}\|\Ii^-H_k\|_{R_k}
\end{equation}
and Lemma \ref{lemma estimates on H step elim} yields $g_{k+1}$.
Therefore, $\Ii^-H_{k+1} = \Ii^-H_k\circ g_{k+1} - \Ii^-(H_k+\{H_k,G_{k+1}\})$ and 
\begin{equation}
\begin{split}
\|\Ii^-H_{k+1}\|_{R_{k+1}} 
& \leq 
\|H_k\circ g_{k+1}-H_k-\{H_k,G_{k+1}\}\|_{R_{k+1}}
\\
& \leq
2 \left(\frac{\|\Ii^-H_k\|_{R_k}}{\varepsilon'}\right)^2 \|H_k\|_{R_k} \\
& \leq
\left(\frac{4\|H\|_R}{{\varepsilon'}^2}\right)^{2^{k+1}-1} \|\Ii^-H\|_R^{2^{k+1}}.
\end{split}
\end{equation}
Similarly,
\begin{equation}
\begin{split}
\|H_{k+1}-H_k\|_{R_{k+1}} \leq &
\|\nabla H_k\|_{R_{k+1}}\|g_{k+1}-\id\|_{R_{k+1}}
\\
\leq &
\frac{2\pi+1}{4\delta_{k+1}\varepsilon'} \|\Ii^-H_k\|_{R_k} \delta_{k+1} \|H_k\|_{R_k}
\\
\leq &
\frac4{\varepsilon'} \|\Ii^-H_k\|_{R_k} \|H\|_R.
\end{split}
\end{equation}
Finally, making use of the above inequality,
\begin{equation}
\begin{split}
\|H_{k+1}\|_{R_{k+1}} 
& \leq
\|H\|_R+
\sum_{i=1}^{k+1}\|H_i-H_{i-1}\|_{R_{k+1}} \\
&\leq
\|H\|_R + \frac{4\|H\|_{R}}{\varepsilon'} \sum_{i=1}^{k+1} \|\Ii^-H_{i-1}\|_{R_{i-1}}  \\
&\leq
\|H\|_R + 
\|H\|_R 
\sum_{i=1}^{k+1} 
\left(\frac{4\|H\|_R\|\Ii^-H\|_R }{{\varepsilon'}^2}\right)^{2^{i-1}} 
\\
& \leq 
\left(1+\frac12+\sum_{i=1}^{k}\frac1{4^i}\right)
\|H\|_R
<
2\|H\|_R.
\end{split}
\end{equation}
\end{proof}

Theorem \ref{theorem existence of g} will now be a consequence of the following result.

\begin{theorem}
If 
\begin{equation}
\|H-H^0\|_R<\varepsilon=\frac{\varepsilon'^2}{12\|H^0\|_R} 
\leq \frac{\varepsilon'^2}{8\|H\|_R},
\end{equation}
then there exists $g=\lim_{k\to+\infty}g^{(k)}\in\GG_{R'}$ such that $\Ii^-H\circ g=0$ on $\DD_{R'}$.
Furthermore, the maps $\fG\colon H\mapsto g$ and $\UU\colon H\mapsto H\circ g$ are analytic, and
\begin{eqnarray}\label{final estimates for g and H' 1}
\|g-\id\|_{R'} & \leq  & \frac1{\varepsilon} \|\Ii^-H\|_R  
\\ \label{final estimates for g and H' 2}
\|H\circ g-H^0\|_{R'} & \leq  & \left(1+\sqrt\frac{12\|H^0\|_R}{\varepsilon}\right) \|H-H^0\|_R.
\end{eqnarray}
\end{theorem}

\begin{proof}
Lemmas \ref{estimates G and g} and \ref{estimates on Hk and gk} imply that the sequence $g^{(k)}$ converges to a map $g\colon\DD_{R'}\to\DD_{R}$ which is analytic and symplectic, and $H_\infty=\lim_{k\to+\infty} H_k=H\circ g$.
Moreover, $\Ii^-H\circ g=\Ii^-H_\infty=0$.
Since the convergence is uniform, the maps $H\mapsto g$ and $H\mapsto H\circ g$ are analytic.

Notice that
\begin{equation}\label{ineq for sums}
\begin{split}
\sum_{i=1}^{+\infty} 
\left(\frac{4\|H\|_R\|\Ii^-H\|_R }{{\varepsilon'}^2}\right)^{2^{i-1}}
& \leq
\frac{4\|H\|_R\|\Ii^-H\|_R }{{\varepsilon'}^2}
+\sum_{i=1}^{+\infty} 
\left(\frac{4\|H\|_R\|\Ii^-H\|_R }{{\varepsilon'}^2}\right)^{2i}
\\
& \leq 
\left(1+
\frac{16\|H\|_R}{3{\varepsilon'}^2}\|\Ii^-H\|_R\right)
\frac{4\|H\|_R}{{\varepsilon'}^2} \|\Ii^-H\|_R
\\
&\leq 
\frac{20\|H\|_R}{3{\varepsilon'}^2} \|\Ii^-H\|_R 
\leq \frac1\varepsilon \|\Ii^-H\|_R.
\end{split}
\end{equation}

The inequality in \eqref{final estimates for g and H' 1} follows by taking the limit $k\to+\infty$ in \eqref{final est on gkk}.
That is,
\begin{equation}
\|g-\id\|_{R'}  \leq
\sum_{i=1}^{+\infty} 
\frac{\delta_i}{\varepsilon'} \|\Ii^-H_{i-1}\|_{R_{i-1}}
\leq
\frac1{\varepsilon}\|\Ii^-H\|_R.
\end{equation}

Now,
\begin{equation*}
\begin{split}
\|H_\infty-H^0\|_{R_k} & \leq \|H-H^0\|_R +\sum_{i=1}^{+\infty}\|H_i-H_{i-1}\|_{R_i}
\\
& \leq \left(1+\sqrt\frac{12\|H^0\|_R}{\varepsilon}\right) \|H-H^0\|_R,
\end{split}
\end{equation*}
where we have used Lemma \ref{estimates on Hk and gk} and the fact that $\|\Ii^-H\|_R\leq\|H-H^0\|_R$.
\end{proof}

\end{appendix}

\section*{Acknowledgements} 

We would like to express our gratitude to D. Kleinbock, H. Koch, R. S. MacKay, G. Margulis, Ya. Sinai and J.-C. Yoccoz for useful discussions and comments.

JLD was supported by Funda\c c\~ao para a Ci\^encia e a Tecnologia, and
JM by an EPSRC Advanced Research Fellowship.
We would also like to thank the Isaac Newton Institute, CAMGSD/IST and Cemapre/ISEG (through FCT's Program POCTI/FEDER) for travel support.


\end{document}